\documentclass[final,1p,times,authoryear]{elsarticle}

\usepackage{amssymb}
\usepackage{amsmath}

\usepackage{lineno}
\usepackage{siunitx}
\usepackage{url}
\newcommand{\vect}[1]{\boldsymbol{#1}}

\usepackage{pdflscape}

\journal{Icarus}

\begin{document}

\begin{frontmatter}




\title{A Statistical Approach to Quantifying Uncertainty in Meteoroid Physical Properties}


\author[UWO,WS]{Maximilian Vovk\corref{cor1}}\ead{mvovk@uwo.ca}

\author[UWO,WS]{Denis Vida}\ead{dvida@uwo.ca}

\author[UWO,WS]{Peter G. Brown}\ead{pbrown@uwo.ca}

\cortext[cor1]{Corresponding author}

\affiliation[UWO]{organization={Department of Physics and Astronomy, University of Western Ontario},
            addressline={1151 Richmond Street}, 
            city={London},
            postcode={N6A 3K7}, 
            state={Ontario},
            country={Canada}}

\affiliation[WS]{organization={Institute for Earth and Space Exploration, University of Western Ontario},
            addressline={Perth Drive}, 
            city={London},
            postcode={N6A 5B8}, 
            state={Ontario},
            country={Canada}}

\begin{abstract}

\paragraph*{\textbf{Importance:}}
Meteoroid bulk density is a critical value required for assessing impact risks to spacecraft, informing shielding and mission design.

\paragraph*{\textbf{Research Gap:}}
Direct bulk density measurements for sub-millimeter to millimeter-sized meteoroids are difficult, often relying on forward modeling without robust uncertainty estimates. Methods based solely on select observables can overlook noise-induced biases and non-linear relations between physical parameters.

\paragraph*{\textbf{Objective:}}
This study aims to automate the inversion of meteoroid physical parameters from optical meteor data, focusing on bulk density and its associated uncertainties.

\paragraph*{\textbf{Methodology:}}
We compare an observables-based selection method (PCA) with an RMSD-based approach used to select among millions of ablation model runs using full light and deceleration curves as constraints. After validating both approaches on six synthetic test cases, we apply them to two Perseid meteors recorded by high sensitivity Electron-Multiplied CCD (EMCCD) cameras and high precision mirror-tracked meteors detected by the Canadian Automated Meteor Observatory (CAMO).

\paragraph*{\textbf{Key Findings:}}
Our results show that relying only on observables, as in the PCA approach can converge to wrong solutions and can yield unphysical solutions. In contrast, the RMSD-based method offers more reliable density constraints, particularly for bright and strongly decelerating meteor. Small relative measurement precision in brightness and lag relative to the full range of observed lag and luminosity is the key to tight solution.

\paragraph*{\textbf{Implications:}}
We provide the first objectively derived uncertainty bounds for the physical properties of meteoroids. Our approach solves the solution degeneracy problem inherent in forward modelling of meteors.  This strategy can be generalized to other showers, paving the way for improved meteoroid models and enhanced spacecraft safety.

\end{abstract}

\begin{graphicalabstract}
\includegraphics[width = \textwidth]{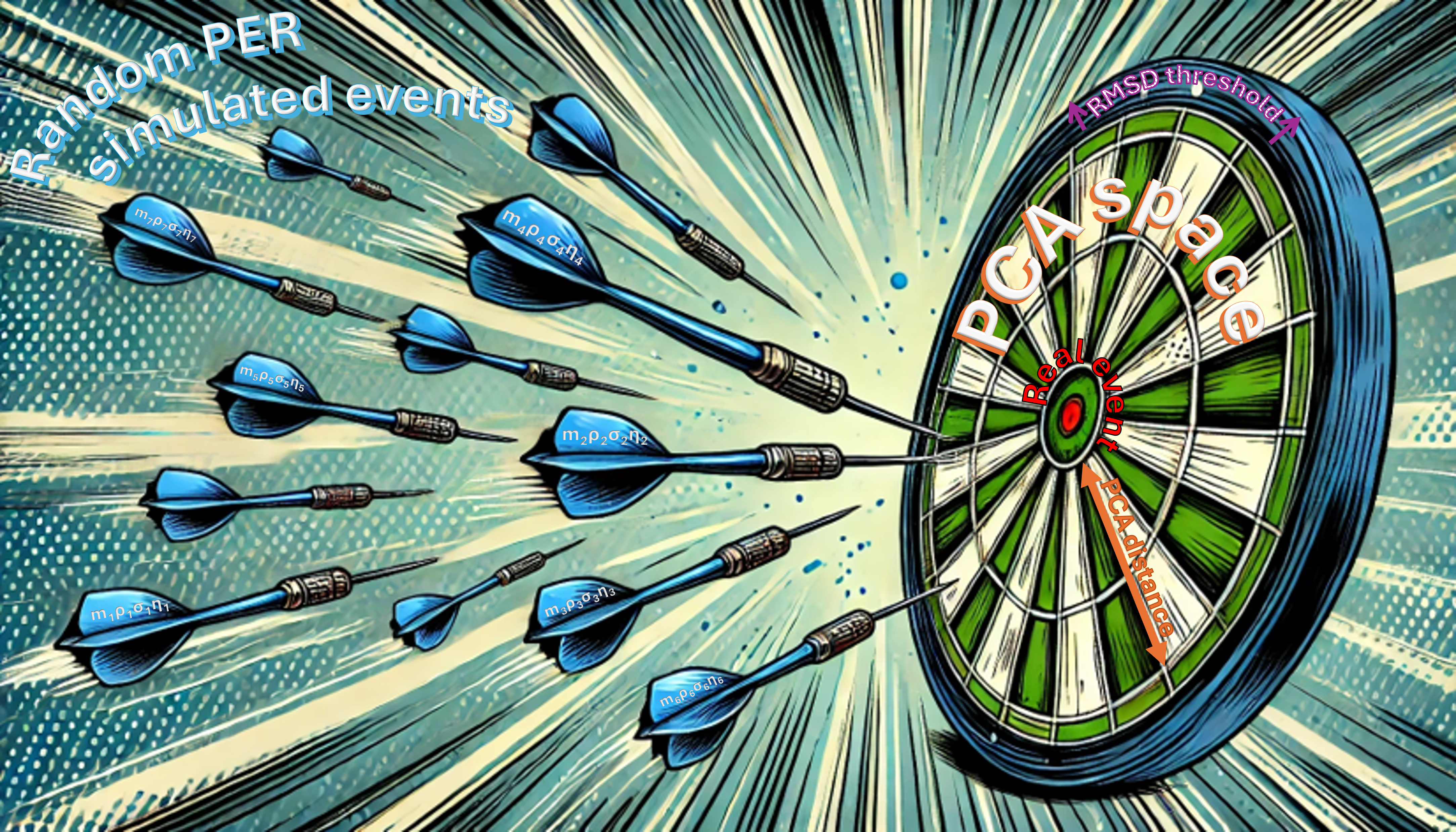}
\end{graphicalabstract}

\begin{highlights}
\item New statistical approach applied to faint meteor observations to invert physical properties of sub-mm meteoroids to improve spacecraft risk assessment
\item Demonstrated that relying solely on a linear combination of observables is insufficient for accurate bulk density estimation
\item Automated meteor ablation model used to estimate meteoroid bulk density through fits to meteor light-curve and deceleration measurements
\item First objective uncertainty bounds from individual meteor observations applied to recover original meteoroid physical characteristics applied to Perseid meteors
\end{highlights}


\end{frontmatter}


\section{Introduction}\label{sec:introduction}




Meteoroid impacts on spacecraft are a serious hazard to long-term space operations. A typical operational satellite will be hit by a larger meteoroid ($>0.2$~mm in size) every few years, potentially leading to severe damage as such impacts carry the equivalent kinetic energy to a small caliber rifle bullet. Such impacts can sever wires, damage spacecraft components, and even compromise spacesuit integrity \citep{moorhead2019meteoroid}.

The penetration ability of a hyper-velocity meteoroid impact is quantified through use of ballistic-limit equations \citep{Christiansen2001}. Examining widely used Ballistic Limit Equations (BLEs) it is meteoroid velocity, mass and bulk density which most determine resulting damage \citep{moorhead2020nasa}. Accurate measurement of these variables, along with an understanding of their associated uncertainties, is vital for developing effective protective measures against meteoroid impacts.

Sub-mm to mm-sized meteoroids are too small to be observed directly in interplanetary space, making in-situ measurements challenging. Instead, their speed and density must be inferred from the faint light and electrons produced as they enter Earth's atmosphere \citep{Ceplecha1998e}. 


Estimating the pre-atmospheric speed of a meteoroid is commonly done through the measurement of its entry trajectory. This can be achieved through common optical observations at widely separated stations \citep{Vida2020theory}. Among the many optical systems employed worldwide for this purpose \citep{koten2019meteors}, the Canadian Automated Meteor Observatory (CAMO) system is among the most precise in that it achieves velocity measurement uncertainties of order 10 m/s and trajectory residuals under a meter \citep{vida2021high} for mm to cm-sized meteoroids.

In contrast to the relatively high precision with which velocity can be recorded, estimating the bulk density of meteoroids presents a more complex challenge because it cannot be directly observed. Instead, it must be inferred through ablation models that fit light curve and velocity data \citep{popova2019modelling}. 

Meteoroid ablation at mm sizes and larger is driven by fragmentation. High-resolution imagery of small meteors shows that $\geq$90\% of faint meteors \citep{Subasinghe2016} show explicit evidence of fragmentation. The remaining small minority likely also experience fragmentation but at levels below the instrument detection floor. These modern observations refute the classical notion that meteoroid ablation proceeds on an atom-by-atom basis through pure vaporization or melting \citep{opik1958physics}.

Most modern numerical small meteoroid ablation models are based on the classic dust-ball fragmentation model of \cite{Hawkes1975}. These include the thermal erosion model of \cite{Campbell-Brown_Koschny_2004} (where all grains are released at once), and the erosion fragmentation model of \cite{borovivcka2007atmospheric} (quasi-continuous release of grains over time). Both of these variants of the dust-ball model attempt to explicitly incorporate fragmentation. These models are typically fit through forward modeling of ablation parameters to match measurements. However, due to the complexity of the model, rigorous uncertainty quantification is difficult to achieve \citep{vida2024first,buccongello2024physical}. Thus far, this approach has not produced a simple means to estimate uncertainty in the inverted meteoroid physical quantities which include mass, bulk density, ablation and erosion coefficients and grain distributions. Most importantly past approaches do not address the issue of solution uniqueness and degeneracy.


Although many past efforts have attempted to apply numerical model fitting to measured meteor data (see \citep{popova2019modelling} for a summary) very few have attempted to automate the process. Automation of the fitting process has the potential to significantly enhance the precision of uncertainty quantification and ensure the uniqueness of fits. However, automation is challenging due to the complexity of the problem which involves choosing an appropriate cost function, a powerful optimization algorithm, and requires significant computational resources.

Only a few prior works have attempted automated meteoroid ablation model fits. Among the earliest was that of \cite{Kikwaya_2011} who employed a brute-force grid search approach using the \cite{Campbell-Brown_Koschny_2004} thermal ablation model to fit observed light curves and deceleration data of meteors. This method struggled with accurately modeling grain distributions, particularly as it was based on observational data of lower metric accuracy than what is available today, where in many cases deceleration could not be measured at all. The simple grid search approach to optimization, while in theory rigorous, was very slow and prone to finding local minima due to the sparse coverage of the multidimensional parameter space.

More recently, \cite{Henych_Borovička_Spurný_2023} and \cite{Tarano_Wheeler_Close_Mathias_2019} introduced global optimization tools using genetic algorithms (GA) for semi-automatic modeling and estimation of fireball fragmentation and physical characteristics. However, these methods were primarily designed to work on fireballs and not on mm-sized meteoroids. 

Building on these advancements, \cite{vida2024first} developed an automatic local optimization method based on the erosion fragmentation model of \cite{borovivcka2007atmospheric}. The method relies on a good initial guess for the meteoroid properties and uses a Nelder-Mead optimizer for iterative refinement to achieve an optimal fit. However, it still depends on a manual and subjective step of performing an initial parameter estimate and faces challenges with local minima and the construction of a reliable cost function.

Broadly speaking, despite all previous advancements, no focused research has solely been dedicated to rigorously defining the uncertainties in the physical parameters of meteoroids. Our study addresses this critical gap by being the first to specifically target and quantify uncertainties in the physical meteoroid parameters estimated through ablation model comparison to observations.

Previous work, such as that conducted by \cite{buccongello2024physical}, employed high-resolution optical measurements by the Canadian Automated Meteor Observatory's (CAMO) telescopic tracking system \citep{weryk2013canadian, vida2021high} as inputs for modeling. Despite the high measurement accuracy, estimating uncertainties proved challenging due to the manual and subjective forward modeling approach used. The reliance on manual fitting by a few analysts led to relatively coarse uncertainty estimates, highlighting the need for more precise methods.

In this paper, we present two novel probabilistic methods for estimating uncertainties of the physical characteristics of meteoroids. The first method utilizes Principal Component Analysis (PCA) \citep{dunteman1989PCA} to infer physical meteoroid properties based on observable variables, while the second is a brute-force approach that fits the full light-curve and deceleration profiles.  
We use Electron Multiplying Charge-Coupled Device \citep[EMCCD; ][]{vida2020new, gural2022development, Mazur2023} cameras and the mirror tracking system of CAMO as data sources. EMCCD cameras provide video observations of meteors down to a detection limit of magnitude +8. Complementing the high sensitivity of EMCCD observations are the high temporal and spatial resolution of the mirror tracking system of CAMO. When combined, these two optical systems provide ultra sensitive, high cadence and high precision measurements of individual meteors.

To estimate the physical properties of a meteoroid, we first generate synthetic meteor observations using the ablation model of \cite{borovivcka2007atmospheric} and known observational parameters from the EMCCD cameras. Observed quantities such as speed and zenith angle are held fixed to match the distribution of real meteor data collected by the EMCCD systems, while the unknown meteoroid characteristics (e.g., density, ablation coefficients) are sampled over a broad parameter space. In the brute-force (RMSD) method, we generate simulations until a prescribed number of them remain under a specified Root Mean Square Deviation (RMSD) threshold measured between model predictions and observations of meteor light curve and deceleration profiles. 

In the PCA method, we fit each synthetic simulation’s light curve and deceleration profile through parametrization of these “observables” that characterize their shapes. Principal Component Analysis (PCA) then identifies synthetic cases that lie close (in PCA space) to the real meteor, normalized by each observable’s measurement uncertainty. In both approaches, the set of accepted synthetic meteors represents plausible solutions for the true meteoroid parameters within measurement uncertainty. By resampling noise multiple times, we identify the full range of solutions in physical-parameter space consistent with the observations.


The validation of the two methods has been performed on six representative model generated, synthetic Perseid meteors for which physical properties are known. The synthetic meteors were simulated to represent a range of Perseid speed, masses, and entry angles consistent with our EMCCD observational datatset. Upon successful validation, we applied our method to 2 real observed Perseids captured simultaneously by EMCCD and CAMO mirror tracking cameras for which detailed measurement uncertainty could be directly recovered from observations.

We use the Perseid meteor shower as a case study. The shower was selected due to the availability of extensive observational EMCCD data, recently constrained density estimates \citep{buccongello2024physical} using the same ablation model, and the threat that they pose to spacecraft in orbit due to their high activity and kinetic energies \citep{Moorhead2019a}.

This paper is organized as follows: in Section \ref{sec:Methods}, we outline the methodology and show the validation of the two methods, Section \ref{sec:results} presents the results of applying the method to real Perseids and for the first time demonstrates statistically robust uncertainties of meteoroid density measurement. In Section \ref{sec:discussion}, we examine the implications of these findings, including the rationale for considering observable or curve fitting as a mean of physical uncertainties estimation, and gave a rule-of-thumb to define when the density can be estimated and when not. Finally, in Section \ref{sec:conclusions}, we summarize the key conclusions of the study and propose directions for future research.

\section{Methods} \label{sec:Methods}





Just like all previous forward-modeling approaches, our methods are based on the core assumption that a meteor’s measurable behavior in the atmosphere is uniquely related to the key physical characteristics of the meteoroid. In a further step, we assume that the observables such as the meteor heights (beginning, peak, terminal), speed, light curve shape, and deceleration have a direct connection to the meteoroid's mass, bulk density, ablation characteristics, and structure. We recognize that this connection is not straightforward nor is it guaranteed to be unique and thus we employ advanced algorithms to explore the statistical space of allowable solutions.

\begin{figure}[h!]
\centering
\includegraphics[width=0.8\linewidth]{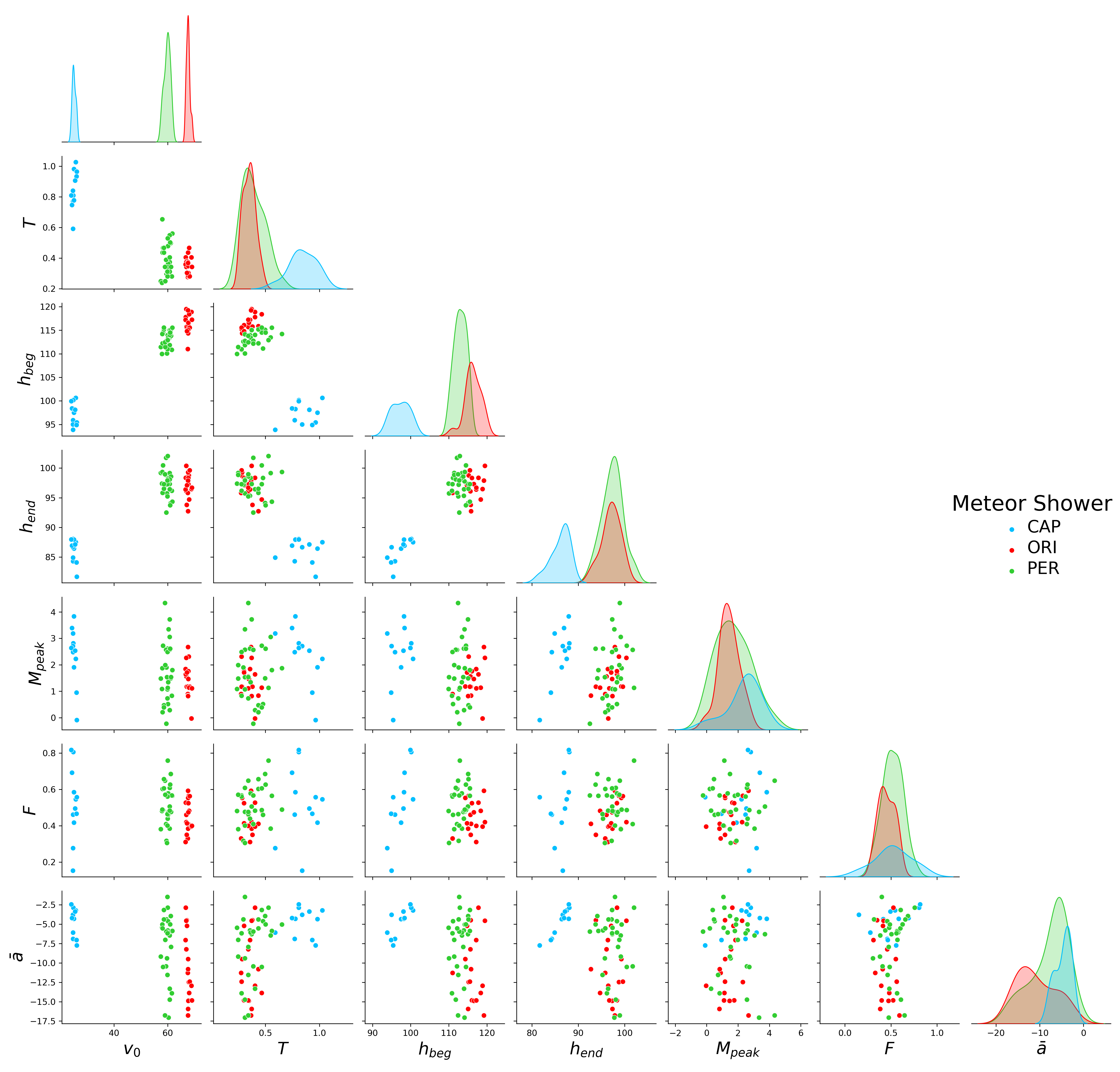}
\caption{Meteors as recorded by the CAMO EMCCD systems associated with the Perseids, Orionids, and Alpha Capricornids and some of their observable characteristics: initial velocity $v_0$ (km/s), total duration $T$ (s), beginning height $h_{beg}$ (km), end height $h_{end}$ (km), peak absolute magnitude $M_{peak}$, the F parameter, and the average deceleration $\bar a$ (km/s$^2$).}
\label{img:EPER_CAP_ORI}
\end{figure}

Figure \ref{img:EPER_CAP_ORI} shows a correlation matrix demonstrating how some of these measured parameters from our EMCCD observations vary among individual meteors from three select meteor showers. For example, although both are fast showers, the Orionids and the Perseids, have drastically different decelerations. On average, the Orionids decelerate twice as much as the Perseids, despite in addition having similar peak magnitudes and total durations. The Orionids also have higher beginning heights. Taken together these are strong indications that Orionid meteoroids are more fragile and ablate more vigorously than the Perseids. Alpha Capricornids, which are slower than the other two showers, decelerate the least but also have much lower beginning and end heights, opening up questions about how to calibrate the observed values in an absolute sense and provide unbiased estimates of meteoroid physical properties.

That there is distinct distribution of the observables among stream members is a result known from many earlier studies \citep[e.g][]{Jenniskens2023}. This is fundamental in supporting our core assumption: that from the observable characteristics of individual meteors we can estimate each individual meteoroids allowable range of physical properties such as mass, density, and internal structure consistent with measurement uncertainty.

The same assumption about the connection of observable and relative physical properties of meteoroids was made previously by other authors through the introduction of parameters like $k_B$, PE, and $k_c$. The $k_B$ parameter by \citet{ceplecha1967classification} was one of the early attempts to link the beginning height of meteors to their physical strength. This parameter, defined as 
\begin{equation}\label{eq:kB_eq}
    k_B = \log \rho_B + \frac{5}{2} \log v_\infty - \frac{1}{2} \log \cos z_R,
\end{equation}
combines the atmospheric density ($\rho_B$) at the meteor's beginning height, pre-deceleration speed ($v_\infty$), and the zenith angle ($z_R$). By analyzing the distribution of the $k_B$ parameter, \citet{ceplecha1967classification} identified groups that suggested differences in meteoroid strength and structure, although the sensitivity of the observational system biases the parameter's absolute value \citep{borovivcka2019physical}.

Similarly, \citet{ceplecha1976fireball}’s PE parameter focused on large meteoroids by analyzing their end heights and linking them to physical properties like density and strength. The PE parameter is defined as 
\begin{equation}\label{eq:PE_eq}
    PE = \log \rho_E - 0.42 \log m_\infty + 1.49 \log v_\infty - 1.29 \log \cos z_R,
\end{equation}
where $\rho_E$ is the density of the atmosphere at the fireball's end height, $v_\infty$ is the entry speed, and $m_\infty$ is the initial mass. This parameter allowed for the classification of fireballs into different types, suggesting variations in their structural strength and density.

\citet{jenniskens2016cams_kc} refined the $k_b$ concepts by introducing the $k_c$ parameter, which aimed to improve the classification of meteoroids based on their beginning height and entry speed. The $k_c$ parameter, defined as 
\begin{equation}\label{eq:kc_eq}
    k_c = H_b + \frac{2.86 - 2.00 \log v_\infty}{0.0612},
\end{equation}
where $H_b$ is the beginning height and $v_\infty$ is the entry speed, providing a cleaner separation of different meteoroid groups, particularly for lower-speed meteors. Unlike the $k_B$ parameter, which assumes the meteoroid surface temperature follows the theoretical $v^{2.5}_\infty$ dependence, the $k_c$ parameter follows a $v^2_\infty$ dependence, closer to the observed dependence.  

These quantitative classification systems are well established and have been used to normalize the observables and roughly sort meteoroids into compositional groups (e.g. asteroidal, carbonaceous, cometary). In this paper, we aim to push this assumption further by directly inverting absolute values of physical properties of individually observed meteoroids solely based on these observables through a rigorous numerical approach.

In our observation-driven model, we employ an unconventional use of Principal Component Analysis (PCA), which is commonly applied in planetary science for asteroid spectra. In contrast to that approach which only looks at a 1D series of data to find spectra that match best, we use a multidimensional PCA approach to explore the variance in a dataset of synthetic meteors with different physical parameters—such as density, mass, and ablation coefficient—and their impact on observable characteristics like light curves and deceleration profiles. By maximizing the variance explained by each principal component, we reduce the dimensionality of input data, generating new linear combinations of variables that help us identify the most likely physical parameters of a given meteor. This restricts the phase space we need to search to find good solutions for a particular event. 

This approach results in a metric function which gives a single number which indicates how similar two meteors are, enabling an automated search through thousands of simulations to identify the ones which most closely match the observed meteor, without the need to manually define a cost function. PCA helps to identify which simulated meteors are statistically most similar based on key observable parameters. By selecting the principal components that explain the most variance, the simulations are refined by improving the accuracy of the explored physical parameters. We will also compare this approach with a brute-force match of observed vs. modelled individual meteor light curves and decelerations through an RMSD value.

\subsection{Equipment, Data Collection and Reduction} \label{subsec:obs}

\begin{figure}[h!]
\centering
\includegraphics[width=0.45\linewidth]{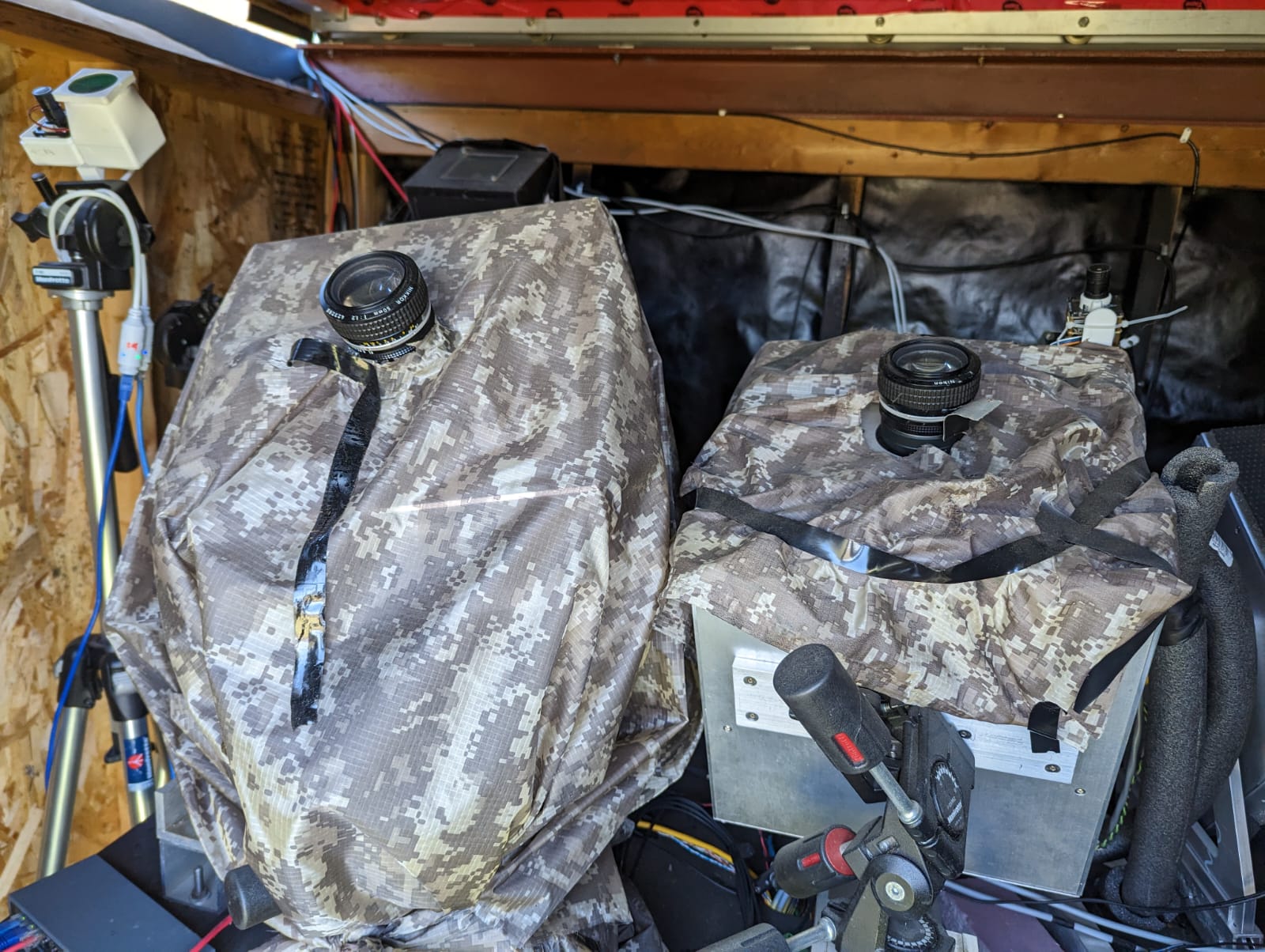}
\caption{Two EMCCD cameras at Elginfield observatory positioned at 40-degree (left) and 70-degree (right) elevations.}
\label{img:EMCCDcamera}
\label{img:CAMOcamera}
\end{figure}

For this study, we make use of two pairs of Electron Multiplying Charge-Coupled Device (EMCCD) cameras which are part of the Canadian Automated Meteor Observatory facility. The Western Meteor Physics Group (WMPG) operates four N\"uv\"u HN\"u1024 EMCCD cameras\footnote{https://www.nuvucameras.com/products/hnu-1024/} at two sites (Elginfield 43.19279$^{\circ}$ N, 81.31565$^{\circ}$ W; and Tavistock 43.26420$^{\circ}$ N, 80.77209$^{\circ}$ W), both located in Southwestern Ontario and separated by $\sim45$~km. The EMCCD cameras are binned 2x2 resulting in a video resolution of $512 \times 512$~px. Each camera is equipped with a 50 mm f/1.2 Nikkor lens, providing a $15^\circ \times 15^\circ$ field of view. They are operated at 32 frames per second and are capable of detecting faint meteors with a limiting peak magnitude of +6 and a per-point detection limit of +8 \citep{vida2020new}. As shown in Figure \ref{img:EMCCDcamera}, the camera pair labeled "F" is positioned at a 70-degree elevation to capture meteors above 90 km, while the "G" set, angled at 40 degrees, targets meteors between 70 km and 120 km in altitude. 

Operationally, the EMCCD cameras automatically schedule data capture in circular video buffers every night when sky conditions are good (dark, no clouds, no moon). An automated meteor detection algorithm is applied to the recorded video files after nighttime observations using a hybrid threshold/matched filter algorithm as described in \citet{gural2022development}.

Between 2018 and 2023, the EMCCD cameras detected 404 multi-station Perseids (PER). From the initial automated solutions associated with these detections a more refined solution with covariances was produced using the open source software library\footnote{WesternMeteorPyLib: \url{https://github.com/wmpg/WesternMeteorPyLib}} which is the implementation of the Monte Carlo trajectory solver by \cite{Vida2020theory}. 

An examination of trajectory solutions computed using the measurements produced by the automated matched filter detector showed that the algorithm sometimes produced unphysical deceleration profiles due to it forcing an empirical model of meteor motion across the field of view. We therefore manually reduced all meteors used in our study  using the SkyFit2 software\footnote{Raspberry Pi Meteor Station library: \url{https://github.com/CroatianMeteorNetwork/RMS}}. Manual reduction consisted of estimating the position of the meteor for each observed frame using a weighted centroid approach and then manually masking the visible meteor region for photometric measurement. The astrometric/photometric calibration process is described in \citet{vida2021global}.

\begin{figure}[h!]
\centering
\includegraphics[width=1\linewidth]{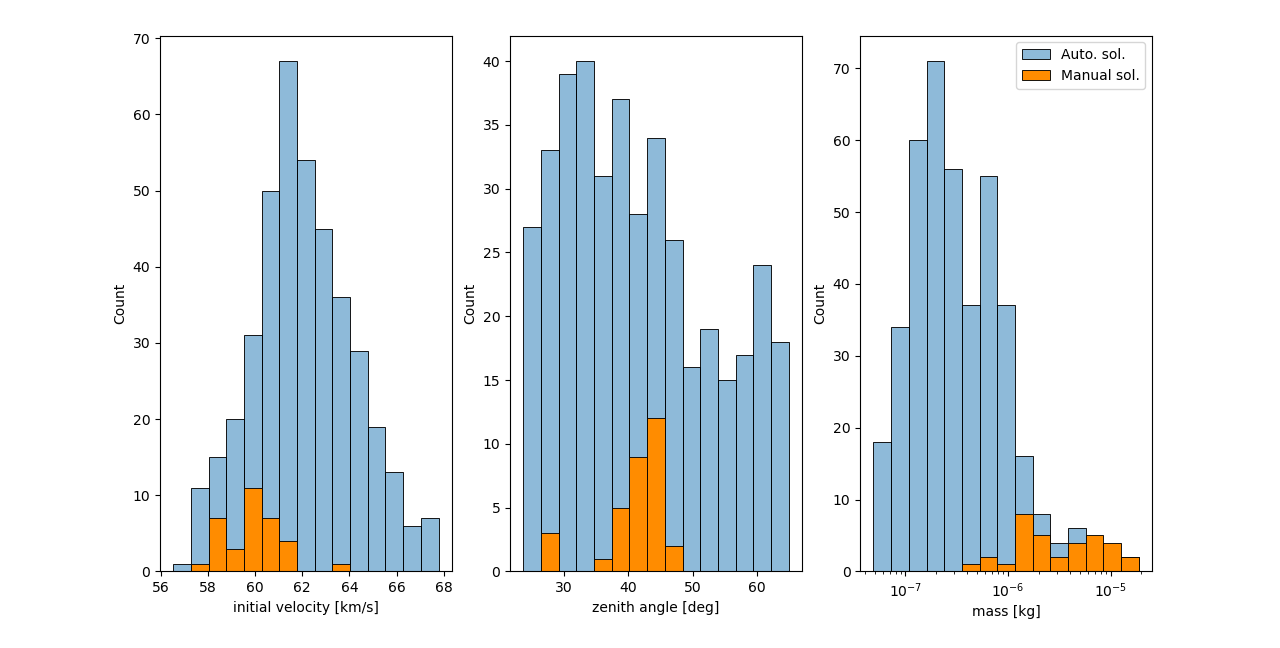}
\caption{Distribution of initial velocity, radiant zenith angle and mass for 408 PER meteors recorded by the EMCCD system between 2018-2023 (blue vertical bars). The photometric mass was computed from light curves measured by the automated meteor detector, using a fixed luminous efficiency of 0.7\%. The 34 PER selected for manual reduction are shown in light orange.}
\label{img:EMCCD_PER}
\end{figure}

A total of 34 Perseids were manually measured. Figure \ref{img:EMCCD_PER} shows our selection of manually reduced meteors compared to all Perseids detected by the EMCCDs between 2018 and 2023. We preferentially selected brighter and longer meteors to have as many data points as possible for accurate model fits.

\subsection{Empirical Lightcurve and Dynamics Models} \label{subsec:empirical_models}

Through trial and error, we empirically found analytical functions which generally fit well to observed meteor dynamics and photometric time series. Specifically, both can be described by piece-wise functions. 

The light curve is modelled as two parabolas to specifically allow characterizing double-peaked, non-symmetrical, and triangular light curves. The functional form of the light curve model is 

\begin{align} \label{eq:magnitude_model}
M(t) = 
    \begin{cases}
        d_1 t^2 + s_1 t + c_1 & \text{if } t < t_{peak} \\
        d_2 (t - t_{0m})^2 + s_2 (t - t_{0m}) + c_2 & \text{if } t \geq t_{peak} \,,
    \end{cases}
\end{align}

\noindent where $t$ is the time and $t_{peak}$ is the time of the peak magnitude.

Meteor velocity in the atmosphere has traditionally been modelled by assuming exponential deceleration in the form of

\begin{align} \label{eq:exponential_v}
    v(t) = v_0 - a_0 k e^{k t} \,,    
\end{align}

\noindent where $v_0$ is the initial velocity, $a_0$ is the initial deceleration, and $k$ is a positive constant \citep{whipple1938photographic, jacchia1957preliminary}. However, the function has been proven to be numerically challenging to fit, with the parameters $\dot v_0$ and $k$ being coupled \citep{egal2017challenge}. Instead, to obtain more meaningful parameters which scale linearly with the magnitude of deceleration, we opt for the following piecewise function

\begin{equation} \label{eq:dyn_v_model}
v(t) = 
    \begin{cases}
        v_0 & \text{if } t < t_{0} \\
        v_0 - 3 |a| (t - t_{0})^2 - 2 |b| (t - t_{0}) & \text{if } t \geq t_{0} \,,
    \end{cases}
\end{equation}

\noindent where $t_{0}$ is the time when the deceleration starts to follow a quadratic form. Furthermore, as the measured point-to-point velocities are often noisy and the initial velocity can be accurately measured by fitting a line to the measurements of length vs. time \citep{Vida2020theory}, we fit this empirical function on the lag. The lag measures how much the observed meteor falls behind a hypothetical non-decelerating body. The lag $\ell(t)$ can thus be expressed as

\begin{equation}
\ell(t) = 
    \begin{cases}
        0 & \text{if } t < t_{0} \\
        -|a| (t - t_{0})^3 - |b| (t - t_{0})^2 & \text{if } t \geq t_{0} \,.
    \end{cases}
\end{equation}

The minimization function optimized three variables: the time $t_{0}$ and the two coefficients (a, b) of the third-order polynomial. The initial guess for $t_0$ was the mean value of the total elapsed detection time, the initial guess for the third-order term, was set based on the average value of lag while the second order term was initially set to zero. The coefficients are forced to be negative in the equations as not to produce apparent accelerations which are unphysical.

\begin{figure}[h!]
\centering
\includegraphics[width=1\linewidth]{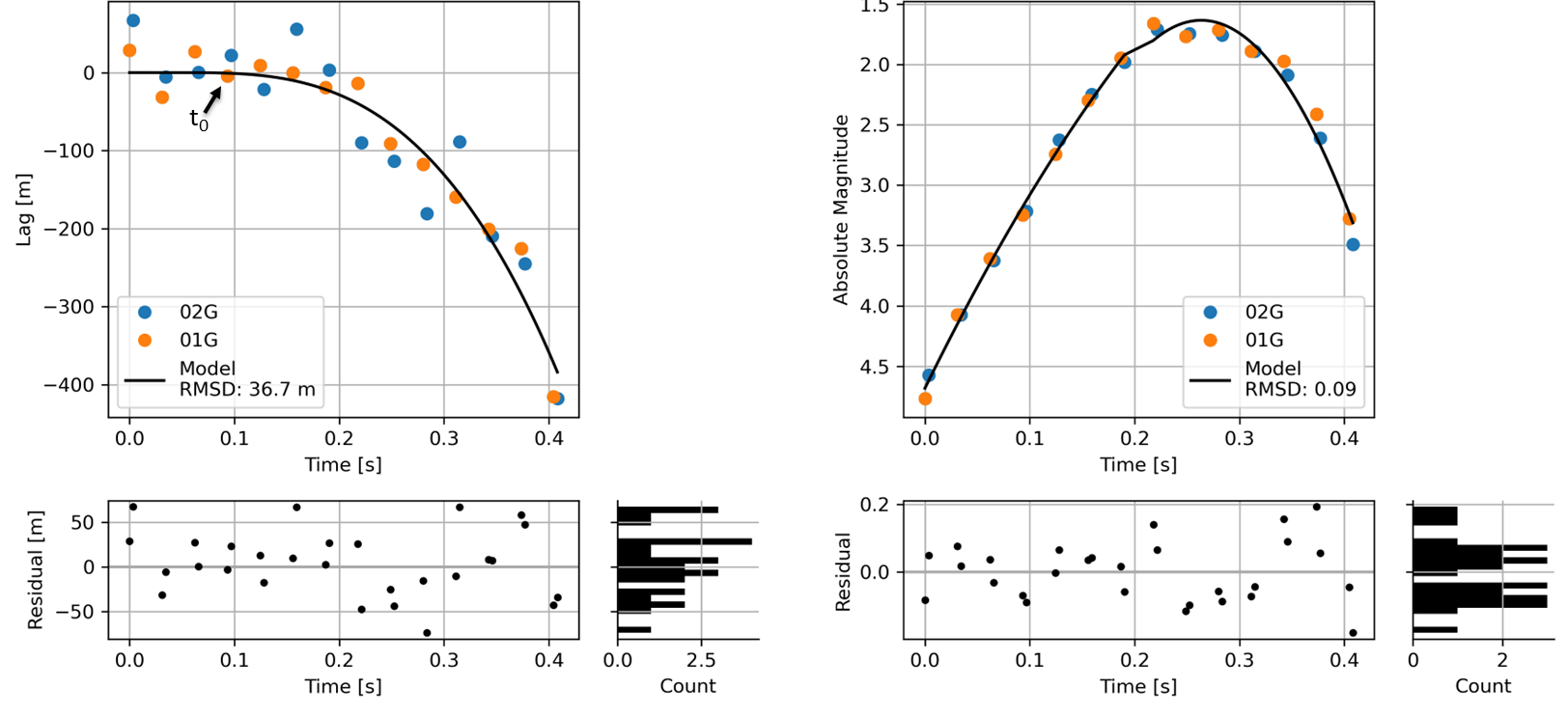}
\caption{An example of the two piece-wise lag and magnitude fits applied to EMCCD measurements for a manually reduced Perseid observed on 2021/08/13T06:14:52 UT. The colored points represent individual measurements from the two different stations (Tavistock - orange, Elginfield - blue). The left plot shows the apparent lag in meters together with the functional fit (solid black line). Here the root mean squared difference (RMSD) between the fit and measurements produces an RMSD value of 36.7~m. The right plot shows the absolute magnitude in the G-band and the polynomial fit (solid line). For each plot the residual distribution between the measurements and functional fits in time are shown together with a histogram of all residual values.}
\label{img:example_stdv}
\end{figure}

An example of these fits on a Perseid meteor is shown in Figure \ref{img:example_stdv}. It can be seen that the empirical function accurately follows the observed behavior of the meteor and the distribution of fit residuals roughly follows a Gaussian distribution. We have confirmed this is the case for all observed meteors used in this study. 

To capture the average measurement uncertainty which will be used to generate test case meteors to validate our approach (more in Section \ref{subsec:validation}), we repeat the procedure on two well-observed Perseids where three different analysts have performed the measurements. These two meteors have been chosen because they have also been observed by the high-resolution Canadian Automated Meteor Observatory's mirror tracking system whose measurements were used to validate EMCCD measurements. The root mean squared differences (RMSD) between our empirical parameterized fits and individual frame measurements of meteor brightness and lag are given in Table \ref{tab:per_meas_rmsd}. Based on these measurements, we adopt an average RMSD of 40~m as being representative of the length measurements. We conservatively adopt a bit higher RMSD of 0.1 magnitudes for the light curve to be in line with the stellar photometry uncertainties found for our instrument \citep{vida2020new}.

However, we note that the limited number of independent manual measurements results in a restricted statistical description, and the observed spread in lag and magnitude values primarily reflects measurement precision rather than true physical variability. Increasing the number of independent measurements would improve the statistical robustness of this assessment.

\begin{table}[h!] 
\caption{Values of the root mean squared difference for two Perseids derived by three different analysis (A, B, C) using subjective measurement methods.}
\begin{tabular}{r r r r} \label{tab:per_meas_rmsd}
Event & Analyst & Lag (m) & Mag \\
\hline \hline
\texttt{20210813\_061452} & A & 42.0 & 0.07 \\
\texttt{20210813\_061452} & B & 34.4 & 0.07 \\
\texttt{20210813\_061452} & C & 36.7 & 0.09 \\
\hline
\texttt{20230811\_082648} & A & 51.9 & 0.04 \\
\texttt{20230811\_082648} & B & 51.0 & 0.04 \\
\texttt{20230811\_082648} & C & 18.8 & 0.07 \\
\hline
\end{tabular}
\end{table}

As a form of dataset size augmentation for the PCA method, we generate 50 realizations with Gaussian noise using this measured RMSD. Each realization is treated as an independent measurement of the same meteor. We then fit the parametric models to each realization to derive the spread in the observables. 


\subsection{Extracting Observables}

The parameterization of meteor light curves and dynamics makes it possible to extract general characteristics of meteors which can uniquely describe them without the need to include their full measurements. 

A total of 25 parameters computed for each observed and synthetic meteor as means for mutual comparison are given in Table \ref{tab:observables}. The parameters are roughly divided into three categories of parameters: general trajectory, dynamics, and light curve. The general parameters describe the observed duration, length, entry angle, and observed heights (begin, peak, end). The dynamic pressure at the peak magnitude is also included as a diagnostic measure. The dynamic pressure is calculated as \( p = c_D \, \rho_{air} \, v^2 \), where \( \rho_{air} \) is the atmospheric density obtained from the NRLMSISE-00 model, \( v \) is the meteoroid velocity, and \( c_D \) is the drag coefficient. The parameters capturing the dynamics include the initial and average velocities and various measures of deceleration. We include many of these additional and redundant parameters as by definition PCA should find the most relevant ones that describe meteors the best. Finally, the light curve is described with the absolute magnitudes at the beginning, peak, and end, the F parameter (a measure from 0 to 1 of the fractional length along the trajectory where the peak brightness occurs), and light curve shape parameters found by fitting Eq. \ref{eq:magnitude_model}.

\begin{table}[htbp]
\caption{Descriptions of observable variables extracted from the meteor trajectory data.}
\label{tab:observables}
\centering
\resizebox{\textwidth}{!}{%
\begin{tabular}{l|l|l}
Variable & Unit & Description \\
\hline \hline
\multicolumn{3}{c}{General Trajectory} \\
\hline
$T$        & s   & Observed duration of the meteor. \\
$L$        & m   & Observed trail length. \\
$z_c$      & deg & Zenith angle at the height of 180 km. \\
$h_{\text{beg}}$  & m   & Height at first detection. \\
$h_{\text{peak}}$ & m   & Height at peak magnitude. \\
$h_{\text{end}}$  & m   & Height at last detection. \\
$k_c$      & km  & CAMS $k_c$ parameter~\cite{jenniskens2016cams_kc}. \\
$Q_{\text{peak}}$ & Pa  & Dynamic pressure at the peak absolute magnitude. \\
\hline
\multicolumn{3}{c}{Dynamics} \\
\hline
$v_0$        & m\,s$^{-1}$ & Initial velocity; measured from the trajectory or taken from the simulation at the beginning height. \\
$v_{\text{avg}}$    & m\,s$^{-1}$ & Average velocity during meteor visibility. \\
$\bar{\ell}$ & m          & Average lag value. \\
$t_0$      & s          & Time from the beginning after which deceleration starts. \\
$\bar{a}_{\text{poly}}$ & m\,s$^{-2}$ & Average deceleration after $t_0$ until the end of the meteor computed using Eq.~\ref{eq:dyn_v_model}. \\
$a_{\text{poly}}(1\,\text{s})$  & m\,s$^{-2}$ & Deceleration after 1 second computed using Eq.~\ref{eq:dyn_v_model}. \\
$\bar{a}$         & m\,s$^{-2}$ & Average deceleration derived by fitting a line through time vs point-to-point velocity. \\
$a_{\text{quad}}(1\,\text{s})$  & m\,s$^{-2}$ & Deceleration after 1 second using a quadratic fit on time vs point-to-point velocity. \\
$a_0 k$          & m\,s$^{-1}$ & Exponential deceleration model term from Eq.~\ref{eq:exponential_v}. \\
\hline
\multicolumn{3}{c}{Light Curve} \\
\hline
$M_{\text{beg}}$  & mag & Absolute magnitude at first detection. \\
$M_{\text{peak}}$ & mag & Peak absolute magnitude. \\
$M_{\text{end}}$  & mag & Absolute magnitude at last detection. \\
$F$        & --  & $F = (h_{\text{beg}} - h_{\text{peak}}) / (h_{\text{beg}} - h_{\text{end}})$. \\
$d_1$      & mag\,s$^{-2}$ & Quadratic coefficient before peak magnitude (Eq.~\ref{eq:magnitude_model}). \\
$s_1$      & mag\,s$^{-1}$ & Linear coefficient before peak magnitude (Eq.~\ref{eq:magnitude_model}). \\
$d_2$      & mag\,s$^{-2}$ & Quadratic coefficient after peak magnitude (Eq.~\ref{eq:magnitude_model}). \\
$s_2$      & mag\,s$^{-1}$ & Linear coefficient after peak magnitude (Eq.~\ref{eq:magnitude_model}). \\
\end{tabular}
}
\end{table}

\subsection{Generating Synthetic Meteor Observations} \label{subsec:abl_gen}

The fundamental idea behind our approach is to match the observable characteristics of real meteors and their synthetic counterparts which have known physical properties. The main challenge of this process is the computation time needed to generate the simulations and test meteoroids with physical properties across a large multidimensional parameter space.

We use the \cite{borovivcka2007atmospheric} erosion model to generate synthetic meteors. Briefly, this model follows the classical single-body equations from the top of the atmosphere (180~km) until the moment $\mu$m-sized grains start to be shed from the surface of the meteoroid. This process is called erosion. The grain masses follow a power-law and it is assumed that the grains also ablate following single-body theory, and can be represented as having a chondritic grain density. The ratio of the meteoroid bulk density to grain density defines the porosity of the body, i.e. the grain density is assumed to be that of silicates ($\sim3000$kg~m$^3$) while the bulk density is normally much smaller, e.g. $\sim300$kg~m$^3$ for cometary meteoroids \citep{vida2024first}, resulting in a $90\%$ porosity. The meteor light curve is the sum of the light production of the main body and all the released grains, and the deceleration can either be tracked for the leading fragment or the brightest point on the meteor. The luminous efficiency used throughout is the velocity and mass-dependent formulation proposed by \cite{vida2024first}. The atmospheric density is approximated using a 7th-order polynomial fit to the logarithm of the NRLMSISE-00 model output. The remaining fixed parameters used throughout the simulations are summarized in Table~\ref{tab:fixed_params}.

\begin{table}[htbp]
\centering
\caption{Fixed parameters used in the simulations. These values remain constant across all simulation runs.}
\renewcommand{\arraystretch}{1.2}
\begin{tabular}{|l|c|c|}
\hline
Parameter & Symbol & Value \\
\hline
Time step & $\Delta t$ & 0.005~s \\
Grain density & $\rho_\mathrm{grain}$ & 3000~kg\,m$^{-3}$ \\
Drag coefficient & $c_D$ & 1.0 \\
Shape factor & $\Gamma$ & 1.21 \\
Grain mass fragments binning & --- & 10 bins\\
\hline
\end{tabular}
\label{tab:fixed_params}
\end{table}

The eleven input parameters of this model can roughly be divided into two groups, bulk meteoroid properties and fragmentation parameters. The bulk properties include the meteoroid mass, speed, entry angle, ablation coefficient, and bulk density. The fragmentation properties are the erosion start height, erosion coefficient, grain density, grain mass distribution index , and the range of grain masses. The erosion coefficient controls how much of the meteoroid's mass is shed into grains per unit of received energy in every 0.005~s time step of integration.

To find the best match between observed meteors and the model while keeping the computation time manageable, we initially create 10,000 synthetic meteors with the mass, speed, and entry angle close to the observed values:

\begin{itemize}
    \item Initial meteoroid mass $m_0$: Range set to $\pm$ half an order of magnitude around the observed photometric mass computed using a fixed luminous efficiency of 0.7\%.
    \item Initial velocity $v_0$: Gaussian distribution with $\sigma = 50~m~s^{-1}$ around the observed value.
    \item Zenith angle $z_c$: Corrected for Earth's curvature and extrapolated to a height of 180~km \citep{vida2024first}. Sampled from a Gaussian distribution with $\sigma = 0.01^{\circ}$ about the observed value.
\end{itemize}

These first three parameters are derived directly from observations and hence are fairly well determined. In contrast, the priors for physical parameters were kept wider and flat, as we treat them as unknowns. Specifically for the Perseids, we explore the ranges shown in the Table \ref{tab:fixed_ranges} which were informed by numbers found by other authors.


\begin{table}[htbp]
\centering
\caption{Fixed prior Parameter range for all simulations}
\begin{tabular}{|l|c|c|}
\hline
\textbf{Parameter} & \textbf{Min} & \textbf{Max} \\
\hline
Meteoroid bulk density $\rho$ [kg~m$^{-3}$] & 100 & 1000 \\
\hline
Ablation coefficient $\sigma$ [kg~MJ$^{-1}$] & 0.008 & 0.03 \\
\hline
Erosion coefficient $\eta$ [kg~MJ$^{-1}$] & 0.0 & 1 \\
\hline
Mass distribution index  $s$ & 1.5 & 2.5 \\
\hline
Minimum erosion mass $m_l$ [kg] & $5 \times 10^{-12}$ & $1 \times 10^{-10}$ \\
\hline
Maximum erosion mass $m_u$ [kg] & $1 \times 10^{-10}$ & $5 \times 10^{-8}$ \\
\hline
\end{tabular}
\label{tab:fixed_ranges}
\end{table}

The observed begin and peak brightness heights determined the range of model sampled erosion starting height $h_{e}$. Based on results by previous authors \citep{borovivcka2007atmospheric, vojacek2019, vida2024first, buccongello2024physical}, we assume that the main rise in the light curve is solely explained by erosion. These authors have found that erosion roughly begins just before the meteor is detected or in the first part of its light curve. Thus, we set the range of erosion starting heights between $h_{\text{e, min}}, h_{\text{e, max}}$ which are defined as 

\begin{align} \label{eq:erosion_heights}
    \Delta h = h_{\text{B}} - h_{peak} \,, \\
    h_{\text{e, min}} = h_{peak} + \frac{1}{2}\Delta h \,, \\
    h_{\text{e, max}} = h_{\text{B}} + \frac{1}{2}\Delta h \,,
\end{align}

\noindent where $h_B$ is the meteor beginning height and $h_{peak}$ is the peak height.

Each simulated meteor was assumed to be detected by two fictional EMCCD cameras with a zero-magnitude meteor power $P_0$ of 935 W (appropriate to the G-bandpass) and a frame rate of 32 fps \citep{brown2020coordinated}. A set of synthetic observations was generated for each of the two cameras and were combined into a single time series by assuming a random phase difference between the observations, with the aim to simulate real-world double-station observations. Meteors were assumed to be ``detected'' if at least 10 frames were above the limiting magnitude. These numbers were chosen based on the real-life performance of the EMCCD cameras and the meteor detector. The meteor detector has a lower minimum duration threshold of 4 frames, but we chose to increase it to ensure that the unique features of each meteor are captured by the simulations. The limiting magnitudes of the first and last detection were taken directly from the observation to achieve an accurate reproduction of the light curve. 

\begin{figure}[h!]
\centering
\includegraphics[width=1\linewidth]{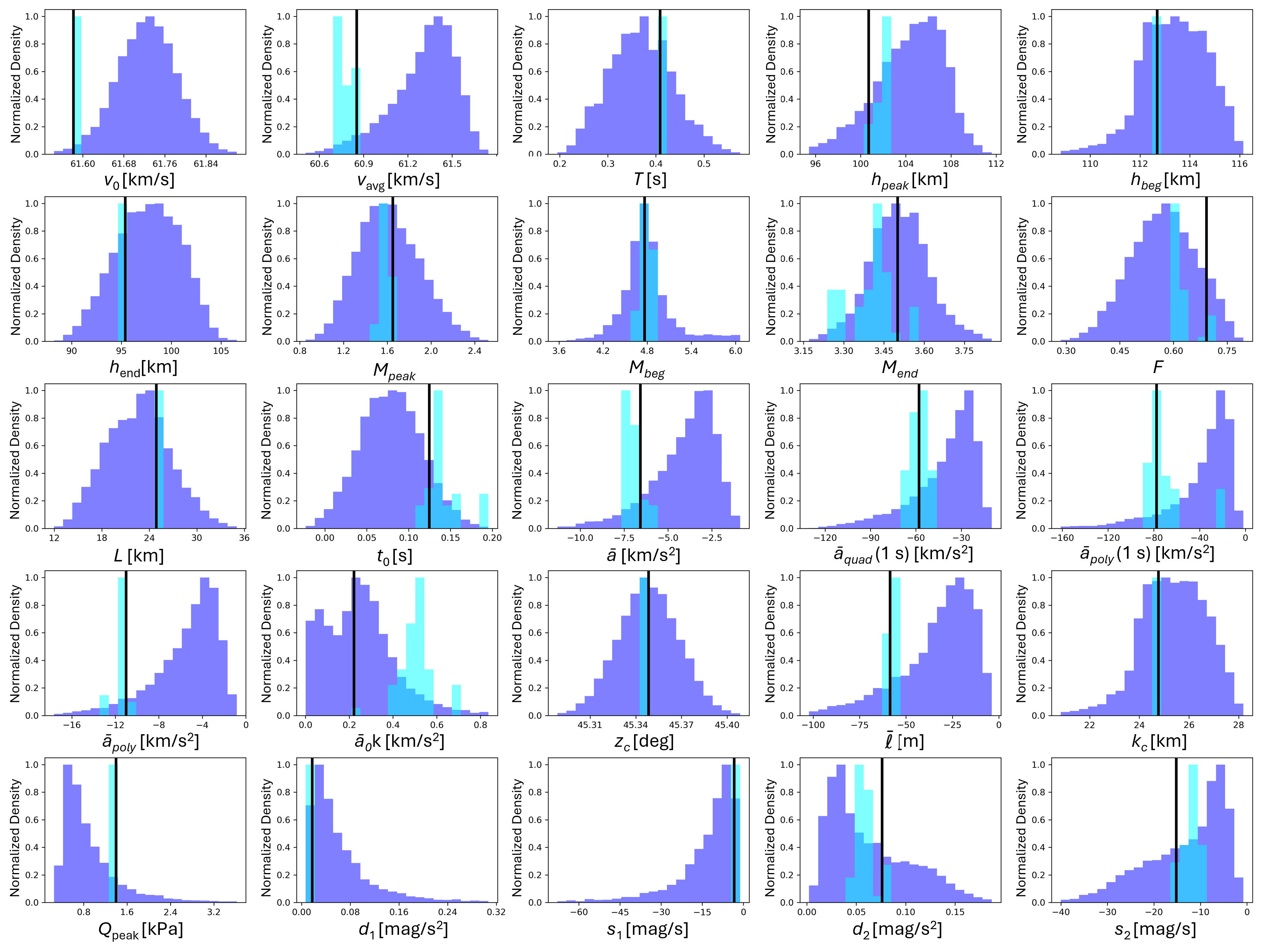}

\caption{Histograms of the 25 observables derived through ablation modeling for the 20210813\_061452 meteor. The black vertical line indicates the values obtained from the noise-free solution. The cyan bars represent the range of values derived from 50 noise realizations, where noise was introduced only in the lag and magnitude profiles. As a result, some observables remain unchanged between the noisy and non-noisy cases.}
\label{img:heavy_observables}
\end{figure}

Figure \ref{img:heavy_observables} shows an example of these 25 observables extracted for the 20210813\_061452 event as measured by analyst C. 
The vertical black lines marks the value of the observables for this measured meteor. The cyan histograms show the range of observables across all 50 realizations generated in the step described in Section \ref{subsec:empirical_models}. The blue histograms show the range of observables derived through ablation modelling and the model priors as given above. We use this plot when modelling each meteor as a means to visually confirm that the simulations encompass the meteor's observed behaviour. If the meteor's values are outside of the range covered by simulations, we modify the priors until a good match is found. As there is redundancy in the observables, not all have to fit perfectly. For example, the exponential deceleration factor $a_0k$ in Figure \ref{subsec:validation} tends to vary significantly between realizations. This variation suggests that it is more susceptible to noise than other parameters, making it a less reliable or well-behaved observable. Conceptually, the application of the Principal Component Analysis in the next step will identify the most important values.

\subsection{Application of Principal Component Analysis}

Assuming the synthetic (model-generated) meteors roughly match the observables of the meteor of interest, the next step is to select the simulations which are the most similar to the observed meteor using a quantitative metric. Manually crafting a metric in a 25-dimensional parameter space is difficult, thus we apply Principal Component Analysis (PCA) to normalize the parameters and find their linear combination which best describes the differences between them, essentially resulting in a distance metric.

Principal Component Analysis (PCA) is a statistical technique that transforms a high-dimensional dataset into a new coordinate system by identifying the axes, or principal components (PCs), that capture the greatest variance in the data. Each principal component is an orthogonal linear combination of the original variables, ranked by the amount of variability it explains. Specifically in our application, a principal component $n$ will have the form of $PC_n = x_{n1} T + x_{n2} L + x_{n3} z_c + x_{n4} h_{beg} + \cdots$, where the method finds the coefficients $x_{ni}$. By projecting the data onto these components, a selection can be made to keep only the top components which when taken together explain at least 95\% of the variance (or more). This way, PCA reduces the complexity of the dataset while preserving the most significant patterns. In essence, it is a dimensionality reduction technique which generally allows one to manipulate a handful of variables (sometimes as few as 2-3) instead of a large multidimensional parameter space.

To correctly condition the inputs (i.e. the synthetic dataset) as required by the method, we first remove outliers with a z-score greater than 3. The z-score, also known as the standard score, measures a value's relationship to the mean of a group of values, expressed in terms of standard deviations from the mean \citep{huck1986overcoming}.

Next, the distributions of the observables are rescaled to have a more Gaussian shape, which is essential for PCA to be effective \citep{dunteman1989PCA}. While we found that the initial velocity $v_i$ and zenith angle $z_c$ are already generally Gaussianly distributed, other variables required transformation using the Yeo-Johnson method \citep{Yeo-Johnson2000}. We also explored other methods, including the Shapiro-Wilk test \citep{ShapiroWilk1965}, logarithmic transformations, and the Box-Cox transformation. We found that the Yeo-Johnson method yielded a distribution closest to Gaussian for this application. The dataset conditioning is completed by standardizing the dataset which involves subtracting the mean from each parameter and dividing it by its standard deviation.

Upon running the PCA algorithm, we analyzed the results and selected the top principal components which when taken together explained more than 99\% of the variance. We empirically found that this number works best for our application (described below), as the traditional value of 95\% tended to yield subpar results. This resulted 11 to 13 principal components, depending on the meteor.

\begin{figure}[h!]
\centering
\includegraphics[width=1\linewidth]{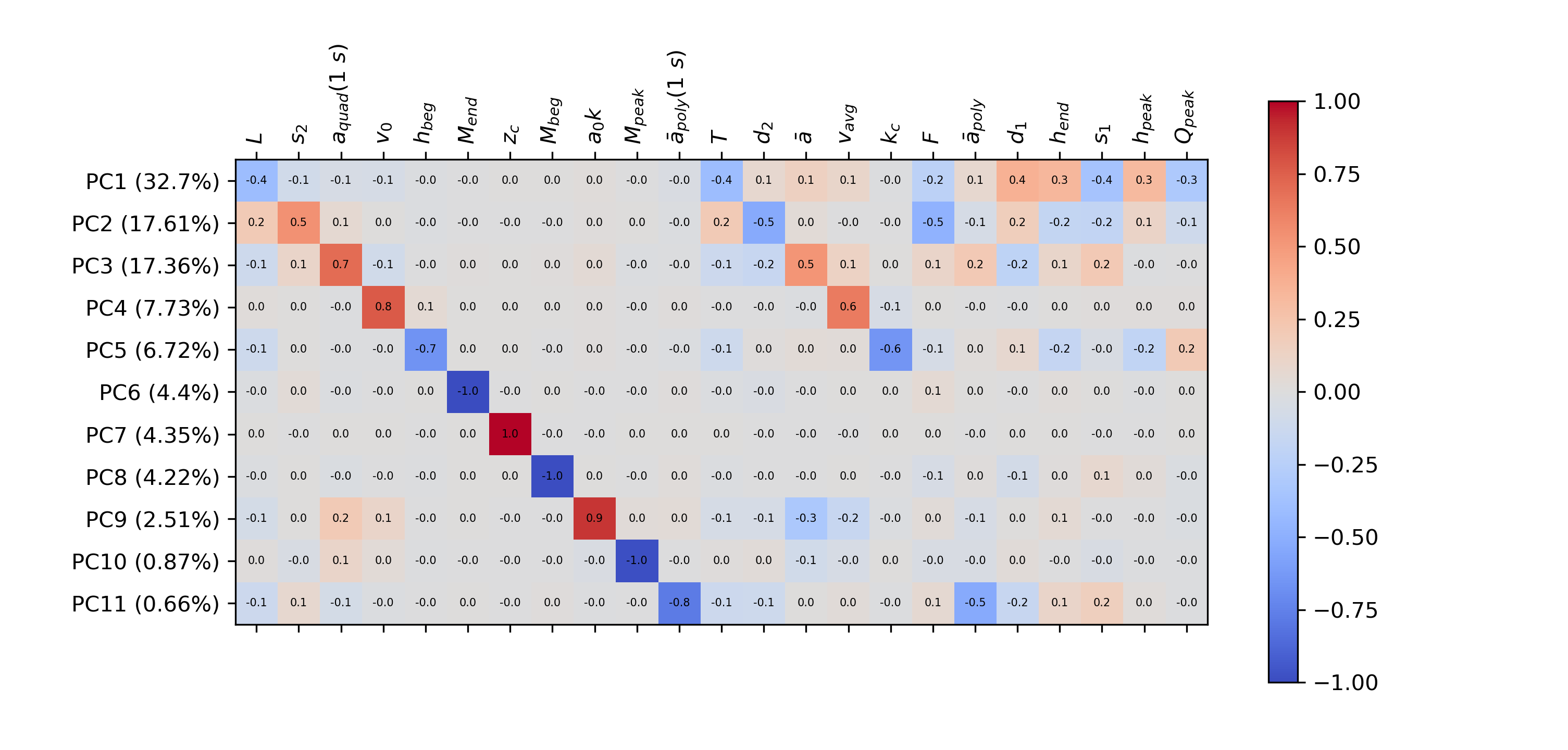}
\caption{PCA correlation matrix of the 20210813\_061452 meteor showing contributions of each observable to the 11 selected principal components. PCs are sorted by their total contribution to the total variability, and the observables are sorted by their greatest contribution to each PC.}

\label{fig:heavy_pca}
\end{figure}

Figure \ref{fig:heavy_pca} shows the principal components found for the 20210813\_061452 meteor. Generally speaking, we find that over 50\% of the variability is captured in only two PCs, but to achieve accurate representation of meteors and capture $>95\%$ of variability, up to 10 components have to be included. We find that the method correctly correlates observables which measure the same property using different methods (e.g. speeds, deceleration measures, total length/duration, begin height and $k_c$). Certain observables always tend to stand alone, such as the zenith angle and the peak magnitude. Even though this particular meteor shows strong correlations for certain PCs, we find significant variability between PCs of different meteors, i.e. no common single observable always dominates the variance for all meteors which show significant individual differences. This is a reflection of the fact that there is no such thing as an average meteor \citep{Ceplecha1998e}.

During the development of our approach we initially assumed that a singular PCA model could be derived for an entire meteor shower, allowing fast evaluation of the model for different meteors. However, we found that the model does not perform well in such a general case and we had to perform ablation model runs for each specific meteor.

With the PCA model in hand, the distance in PC space between two meteors can be defined by using the modified Mahalanobis distance \citep{mahalanobis2018generalized}

\begin{align}\label{eq:pca_dist}
    D_{ij}= \sqrt{(\vect{x_i} - \vect{x_j})^T \vect{S^{-1}}(\vect{x_i} - \vect{x_j})} \,,
\end{align}

\noindent where, $D_{ij}$ is the modified Mahalanobis distance between two sets of principal components $x_i$ and $x_j$ (one for each meteor), with $S$ being the covariance matrix of the dataset. We modified the covariance matrix $S$ scaling it by the inverse of the explained variance of each PC, giving more importance to PCs with higher variance.

\begin{figure}[h!]
\centering
\includegraphics[width=0.7\linewidth]{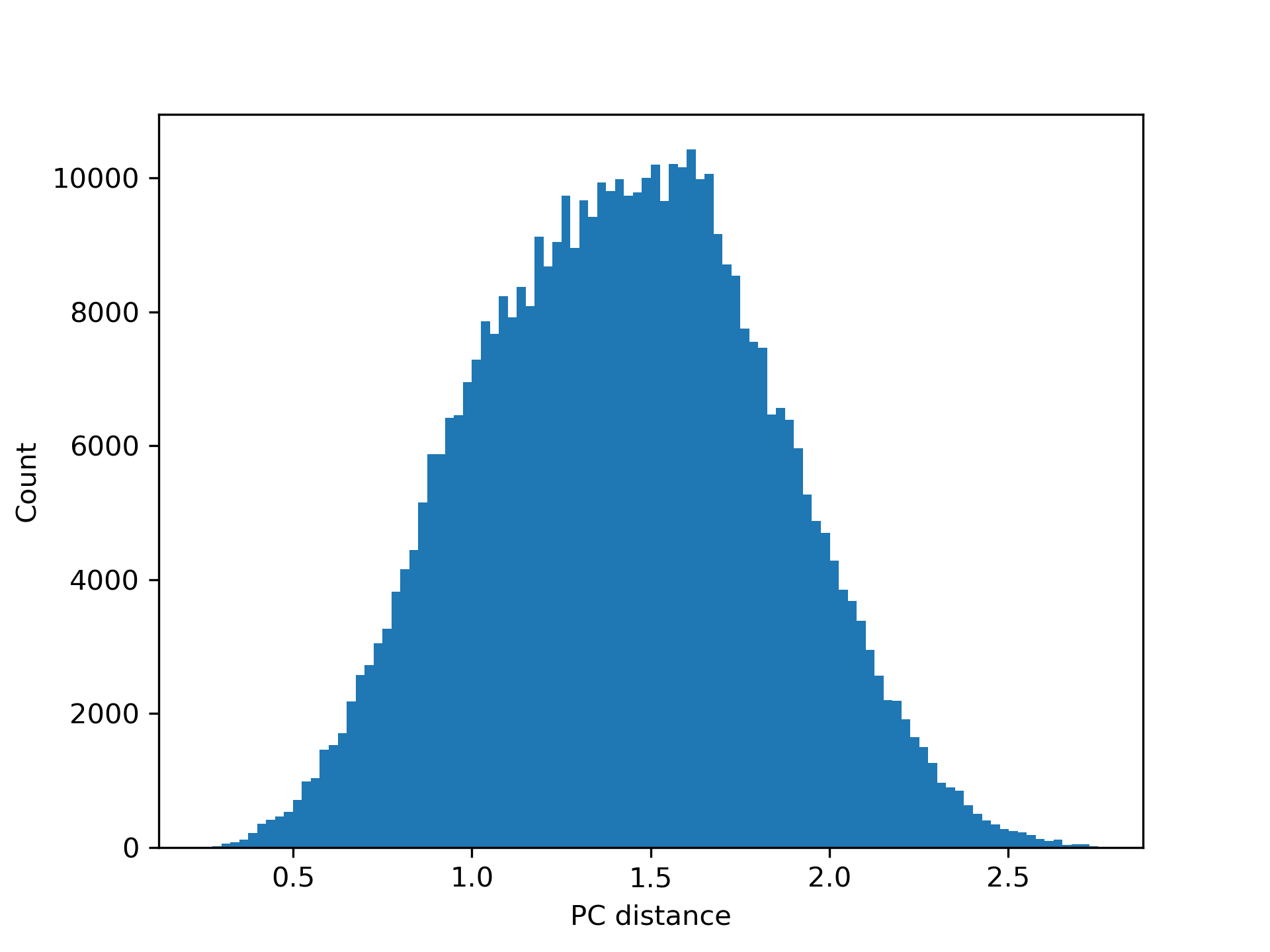}
\caption{The distribution of the modified Mahalanobis distances of all the simulations in the PC space respect to our 20210813\_061452 meteor and the 50 realizations.}
\label{img:distrib_dist}
\end{figure}

Figure \ref{img:distrib_dist} shows the histogram of the Mahalanobis distance of the synthetic meteors created for our example Perseid, including the 50 realizations generated by resampling the reference observation with the measured instrument noise. The distances derived using each realization are plotted together, with the idea that this will propagate the instrumental noise to the PC parameter space and consequently the inverted meteoroid physical properties.

For each of the 10,000 simulations, a total of 51 distances are computed: 50 for each realization and one for the reference observation. From this set, the smallest distance is chosen as the best guess of the distance between the simulation and the reference observation.

In the next step, only the given number of simulations with the smallest distance metric is selected. This threshold is somewhat arbitrary as there isn't a clear number to use to unambiguously select simulations within the observational uncertainty.

\subsubsection{Applying the PCA Method to Real Event} \label{subsubsec:realPCA}

\begin{landscape}
\begin{figure}[h!]
\centering
  \begin{minipage}{0.45\linewidth}
    \centering
    PCA dist. 0.5 percentile \\ 
    \includegraphics[height=\linewidth]{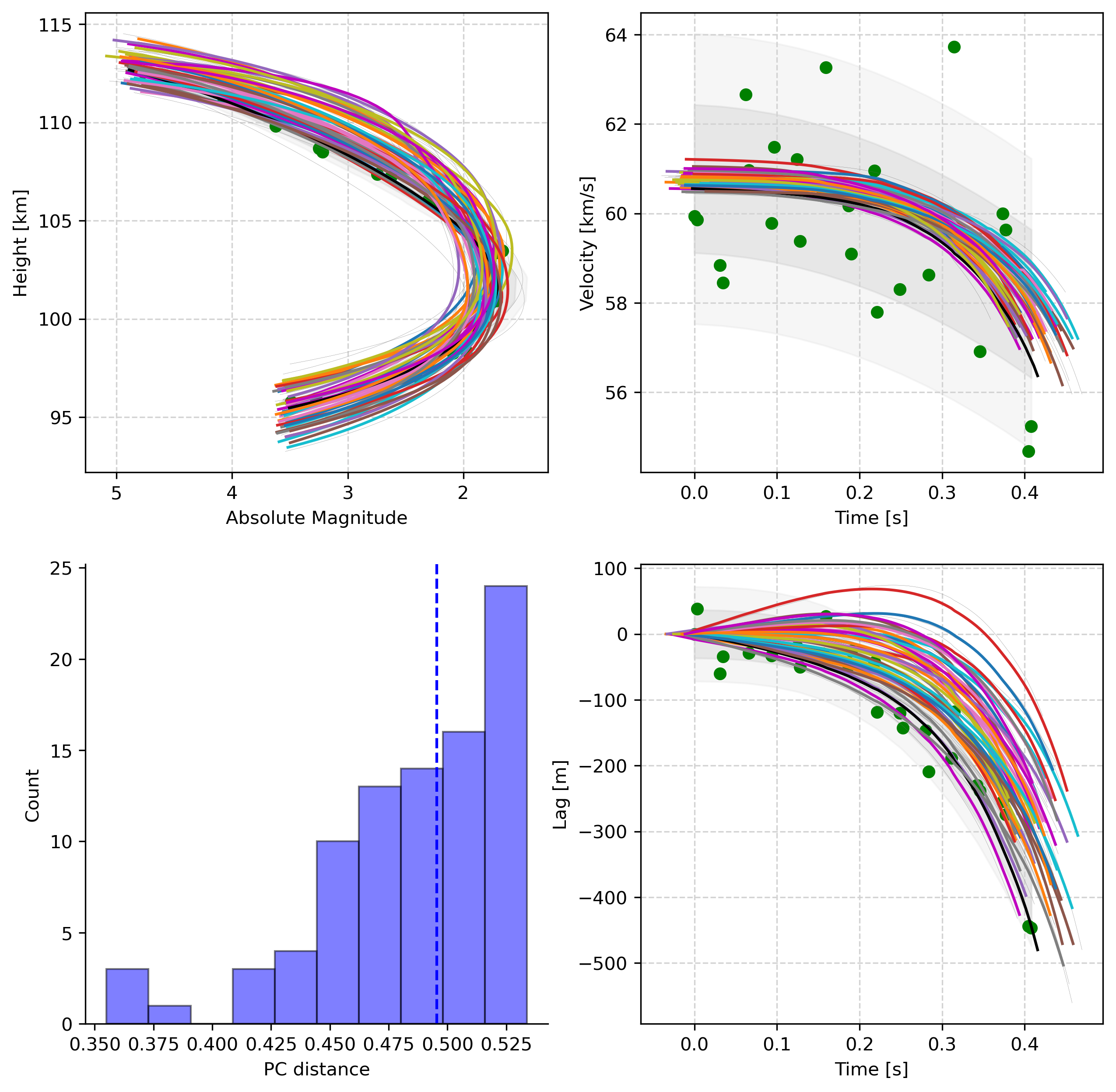}
\end{minipage}
\hfill 
\vrule width 0.5pt 
\hfill 
  \begin{minipage}{0.45\linewidth}
    \centering
    PCA dist. 0.1 percentile \\ 
    \includegraphics[height=\linewidth]{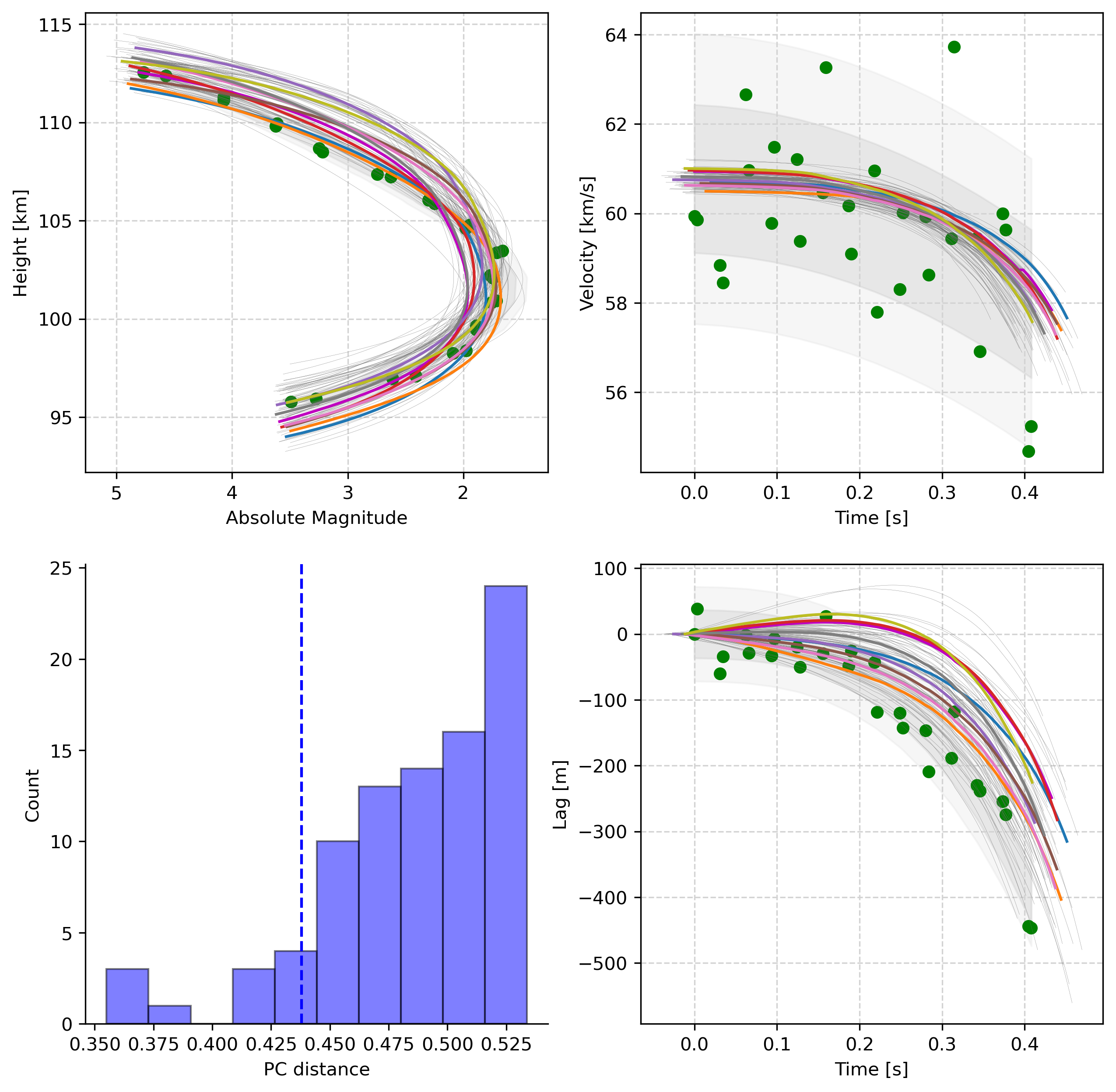}
\end{minipage}
\caption{The plots comparing the absolute magnitude, velocity, and lag measurements for the two cases are shown. The black line in this case is the forward modeling manual fit result (only selected in the 0.5 percentile case), the colored lines are selected simulations, and the thin gray lines are the simulations for the rest on the blue plot. The vertical dashed blue line shows the location of the threshold cut - simulations to the left of this line are selected as being consistent with measurements and are shown as individual colored lines in the lag and magnitude plots. Simulations to the right of this line are outside the threshold values and are shown as grey lines in the plots.}
\label{img:diff_cutsPCA}
\end{figure}
\end{landscape}

To investigate the influence of the choice of the threshold distance,  Figure \ref{img:diff_cutsPCA} shows the cuts at 0.5 and 0.1 percentile for the 20210813\_061452 meteor from Analyst C (see Table \ref{tab:per_meas_rmsd}). It can be seen that the number of selected meteors falls with the reduced threshold even though the simulations with the apparent smaller distances do not clearly fit the data better. The problem is especially pronounced for the dynamics, where some seemingly better solutions are eliminated by using the tighter distance. This may be an indication that the parameter space is undersampled or that the PCA method does not find the optimal distance metric for comparison. This result led us to explore a more brute-force fitting approach between the model and observations.



\subsection{Brute-force method : Direct Application of an RMSD Threshold} \label{subsec:RMSD_sel}

While the PCA approach is designed to be effective in identifying simulations with observable parameters that closely match the reference meteor, from the previous section questions remain about whether the selected light curves and deceleration profiles are a good fit to the observed profiles. To evaluate the accuracy of the PCA distance and our selection threshold, we select synthetic meteors separately based on whether they fall within the measurement uncertainty, skipping the PCA approach altogether. The selection was based on the requirement that the RMSD of the light curve and lag had to be within twice of the mag and the length noise recovered from the polynomial fits.  


This brute-force method offers a contrasting way to estimate uncertainties. Rather than prioritizing a close match to observed variable via PCA, it only requires that the simulated light curve and deceleration “fits well” within the allowed error margins.

To organize the selected simulations in a single ranking, we define one normalized $RMSD_{max}$, for each simulation as follows: 
\begin{align}\label{eq:combined_metric}
    RMSD_{max} = \max \Biggl( 
               \frac{\mathrm{RMSD}_{\mathrm{mag}}}{\mathrm{RMSD_{\mathrm{mag}}^{poly}}}, 
               \frac{\mathrm{RMSD}_{\mathrm{lag}}}{\mathrm{RMSD_{\mathrm{lag}}^{poly}}} 
             \Biggr).
\end{align}

Here, $\mathrm{RMSD}_{\mathrm{mag}}/\mathrm{RMSD_{\mathrm{mag}}^{poly}}$ is the ratio of the meteor's magnitude RMSD to the real magnitude measurement uncertainty defined by the second order polynomial fit. This choice is motivated by the typical structure of meteor light curves, which feature a peak followed by a trailing segment, as discussed in Section~\ref{subsec:empirical_models}. Similarly, $\mathrm{RMSD}_{\mathrm{lag}}/\mathrm{RMSD_{\mathrm{lag}}^{poly}}$ is the analogous ratio for lag to the uncertainty derived from a third-order polynomial fit to lag. Since velocity—the first derivative of lag—is expected to vary smoothly and is often well-approximated by a second-order function, the third-order fit provides a suitable basis for capturing this behavior through differentiation, also detailed in Section~\ref{subsec:empirical_models}.

Sorting by $RMSD_{max}$ lets us see which simulations come closest to the polynomial-fit uncertainties in both magnitude and lag simultaneously while providing an alternative check on the PCA-derived thresholds.

\subsubsection{Applying the RMSD Method to a Real Event} \label{subsubsec:realRMSD}

Using the same range of physical characteristics described in Section \ref{subsec:abl_gen}, we generated simulations until we obtained a set of 100 that meet the RMSD threshold. For meteor 20210813\_061452 from Analyst C (see Table \ref{tab:per_meas_rmsd}), the threshold in RMSD is 73.4 m in lag and 0.18 mag in brightness (i.e. twice the recovered RMSD).

\begin{landscape}
\begin{figure}[h!]
\centering
  \begin{minipage}{0.45\linewidth}
    \centering
    $RMSD_{max}$ 1.8 \\ 
    \includegraphics[height=\linewidth]{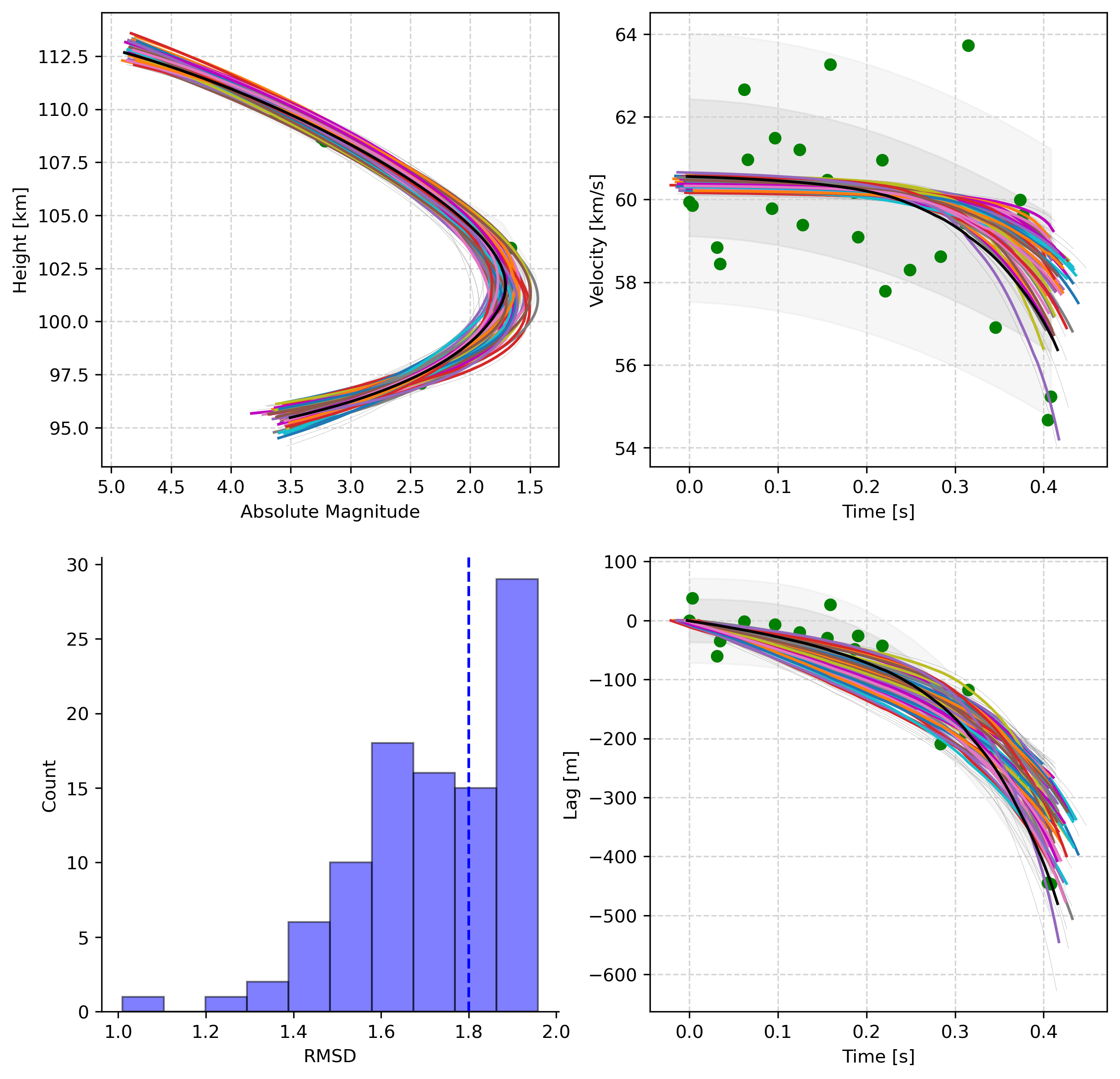}
\end{minipage}
\hfill 
\vrule width 0.5pt 
\hfill 
  \begin{minipage}{0.45\linewidth}
    \centering
    $RMSD_{max}$ 1.5 \\ 
    \includegraphics[height=\linewidth]{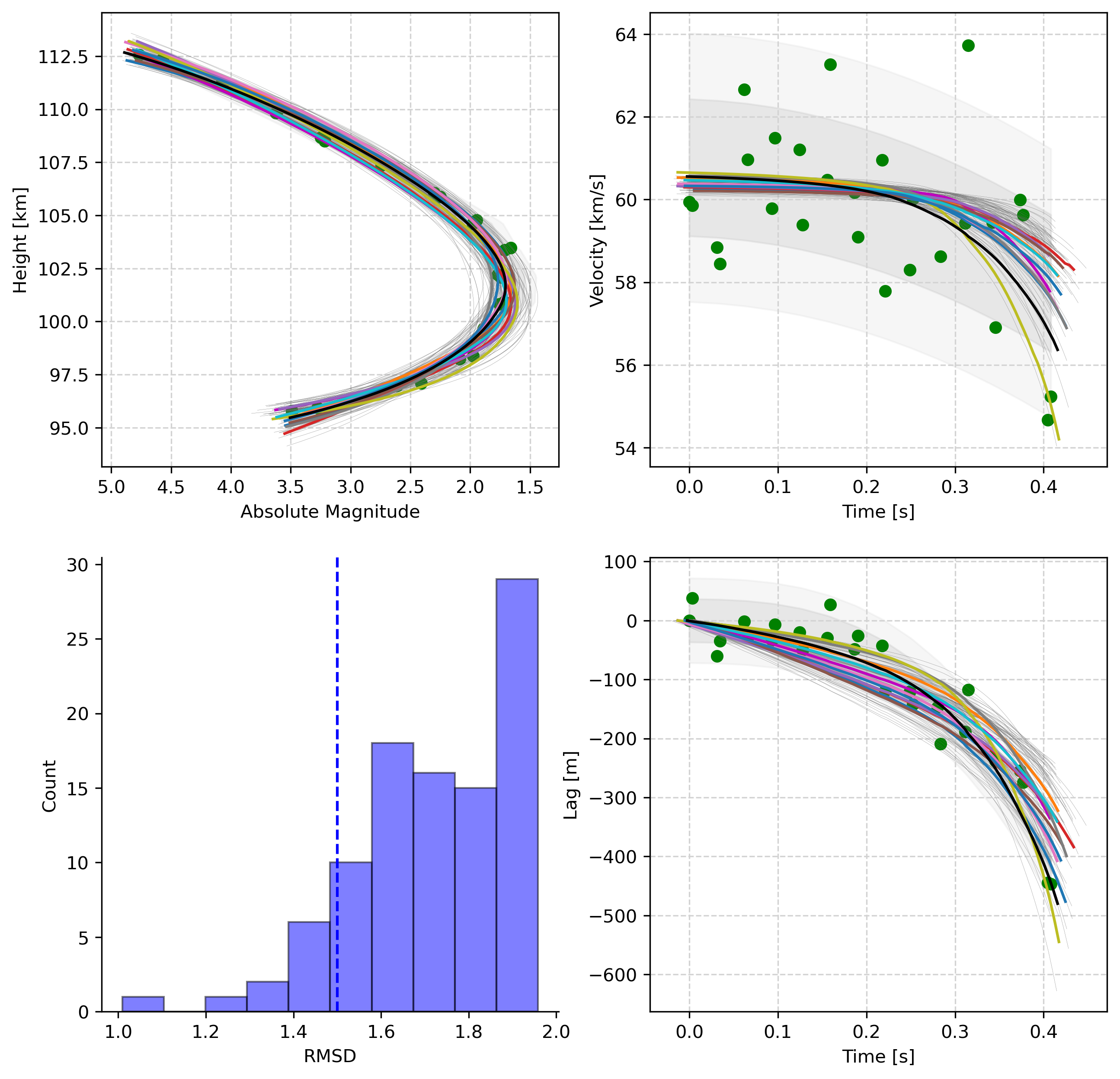}
\end{minipage}
\caption{The difference in the selection of model simulations when the threshold is set at 1.8 (left plots) and 1.5 (right plots) in RMSD for the 20210813\_061452 meteor. The gray lines are the remaining simulations above the RMSD selection (vertical purple line cut) until the end of the blue distance plot.}
\label{img:2021alalystC_RMSD}
\end{figure}
\end{landscape}

To investigate the influence of the chosen RMSD threshold, Figure~\ref{img:2021alalystC_RMSD} illustrates the selection criteria at RMSD values of 1.8 and 1.5 for the meteor 20210813\_061452. As the RMSD threshold is lowered, the number of selected simulations decreases, providing a noticeably better fit to the observed data.

In this subsection and in subsection \ref{subsubsec:realPCA}, we have shown the best fits for meteor 20210813\_061452 for the RMSD and PCA, but the two methods had distinct simulation pools to chose from. Although we have not yet explored how these selected simulations translate into physical parameters, the next section will delve into this analysis. We will compare the performance of both methods using the same pool of simulations, demonstrating how each approach yields physical estimates and associated uncertainties for simulated test cases where the true solutions are known.


\subsection{Validation using Six Model Test Cases} \label{subsec:validation}

Our validation process aims to determine how reliably Principal Component Analysis (PCA) can reproduce physical parameters and predict light curve and deceleration profiles of meteoroids, in comparison to the brute-force (RMSD fit) approach. Specifically, PCA leverages 25 observable values for each meteor, while the brute-force method filters simulations by “goodness of fit” in the measured meteor magnitude and lag. By testing both methods, we can establish which strategy provides more robust uncertainty estimates—whether using PCA-based observables or the simpler RMSD-based selection.

\begin{figure}[h!]
\centering
\includegraphics[width=\linewidth]{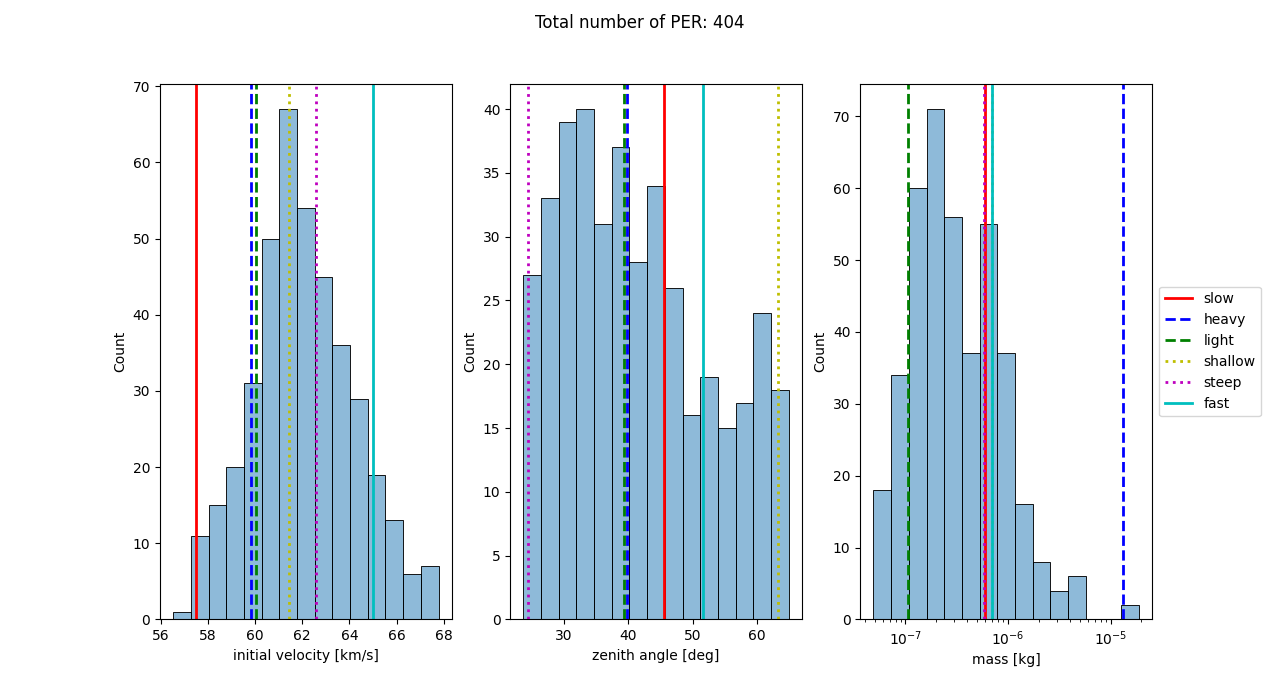}
\caption{The observed distribution of initial speeds (left), radiant zenith angles (middle) and masses (right) for PER meteors recorded by the EMCCD systems (blue bars). Note that meteors are selected as being PER based on radiants being within 3$^\circ$ of the radiant and within 10\% of the shower speed following \citep{Moorhead2021}. The masses here was computed using the measured lightcurve and an assumed fixed luminous efficiency of 0.7\%. This is the same as the distribution shown in Figure \ref{img:EMCCD_PER}. 
From this suite of simulations, six sample meteors were selected to test the method (shown here as different colored vertical lines).}
\label{img:EMCCD_PER_sim}
\end{figure}

To validate the methods, we generated six simulation test cases within our predefined physical parameters for the Perseid analysis, with the aim of quantifying how well our procedure can invert back to the known physical parameters. These test cases were selected to cover a broad range of PER meteoroid characteristics expected to be sampled from our EMCCDs, including initial velocity, zenith angle, and mass, as outlined in Table \ref{tab:init_conditions}. The test cases, labeled Slow, Fast, Steep, Shallow, Heavy, and Light, are representative of various entry scenarios for Perseid meteors. Each simulation was created using the Erosion Fragmentation Model of \cite{borovivcka2007atmospheric}. Figure \ref{img:EMCCD_PER_sim} shows an example of the distribution for some PER observables recorded by the EMCCD systems and the sampling of the model test cases within this observed range.

\begin{table}[h!]
\centering
\caption{The initial conditions for different simulation test cases used for validation. All parameter values are given as they would be at the top of the atmosphere. The name of the test case illustrates which parameter is being sampled near the edge of its distribution while other physical parameters including density are randomly sampled within the predefined range given in Section \ref{subsec:abl_gen}.} 
\begin{tabular}{|c|>{\centering\arraybackslash}p{2cm}|>{\centering\arraybackslash}p{2cm}|>{\centering\arraybackslash}p{2cm}|>{\centering\arraybackslash}p{2cm}|}
\hline
Test Case & $v_0$ [km/s] & $z_c$ [deg] & $m_0$ [kg] & $\rho$ [kg/m$^3$] \\
\hline
Slow & 57.50 & 45.56 & $5.91 \times 10^{-7}$ & 444\\
\hline
Heavy & 59.84 & 39.79 & $1.33 \times 10^{-5}$ & 229\\
\hline
Light & 60.05 & 39.32 & $1.05 \times 10^{-7}$ & 535\\
\hline
Shallow & 61.46 & 63.21 & $6.96 \times 10^{-7}$ & 702\\
\hline
Steep & 62.58 & 24.43 & $5.79 \times 10^{-7}$ & 591\\
\hline
Fast & 65.00 & 51.68 & $7.01 \times 10^{-7}$ & 201\\
\hline
\end{tabular}
\label{tab:init_conditions}
\end{table}

\begin{figure}[h!]
\centering
\includegraphics[width=\linewidth]{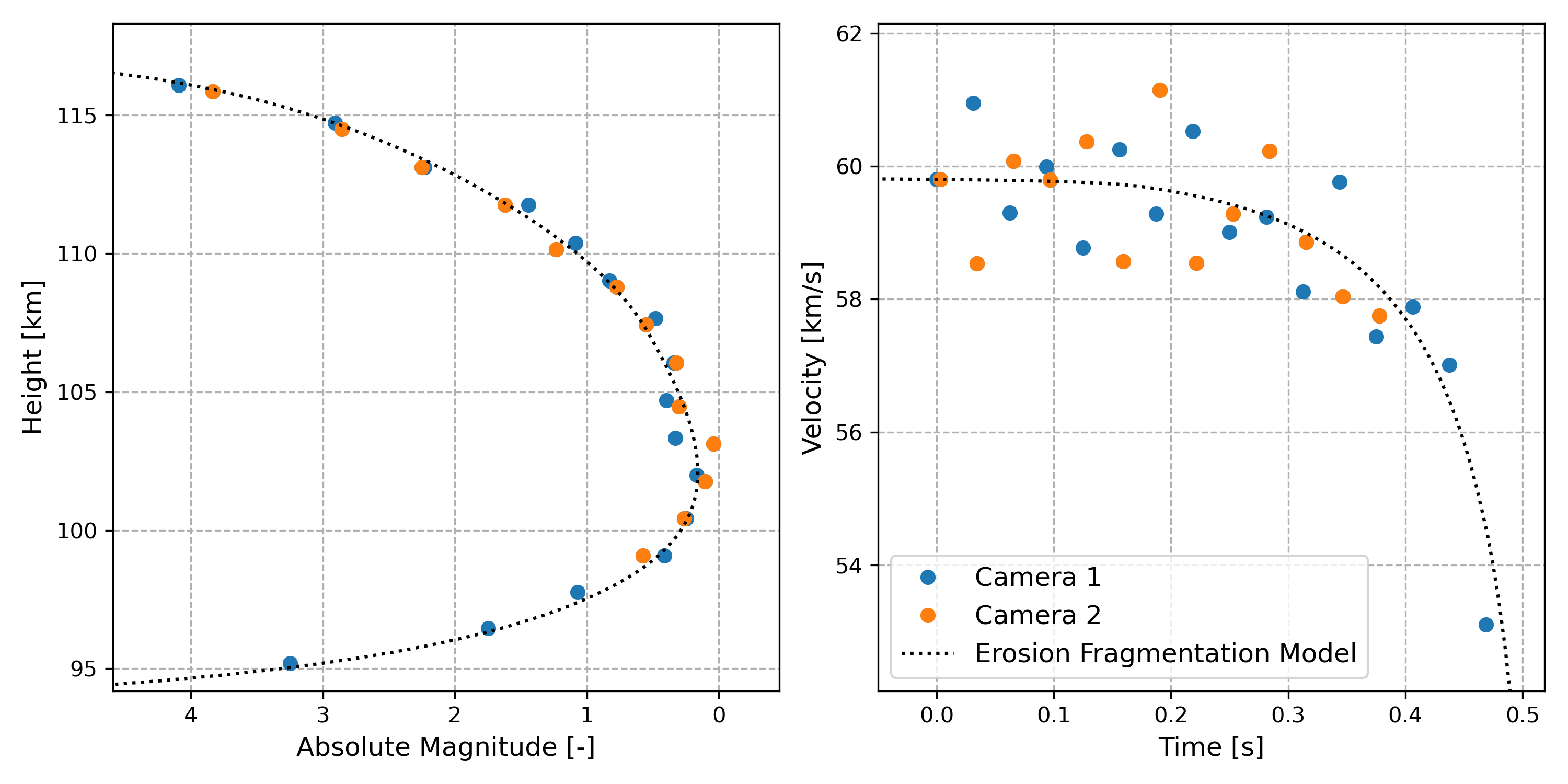}
\caption{An example of the heavy test case showing how noise has been added to the original erosion fragmentation model run (shown as the dashed line) from which it was produced thereby simulating a real observations (represented by the colored points). Here the scatter in synthetic datapoints is produced from Gaussian noise and replicates the average spread in EMCCD measurement uncertainties for PER meteors.}
\label{img:observationsim}
\end{figure}

To simulate real observational conditions, Gaussian noise was introduced to the absolute magnitude and lag data for the six simulation validation cases, with standard deviation values of 40 m for lag and 0.1 for absolute magnitude. This was done to simulate observational uncertainties and match the typical errors observed in real data, where the average PER RMSD for lag is 40 m, and for magnitude is 0.1 (as shown in Table \ref{tab:per_meas_rmsd}). Each simulated meteor was assumed to be detected by two EMCCD cameras, with added randomness in detection times to mimic a real observational setup, as depicted in Figure \ref{img:observationsim}.

For the PCA-based test, we generated 10,000 simulations for each of the six test cases, covering the same range of physical parameters described in Section~\ref{subsec:abl_gen}. Increasing the number of simulations above 10,000 did not significantly alter the outcomes for any test case. Additionally, for each test case, we introduced 50 realizations with noise added to the polynomial fits to mimic the variability expected in real measurements.

For the RMSD-based test, we again used the same parameter ranges but continued generating simulations until at least 100 satisfied the requirement of having both magnitude and lag residuals within $2\times\text{RMSD}$ of the observed data. 

We then combined both sets of simulations (the 10,000 PCA-generated cases and the 100 RMSD-selected cases) to see whether the PCA method consistently identifies the low-RMSD solutions. Finally, we included the noiseless reference simulation in each batch to confirm that both methods correctly identify the ``true'' solution.

Our validation strategy compares the confidence intervals of two different uncertainty estimation methods: the PCA-based approach and the RMSD-based (brute force) approach to determine how well each identifies the ``true'' (reference) physical parameters. Specifically, we verify whether the confidence intervals derived from each method include the real parameter values, and we examine whether the mean or mode of the estimated parameters lie close to those true values.

For the PCA method, we set a threshold in the Principal Component (PC) distance corresponding to the 1st percentile, which, on average, selects 90 out of 10,000 simulations. We then check whether the 10~simulations with the lowest RMSD scores in magnitude and lag also have PC distances that fall under this 1st percentile. If the average PC distance for these top-10 RMSD simulations remains below the 1st-percentile threshold, it indicates that the PCA selection effectively pinpoints models that also fit the data well.

\begin{landscape}
\begin{figure}[h!]
\centering
  \begin{minipage}{0.45\linewidth}
    \centering
    \textbf{PCA} \\ 
    \includegraphics[height=\linewidth]{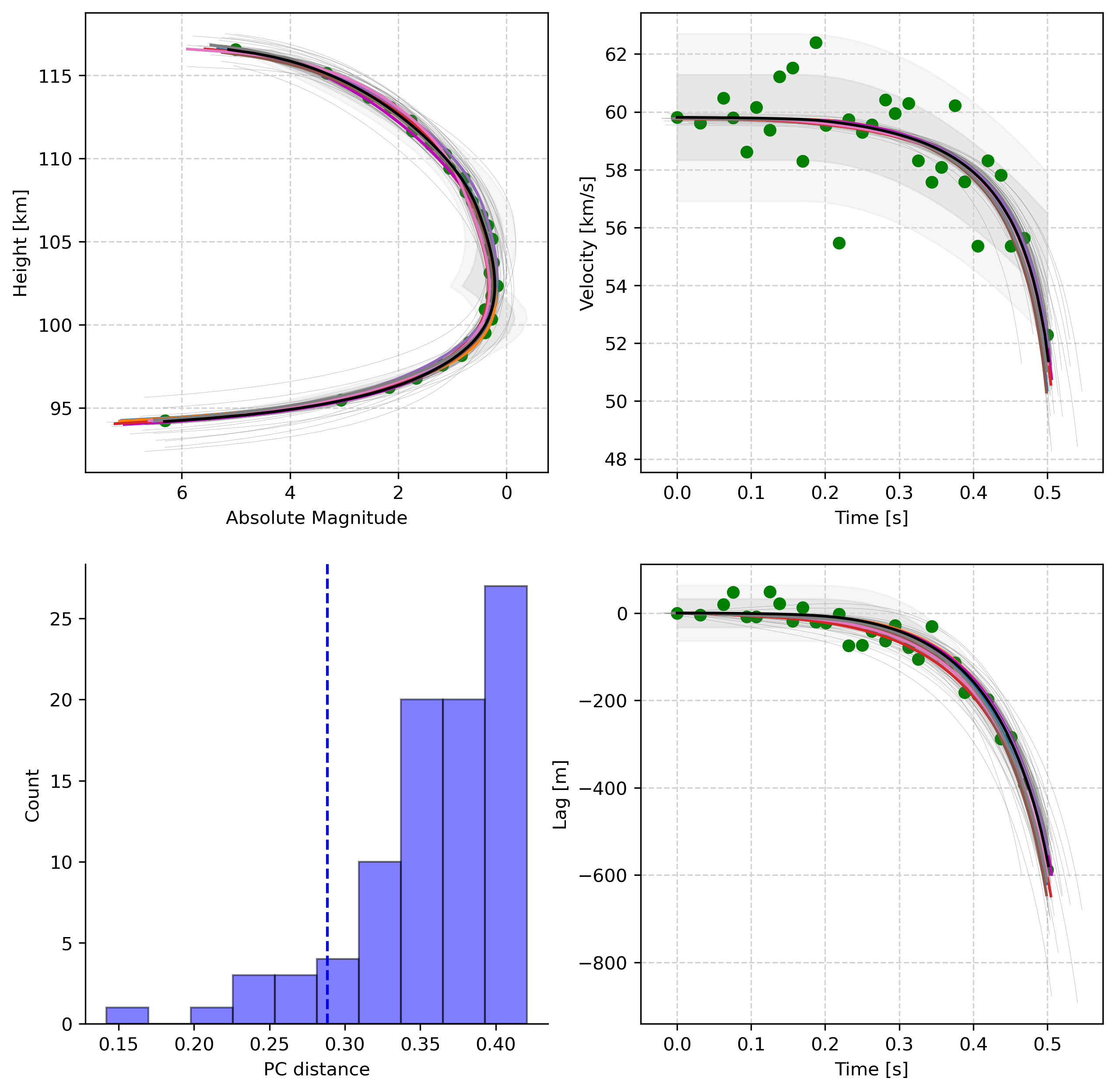}
\end{minipage}
\hfill 
\vrule width 0.5pt 
\hfill 
  \begin{minipage}{0.45\linewidth}
    \centering
    \textbf{RMSD} \\ 
    \includegraphics[height=\linewidth]{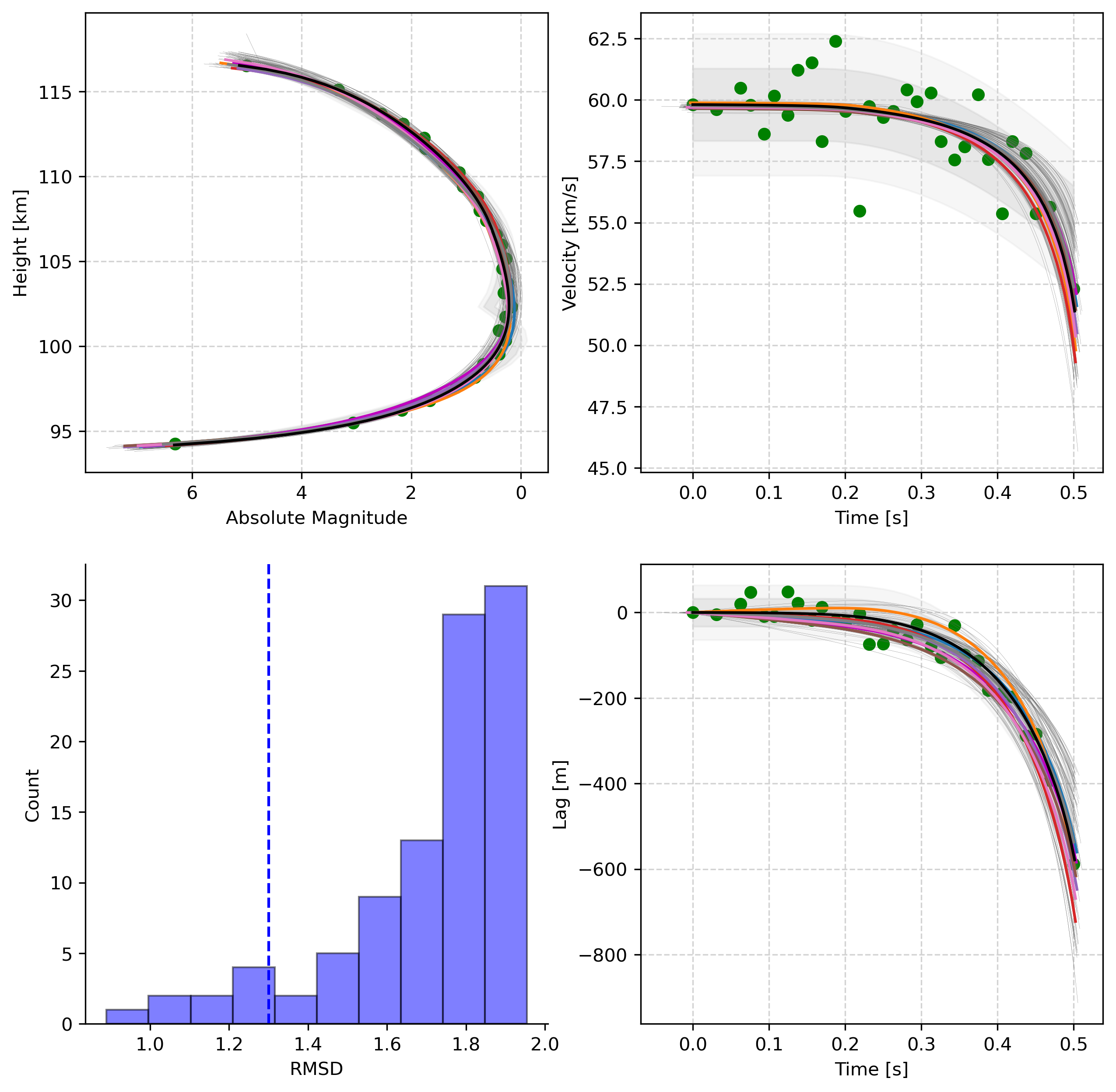}
\end{minipage}
\caption{Light curve, velocity, and lag profiles of the heavy test case for the top 0.1\% of PCA distances (left) and for brute-force simulations below an RMSD threshold of 1.2 (right). The 10 best RMSD solutions in the PCA set have an average PC distance of 0.36 that is below the 1 percentile PC distance  further proving the goodness of the fit. Thin gray lines represent values above the threshold but below 2 RMSD in the brute-force approach, while in the PCA plot, they include all simulations above the threshold and up to the 1st percentile in PC distance.}
\label{img:obs_heavy}
\end{figure}
\end{landscape}

\begin{figure}[h!]
\centering
\begin{minipage}{0.62\linewidth}
    \centering
    \textbf{PCA} \\ 
    \includegraphics[width=\linewidth]{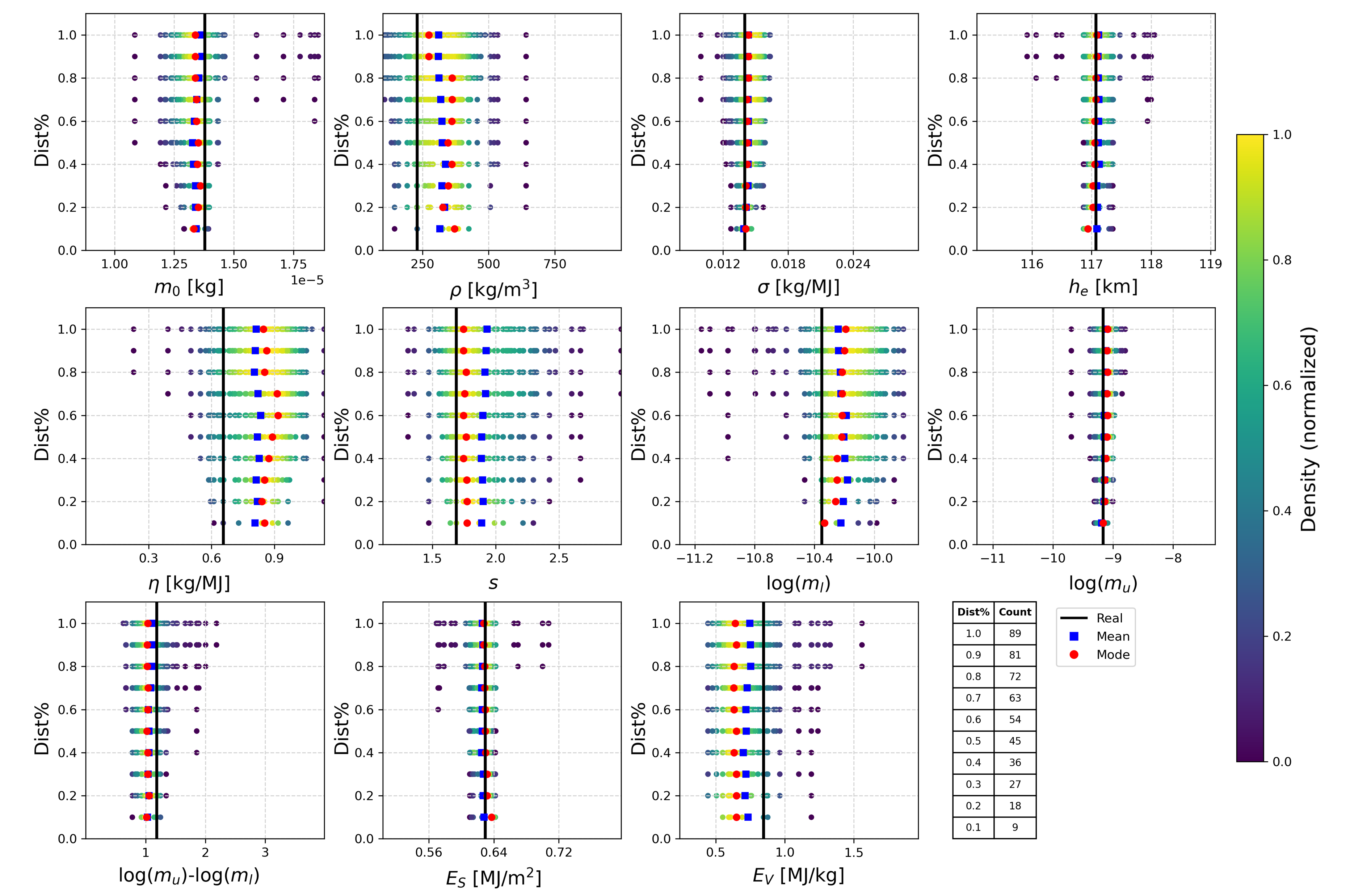}
\end{minipage}
\vspace{1em} 
\begin{minipage}{0.62\linewidth}
    \centering
    \textbf{RMSD} \\ 
    \includegraphics[width=\linewidth]{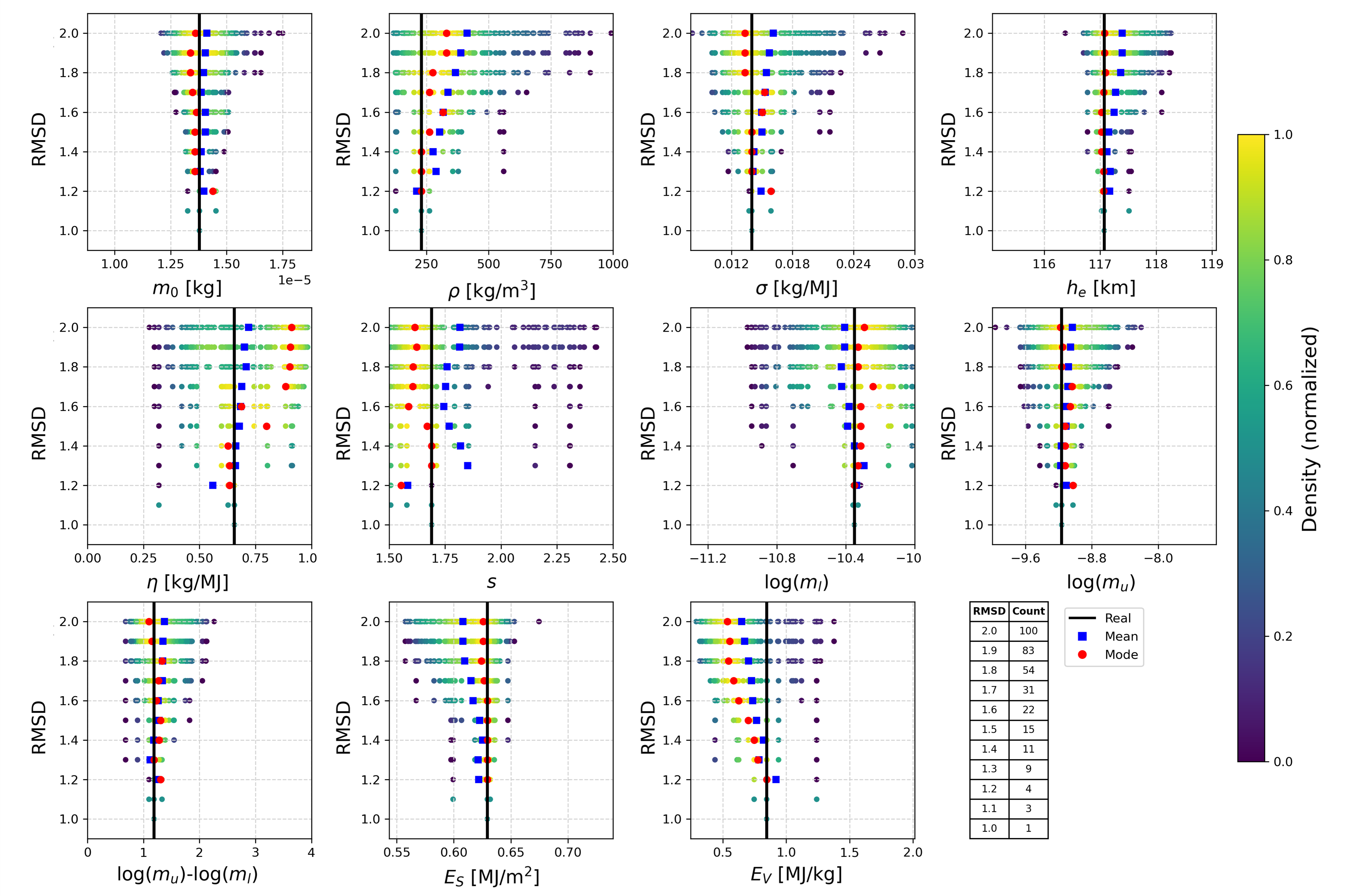}
\end{minipage}

\caption{Waterfall plots for the heavy test case for the PCA method (top) and RMSD approach (bottom). Each colored point represents a selected simulation, colored based on the normalized density of points. In both sets of waterfall plots points near the top of the graphs are furthest from the actual model solution which is given by a vertical black line. This shows how well each approach converges (or not) toward the real solution as the threshold cuts are made more conservative (lower y values) at the expense of fewer solutions. The mean/mode provide a simple visual indicator of the population values as a whole. The number of simulations included in the statistics as a function of threshold value are shown in the table for each approach.} 
\label{img:wat_heavy}
\end{figure}

Both methods achieve their best results for the heavy test case, generating a cluster of simulations that correctly includes the true physical values and aligns well with the observed magnitude and lag data (see Figures~\ref{img:obs_heavy} and \ref{img:wat_heavy}). Notably, the PCA-based approach yields a significantly tighter confidence interval than the RMSD-based approach. Moreover, for the 10~simulations with the lowest RMSD values, the average PC distance is only 0.36, which remains well below the 1\% threshold in PC space.



\begin{landscape}
\begin{figure}[h!]
\centering
  \begin{minipage}{0.45\linewidth}
    \centering
    \textbf{PCA} \\ 
    \includegraphics[height=\linewidth]{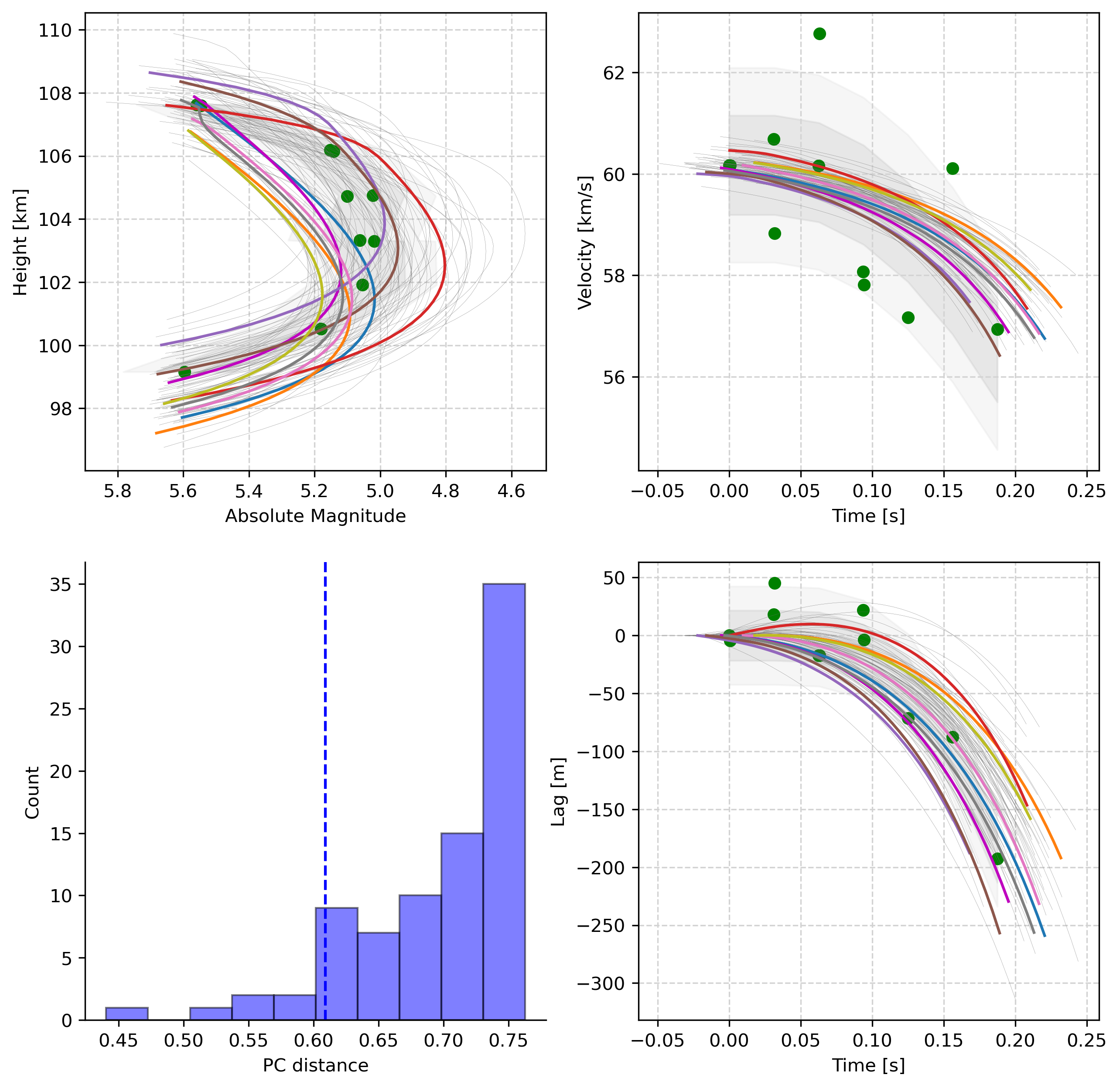}
\end{minipage}
\hfill 
\vrule width 0.5pt 
\hfill 
  \begin{minipage}{0.45\linewidth}
    \centering
    \textbf{RMSD} \\ 
    \includegraphics[height=\linewidth]{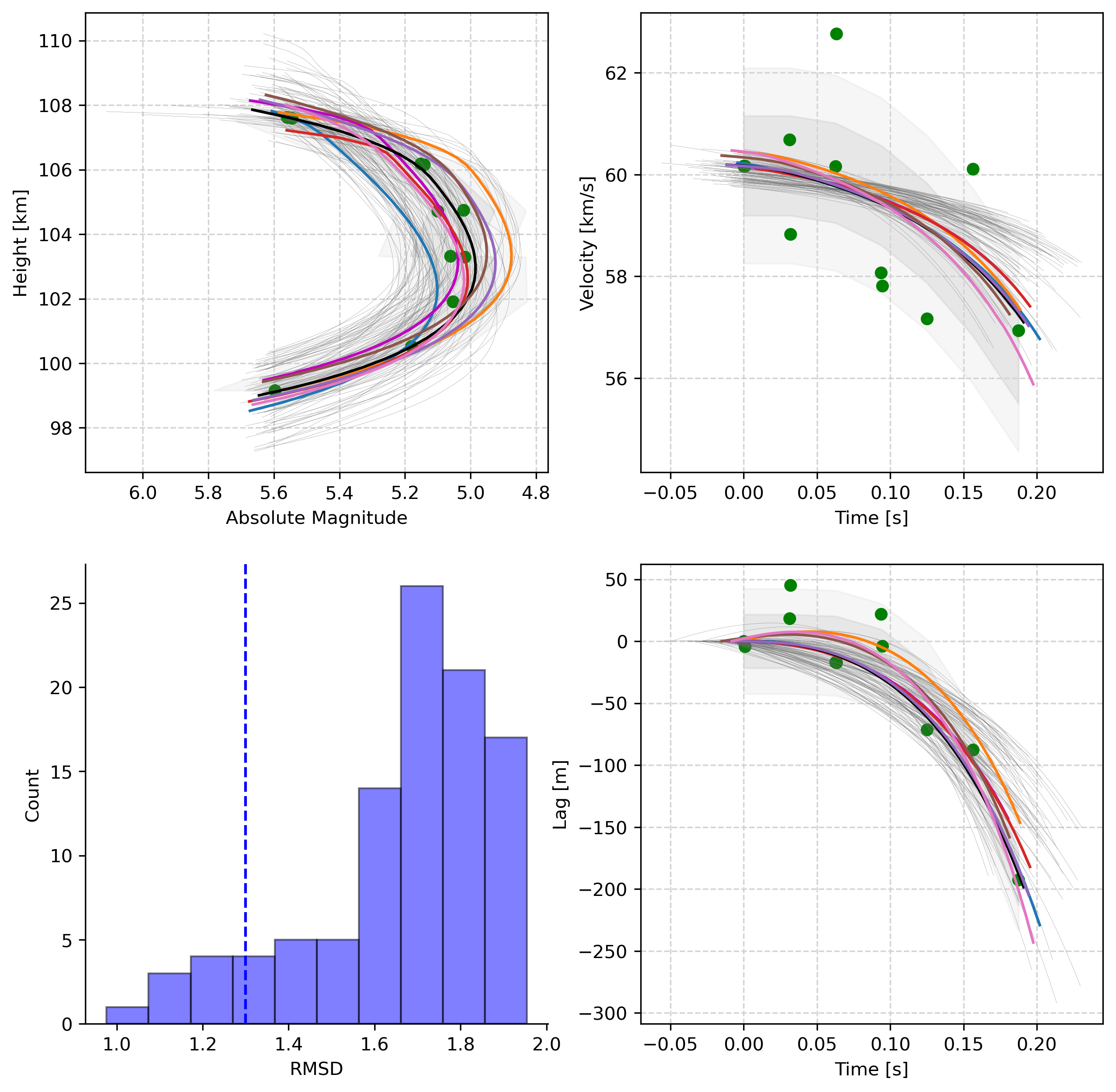}
\end{minipage}
\caption{Light curve, velocity, and lag profiles of the light test case for the top 0.1\% of PCA distances (left) and for brute-force simulations below an RMSD threshold of 1.3 (right). The 10 best RMSD solutions in the PCA set have an average PC distance of 0.86.}
\label{img:obs_light}
\end{figure}
\end{landscape}

\begin{figure}[h!]
\centering
\begin{minipage}{0.8\linewidth}
    \centering
    \textbf{PCA} \\ 
    \includegraphics[width=\linewidth]{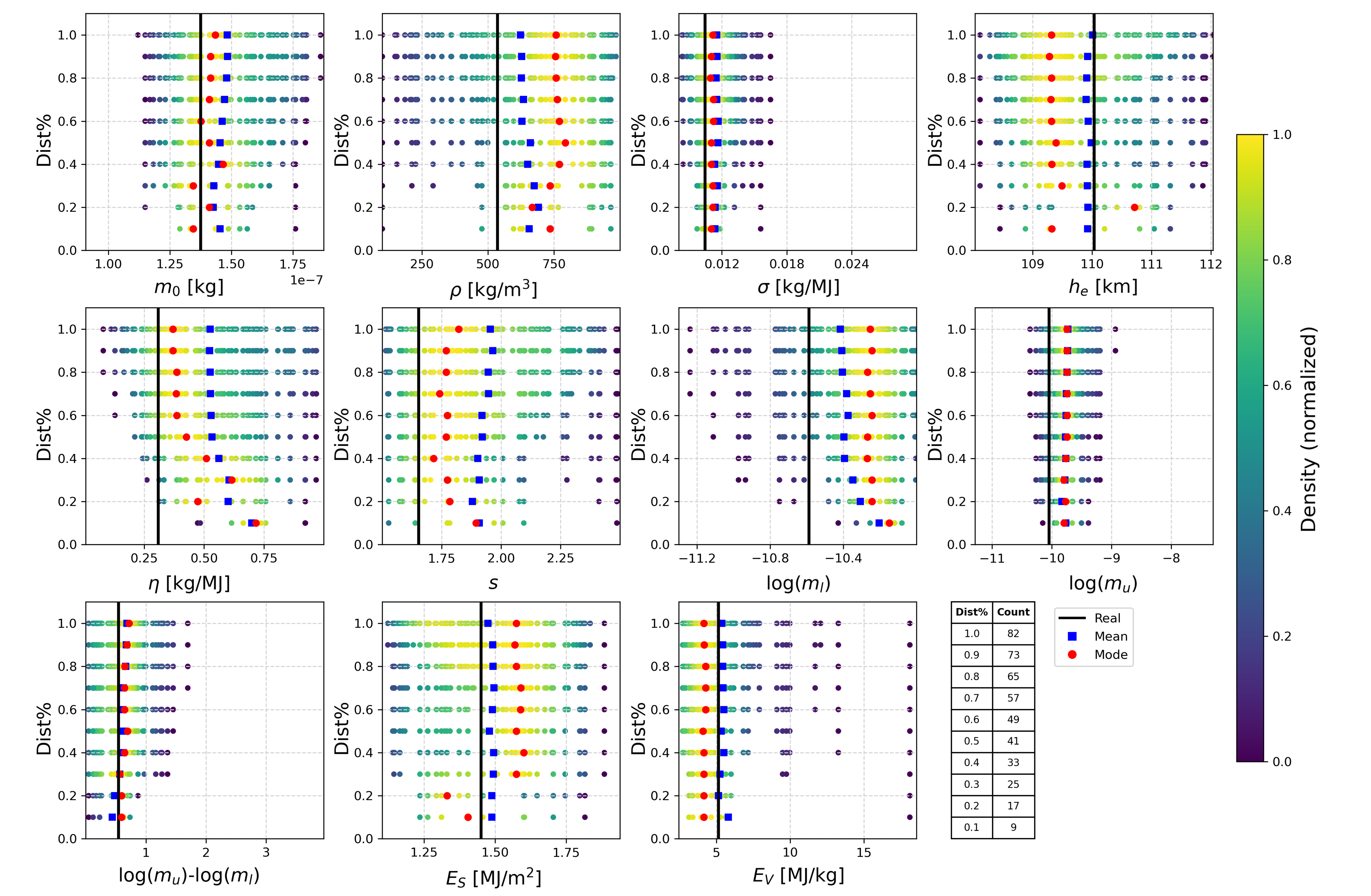}
\end{minipage}
\vspace{1em} 
\begin{minipage}{0.8\linewidth}
    \centering
    \textbf{RMSD} \\ 
    \includegraphics[width=\linewidth]{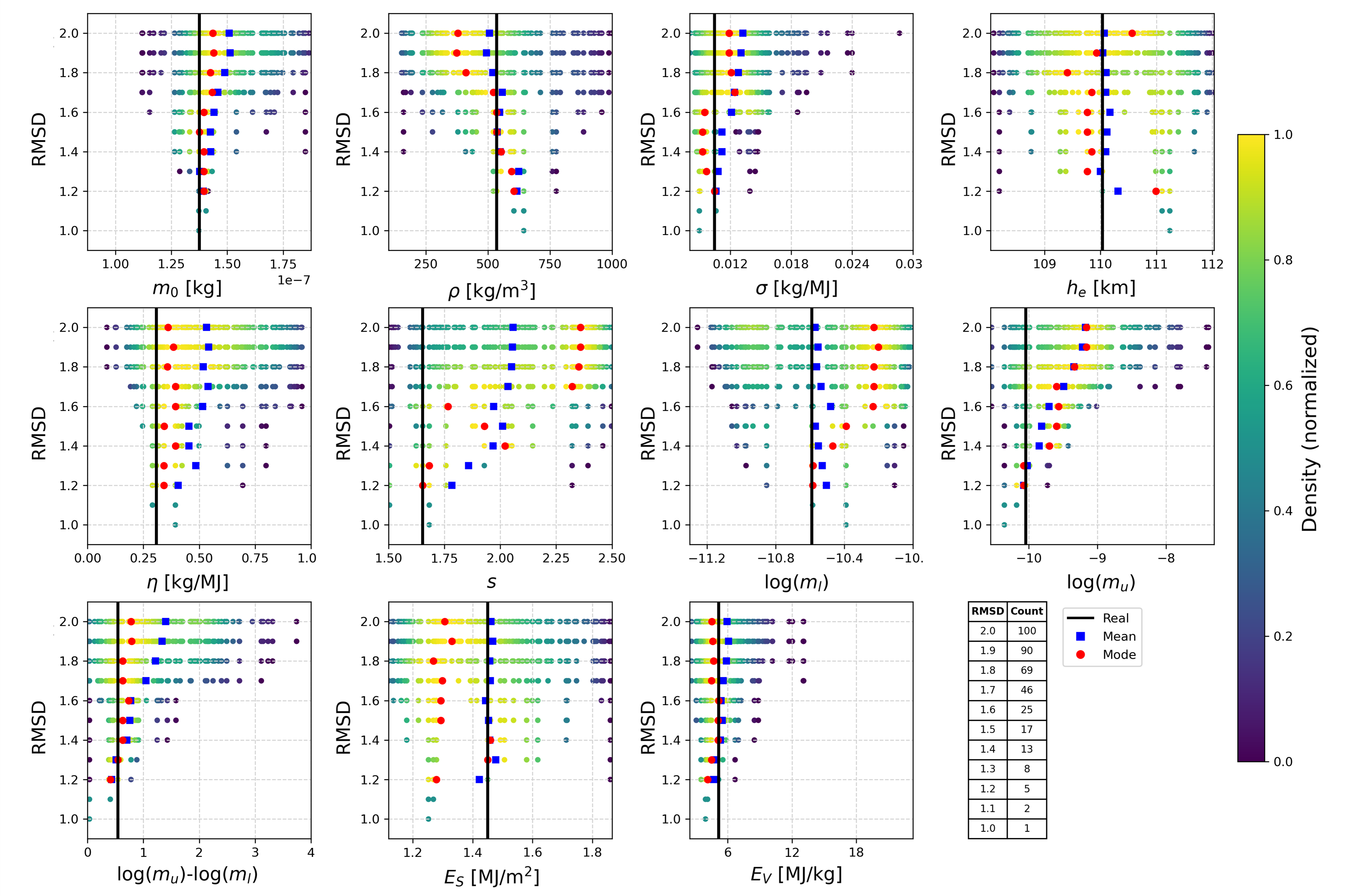}
\end{minipage}

\caption{Waterfall plots for light test case for the PCA method (top) and RMSD results (bottom) for the light test case. In both cases, the mode and mean do not always align with the real physical value but the real values are consistently confined within the uncertainty.}
\label{img:wat_light}
\end{figure}

Among all the test case the worst performing for both approaches was found for the light simulation. The resulting distribution for both methods resulted in a broad distribution of possible physical parameter as shown in Figure \ref{img:wat_light} and a poor fit especially for the PCA approach as shown in figure \ref{img:obs_light}. For this test case the average value for the 10 best RMSD is 0.86 that which is above the 1\% of the PC total distance. This indicates PCA is poor at inverting for the original physical parameters.

The main reason for the difference in fit robustness between these two test cases is simply the number of points (or equivalently the event duration). Both methods struggle when there are few observed points, while convergence is better for longer lasting events (such as the heavy test case).
 


The confidence interval for the physical characteristics derived via the brute-force method consistently includes the true parameter values across all six validation test cases. In contrast, the PCA-based confidence interval fails to capture the true ablation coefficient and mass distribution index for the \emph{steep} case, indicating convergence toward an incorrect solution. This discrepancy is further supported by the poor fit in both magnitude and lag seen in Appendix~\ref{sec:Apx A}. 

\begin{landscape}
\begin{figure}[h!]
\centering
  \begin{minipage}{0.45\linewidth}
    \centering
    \textbf{PCA} \\ 
    \includegraphics[height=\linewidth]{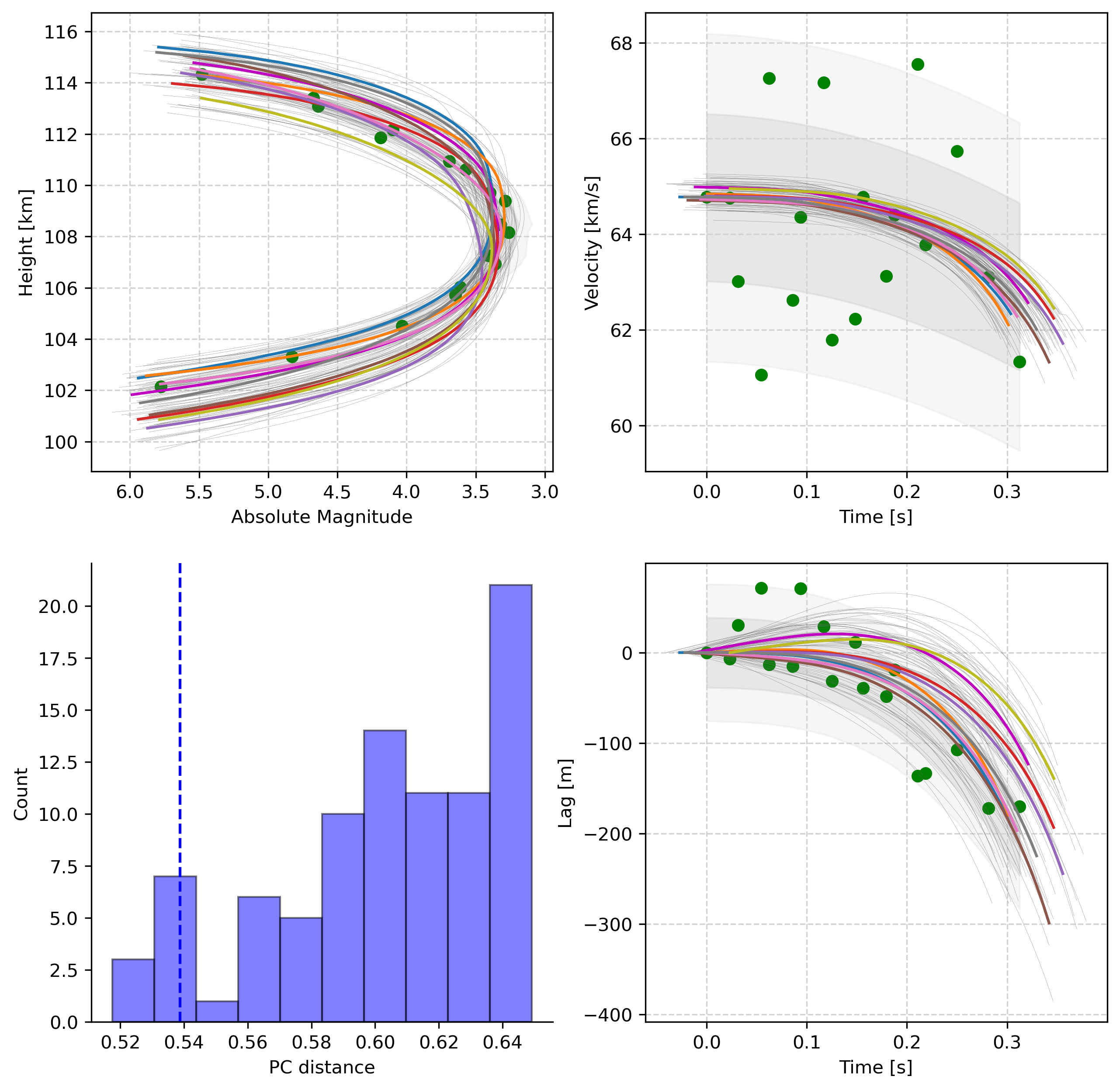}
\end{minipage}
\hfill 
\vrule width 0.5pt 
\hfill 
  \begin{minipage}{0.45\linewidth}
    \centering
    \textbf{RMSD} \\ 
    \includegraphics[height=\linewidth]{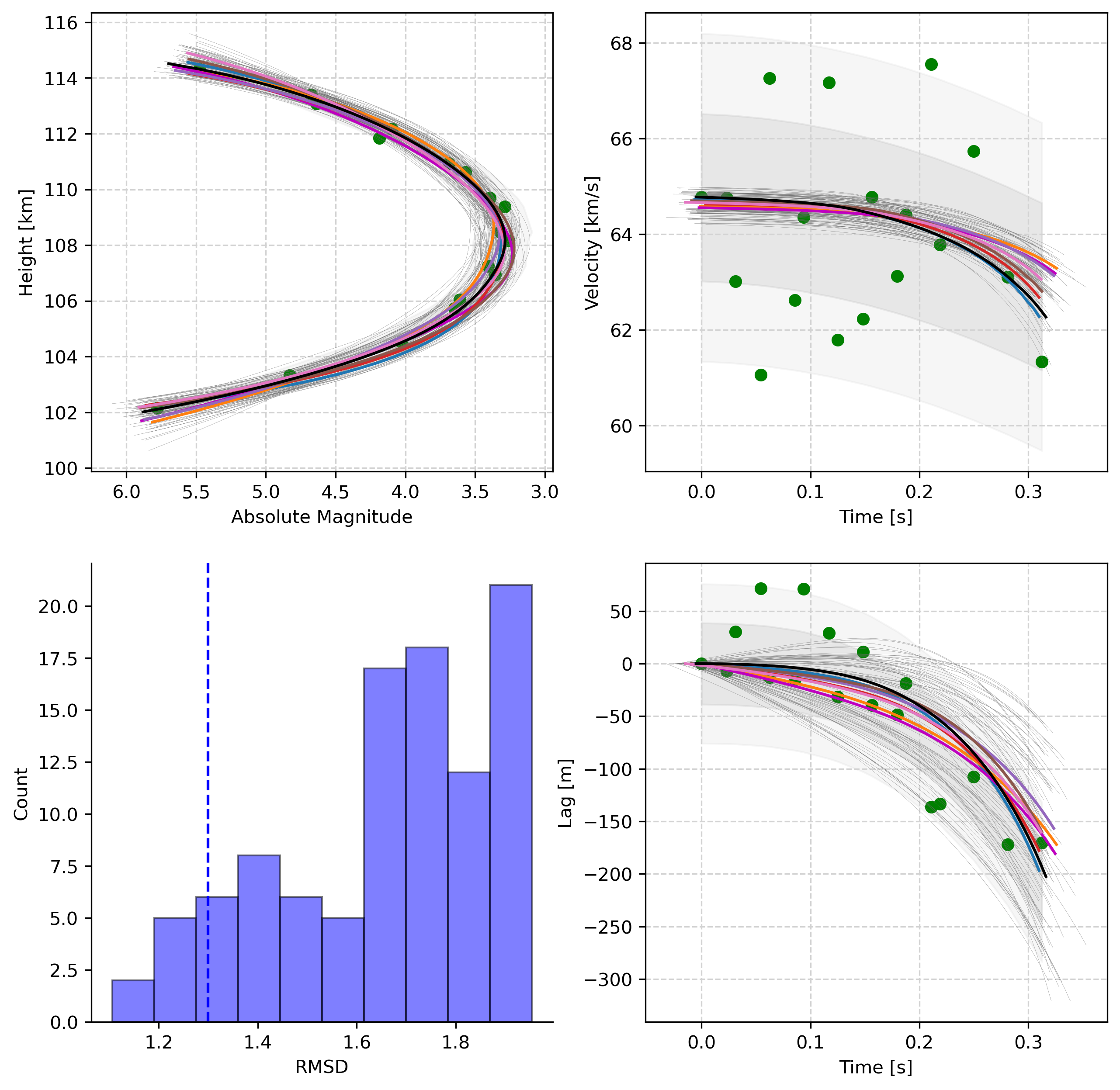}
\end{minipage}
\caption{Light curve, velocity, and lag profiles of the fast test case for the top 0.1\% of PCA distances (left) and for brute-force simulations below an RMSD threshold of 1.3 (right). The 10 best RMSD solutions in the PCA set have an average PC distance of 0.73. It is clear the simulation fits in lag and brightness for PCA, in particular, are poor. The RMSD fits are better but still suffer from the large relative spread in lag due to high measurement uncertainty.}
\label{img:obs_fast}
\end{figure}
\end{landscape}

\begin{figure}[h!]
\centering
\begin{minipage}{0.8\linewidth}
    \centering
    \textbf{PCA} \\ 
    \includegraphics[width=\linewidth]{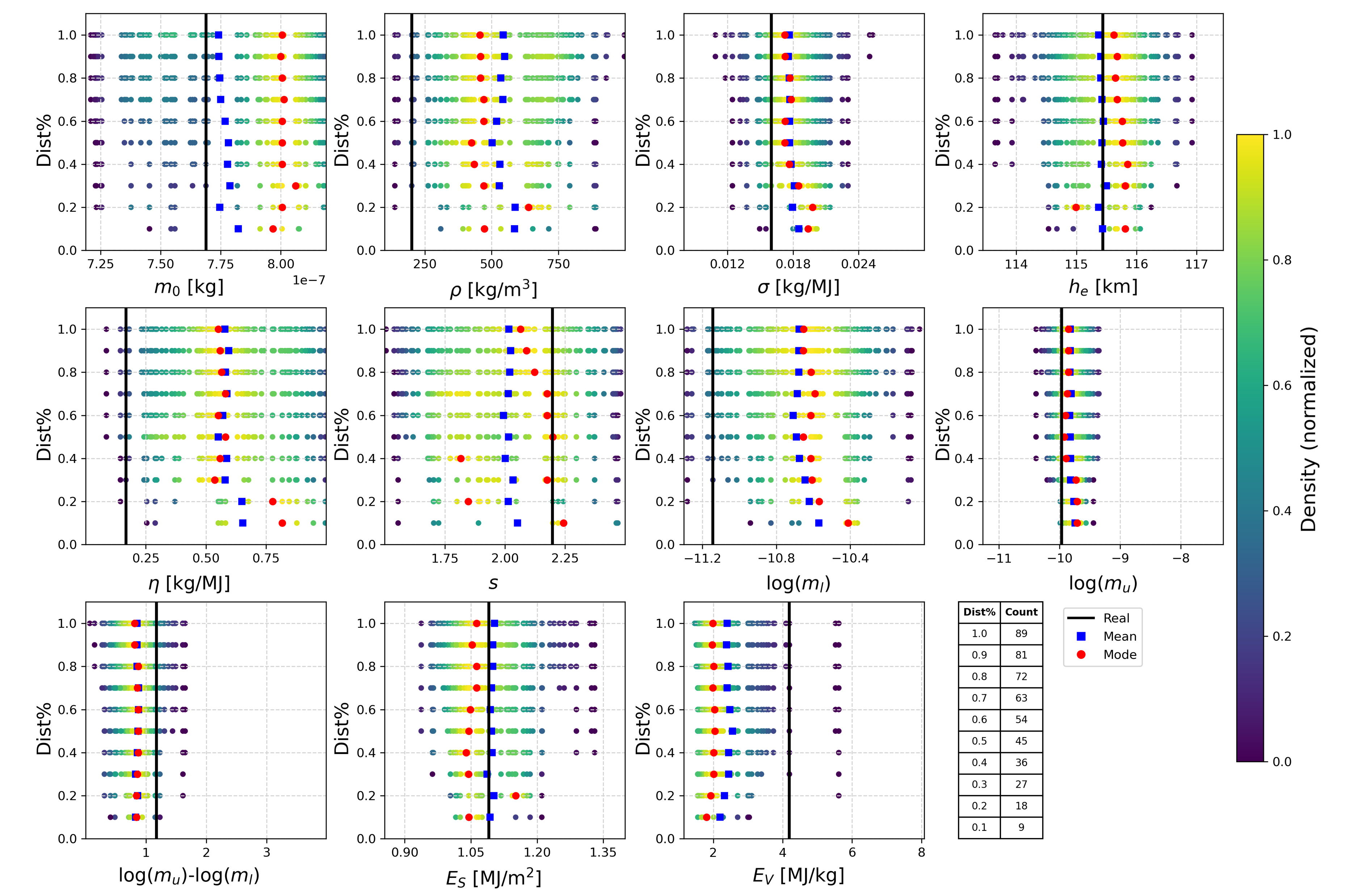}
\end{minipage}
\begin{minipage}{0.8\linewidth}
    \centering
    \textbf{RMSD} \\ 
    \includegraphics[width=\linewidth]{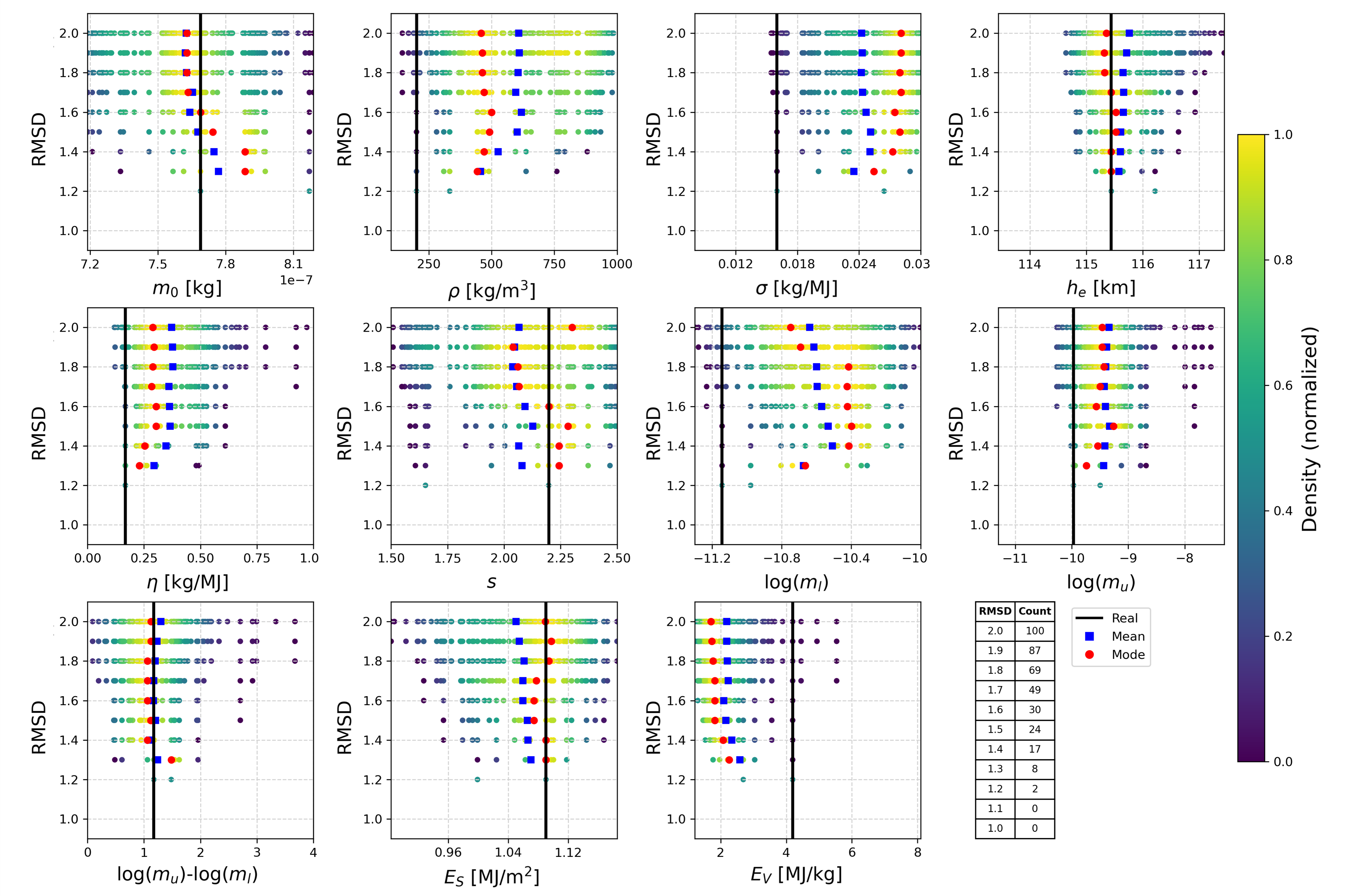}
\end{minipage}

\caption{Waterfall plot of the fast test case for the PCA method (top) and the RMSD results (bottom). In both cases, the mode and mean are quite far from the real density value, though that value is just inside the formal limits of the confidence interval.}
\label{img:wat_fast}
\end{figure}

Additionally, for the \emph{fast} case, the density value lies at the boundary of the confidence interval in both methods (see Figure~\ref{img:wat_fast}). For this same test case, the mean PC distance among the 10~best RMSD simulations is 0.73, which exceeds the 1\% threshold (Figure~\ref{img:obs_fast}). This highlights the limitations of the PCA-based approach.



Appendix \ref{sec:Apx A} provides similarly detailed plots for the remaining test cases and both methods.

The fast, steep, and light test cases show the least deceleration among the six scenarios, with all lag data points remaining below 300 m. Since the lag values cover a relatively narrow range, measurement noise (40 m) in this regime exerts a proportionally larger impact.

One other critical factor affecting the precision of physical parameter estimation in both methods is the introduced noise in magnitude (0.1 mag). 

These noise levels become particularly significant for faint meteors with smaller deceleration. For this reason and due to the larger number of simulated observed points, the heaviest case with a mass of $10^{-5}$kg yields the best results, while the rest of the test cases, with a mass in the order of $10^{-7}$kg, present greater challenges. The comparatively small range in lag and magnitude relative to the noise reduces the method's ability to precisely invert the physical parameters. This underscores that the RMSD and the observable variables are strongly affected by the noise in the data and so cannot substantially improve the precision of the uncertainties in density within the wide initial simulation range.

\subsubsection{Validation with CAMO Lag Noise}\label{subsec:CAMOlag}

\begin{landscape}
\begin{figure}[h!]
\centering
  \begin{minipage}{0.45\linewidth}
    \centering
    \textbf{PCA} \\ 
    \includegraphics[height=\linewidth]{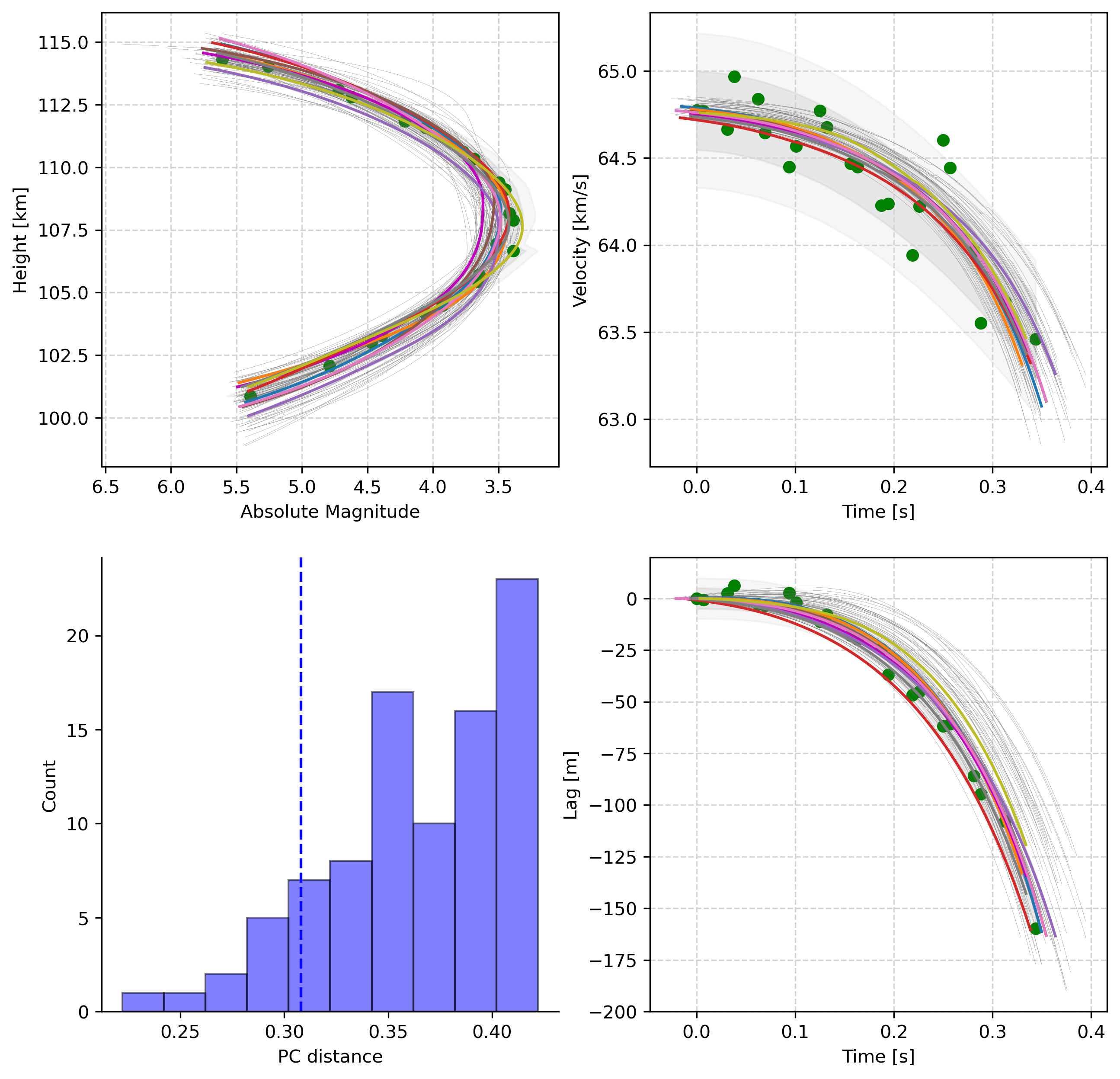}
\end{minipage}
\hfill 
\vrule width 0.5pt 
\hfill 
  \begin{minipage}{0.45\linewidth}
    \centering
    \textbf{RMSD} \\ 
    \includegraphics[height=\linewidth]{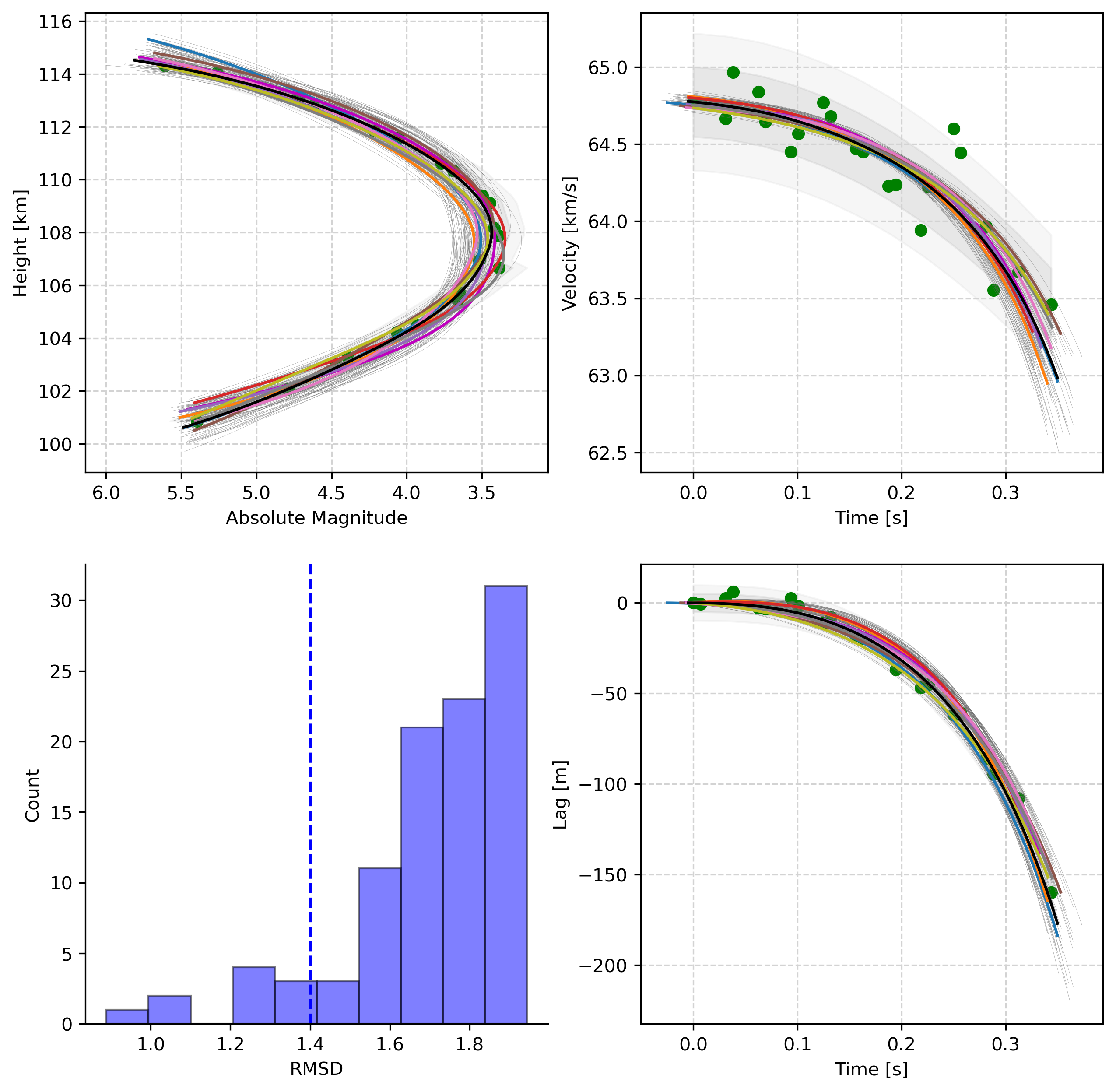}
\end{minipage}
\caption{Light curve, velocity, and lag profiles of the fast test case with CAMO lag noise for the top 0.1\% of PCA distances (left) and for brute-force simulations below an RMSD threshold of 1.4 (right). The 10 best RMSD solutions in the PCA set have an average PC distance of 0.37 that is below the 1\% percentile PC distance showing the effectiveness of PCA to find the best RMSD.}
\label{img:obs_fast_CAMO}
\end{figure}
\end{landscape}

\begin{figure}[h!]
\centering
\begin{minipage}{0.8\linewidth}
    \centering
    \textbf{PCA} \\ 
    \includegraphics[width=\linewidth]{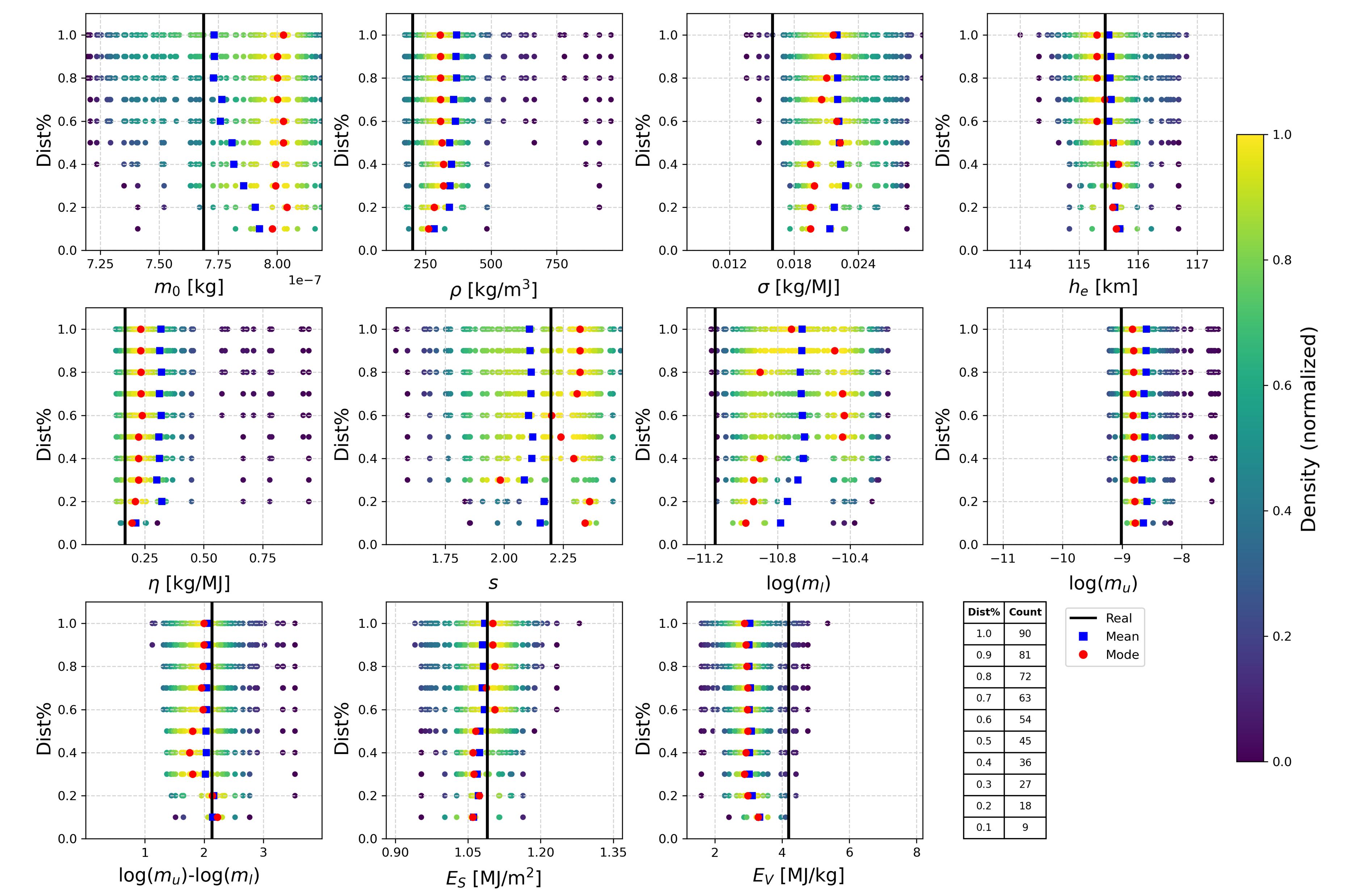}
\end{minipage}
\begin{minipage}{0.8\linewidth}
    \centering
    \textbf{RMSD} \\ 
    \includegraphics[width=\linewidth]{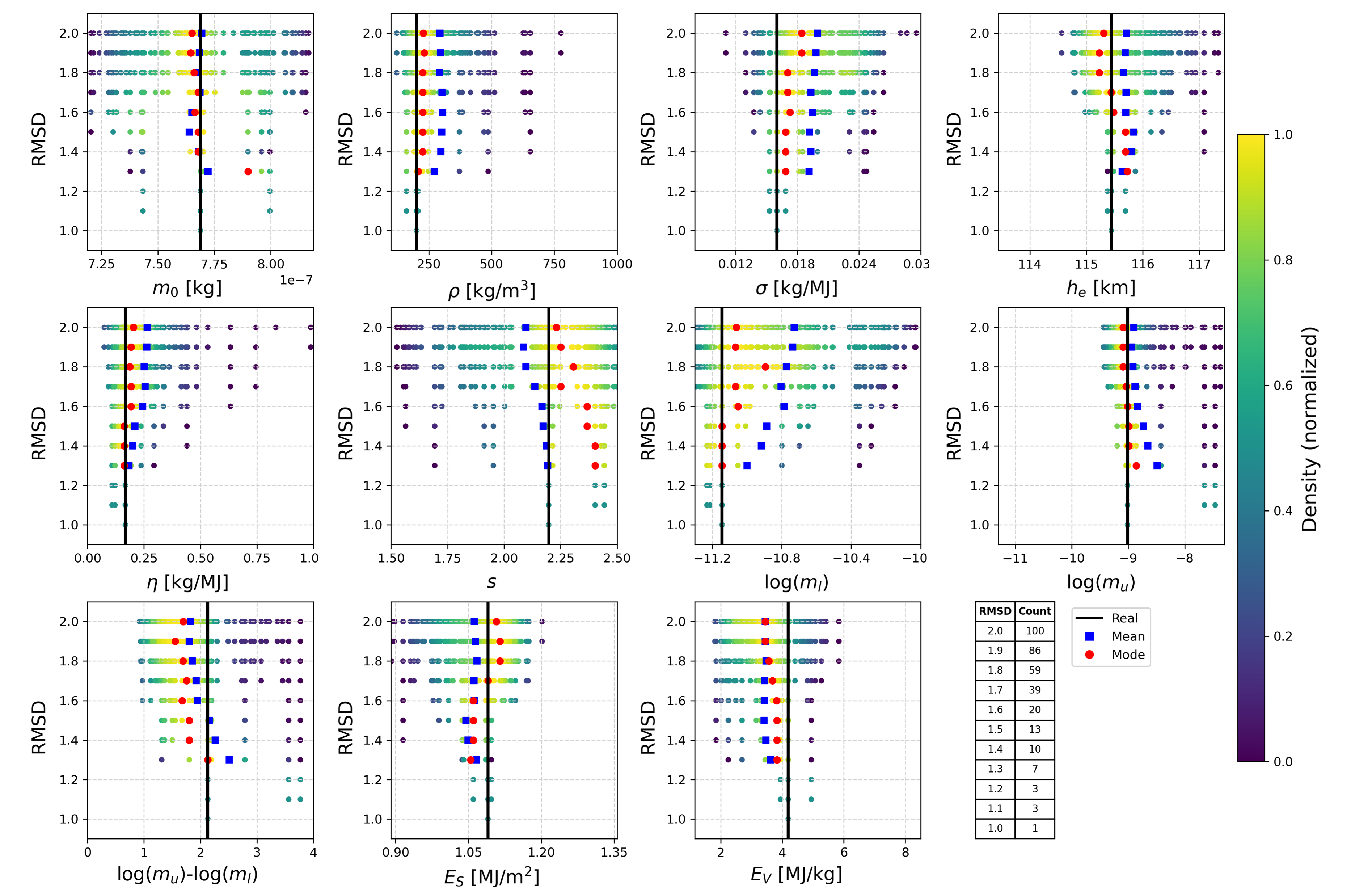}
\end{minipage}
\caption{Waterfall plots for light test case for the PCA method (top) and RMSD results (bottom) for the light test case. In both cases, while the mode and mean do not always align with the real physical value the real value is consistently within the uncertainty bounds defined from the method.}
\label{img:wat_fast_CAMO}
\end{figure}

To address this fundamental limitation and improve density estimates for small meteoroids, we explored how reducing the lag noise from 40\,m to 5\,m among our six validation test cases affects results. This lag error is achievable with the CAMO mirror-tracking system. Although neither our EMCCD nor CAMO cameras can achieve better than 0.1\,mag precision for brightness, CAMO's enhanced lag precision significantly tightens the resulting parameter estimates. As shown in Figures~\ref{img:obs_fast_CAMO} and \ref{img:wat_fast_CAMO}, this reduction in lag noise produces both more precise and accurate density estimations for our simulated test cases. The resulting average PC distance of the 10 best simulations in RMSD is 0.37 well below the 1\% of PC distance. Of the two methods, the brute-force approach consistently yields the best overall performance under these lower-noise conditions.


Overall, our validation tests show that both the PCA-based and brute-force (RMSD) methods can identify plausible initial physical parameters for bright, long-lasting, strongly decelerating meteors using the EMCCD system or for less ideal cases when lag measurements have the precision possible from CAMO.  However, the brute-force approach provides more transparent control over the uncertainty thresholds, since researchers can directly adjust the RMSD cutoff to refine or broaden the range of acceptable simulations. By contrast, while the PCA approach also yields good results, interpreting the PC distance and translating it into a practical uncertainty measure can be less straightforward. Consequently, in terms of ease of application and clarity, the brute-force RMSD method emerges as the more reliable option for general use in meteor modeling.

Relaying solely on a linear combination of observables via the PC approach does not yield the best light curve and deceleration match. Notably, the best performance for both methods are achieved for the most massive and long-lasting meteoroid test case, which had a mass two orders of magnitude greater than our other test cases. Meteoroids with masses as large or larger than this heavy meteoroid type constitutes less than 1\% of the recorded EMCCD PER meteors, underscoring the challenges in constraining the uncertainties in the physical parameters for meteoroids in showers using lower metric resolution systems.

\section{Results} \label{sec:results}

Given that most Perseid meteors are fast and relatively faint, our validation tests highlight the difficulty of estimating their densities when only EMCCD camera data is available. The inherent noise caused by lower temporal and spatial resolution in EMCCD measurements compared to CAMO, particularly for lighter meteoroids with small deceleration, makes it challenging to isolate physically consistent parameters using the RMSD method. Consequently, we sought other strategies to reduce noise in order to improve density estimates for these meteors.

Our initial attempt to derive physical parameters for the Perseid meteors relied on an automated data-reduction pipeline based on  the work described in \citet{gural2022development}. The attraction of this pipeline is that large numbers of meteors have already been automatically measured, reducing the required manual work. However, we found that the automated picks often produced unphysical decelerations, sometimes giving a positive or linear deceleration, for these fast and faint PER meteors.  This prompted us to switch to manual picking of the leading edge in each video frame. We focused on the brightest PER events to make it easier to produce high-quality manual picks.

In an effort to reduce noise, we meticulously adjusted the manual picks to produce smoother lag profiles. Although this method offered more fine-grained control, it also introduced new sources of subjectivity in the picks that lead to systematic biases in our lag measure. Of the 34 bright Perseid events initially processed, only 12 yielded data that fit any physically plausible solution. The remaining events exhibited unrealistic decelerations or light-curve shapes that could not be matched within $2\times\mathrm{RMSD}$.

We attribute these mismatches to the limits of our EMCCD camera data. In particular, striving for sub-pixel accuracy when identifying the meteor’s leading edge can paradoxically degrade the overall fidelity of the measurement. Although such picks may reduce the apparent lag scatter, they push the analysis below the true noise threshold of the camera. As a result, the calculated deceleration slope no longer aligns with the meteor’s light curve, magnifying measurement errors and precluding a physically consistent solution. The result is lag measurements that appear very precise but are in fact not accurate.

To overcome this limitation and constrain the density estimates we searched for EMCCD PER meteors that were also well tracked by the CAMO system. Only two Perseids satisfied this criterion, and we combined the CAMO data with our EMCCD measurements, using CAMO's more accurate picks to guide and refine the EMCCD-based reductions. This hybrid approach preserved the broader brightness range captured by the EMCCD and leveraged CAMO's higher spatial accuracy, mitigating the noise-induced uncertainties that had confounded purely EMCCD-based solutions.

\subsection{Uncertainty distributions of meteoroid physical parameters for two PER EMCCD-CAMO simultaneous events}

This section presents the outcomes of our modeling approach designed for the automated inversion of meteoroid physical parameters using the RMSD threshold method (as described in Section~\ref{subsec:RMSD_sel}). We apply it to two real Perseid (PER) meteors \texttt{20210813\_061452} and \texttt{20230811\_082649} that were observed by both EMCCD systems and the CAMO instrument. This work represents the first instance of deriving probabilistic distributions of physical parameters for individual recorded meteors.

To begin, we relied solely on EMCCD data for each meteor, using manually picked leading-edge positions without further adjustments. Table~\ref{tab:per_meas_rmsd} summarizes the three sets of manual picks performed by different analysts. Despite these efforts, density estimates remained weakly constrained for both meteors, spanning nearly the entire initial density search range. For meteor \texttt{20210813\_061452}, the observed brightness varies between 4.8 and 1.6 magnitudes, while lag extends from 450\,m down to 150\,m among the three analyst picks. Meteor \texttt{20230811\_082649} exhibits a brightness range from 5.5 to 2.6 magnitudes, with lag values between 420\,m and 75\,m. As shown in Appendix~\ref{sec:Apx B}, these broad observational ranges, combined with the more limited EMCCD precision, prevented a decisive convergence to any specific density value during the inversion process.

These findings align with our validation studies, where EMCCD data for lower-mass meteoroids (except the ``heavy'' test case) led to broad, imprecise density estimates. Indeed, the two real meteors studied here each have an estimated mass on the order of $10^{-6}$\,kg, making their density inference particularly sensitive to measurement noise. This result underscores the importance of achieving lower noise levels to constrain physical parameters more effectively. The next section addresses this challenge by fusing the higher-precision CAMO measurements with the more sensitive EMCCD data.

The EMCCD cameras, with their high sensitivity to faint targets, capture the very early onset of meteoroid ablation and thus record the full evolution of the light curve. In contrast, the CAMO narrow-field system, although less sensitive, provides superior deceleration data because of its high temporal and spatial resolution. Consequently, CAMO can generate more precise deceleration profiles, whereas EMCCD data typically reveal earlier portions of the trajectory but with lower positional accuracy.

To capitalize on these complementary strengths, we used the CAMO data to inform and refine the EMCCD picks, thereby reducing lag noise while still ensuring realistic deceleration measurements that also align with the observed light-curve. The results was that the RMSD approach inverted to a much narrower density range. This result highlights the value of fusing EMCCD and CAMO data to better constrain meteoroid density. Both uncertainty distributions indicate values consistent with typical low-density cometary material, corroborating previous findings based on CAMO data from \cite{buccongello2024physical} who found an average density for PER of 365 ± 134 kg/m\(^3\).

As CAMOs high resolution also provides spatially resolved wake, we compared the wake simulations for the selected simulations against CAMO’s recorded wake. We found strong agreement with our best-fit selected simulations (see Appendix~\ref{sec:Apx C}). We describe in detail the measurements, fit and resulting physical parameters for each of these two observed meteors in the following sections.

\subsubsection{Meteor 20210813\_061452}\label{subsubsec:2021}

This meteor, previously analyzed using CAMO data by \cite{buccongello2024physical}, was also captured by our EMCCD cameras. We manually reduced the EMCCD and CAMO data and used our RMSD technique to define a range of possible physical characteristics for this Perseid meteor. For the modeling, we fixed the zenith angle $z_c$ to 45.348°, as derived from the trajectory solution. All other fixed parameters used in the simulation are listed in Table~\ref{tab:fixed_params}.

\begin{figure}[h!]
\centering
\includegraphics[width=\linewidth]{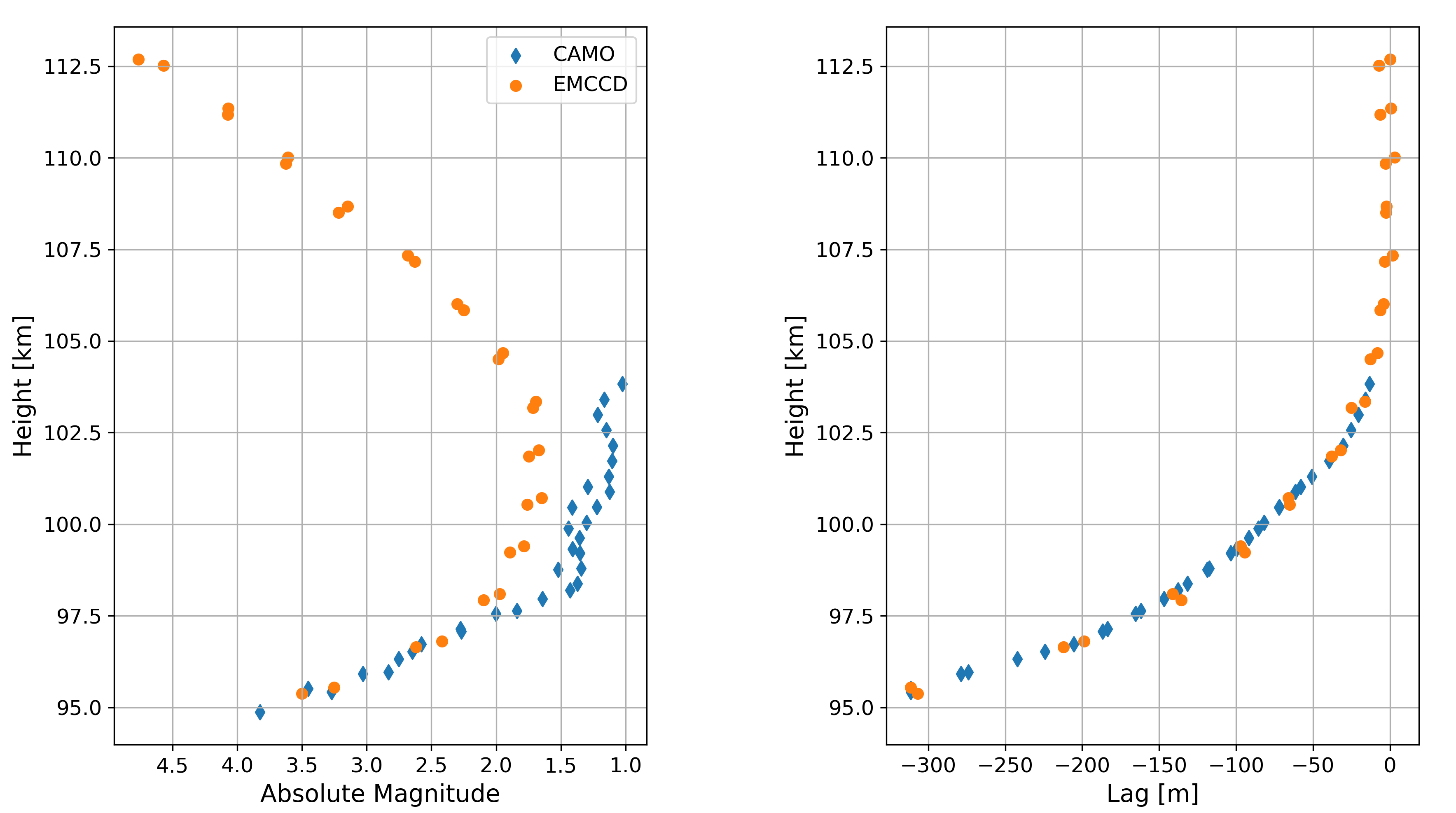}
\caption{A direct comparison of simultaneous measurements from the CAMO narrow-field system and EMCCD data for PER 20210813\_061452. Here the lag as a function of height from the two station EMCCD solution (orange circles) and the two station CAMO narrow field solution (blue diamonds) are directly compared in the left plot. The left plot shows the absolute magnitude measured by the narrow field CAMO and the EMCCD adjusted to the G-band. The apparent offset in brightness between the two datasets at the peak is primarily due to differences in spectral response between the two camera systems. Here the CAMO deceleration profile measured by \cite{buccongello2024physical} was used to inform our EMCCD picks.}
\label{img:CAMO_EMCCD_nick}
\end{figure}

\begin{figure}[h!]
\centering
\includegraphics[width=\linewidth]{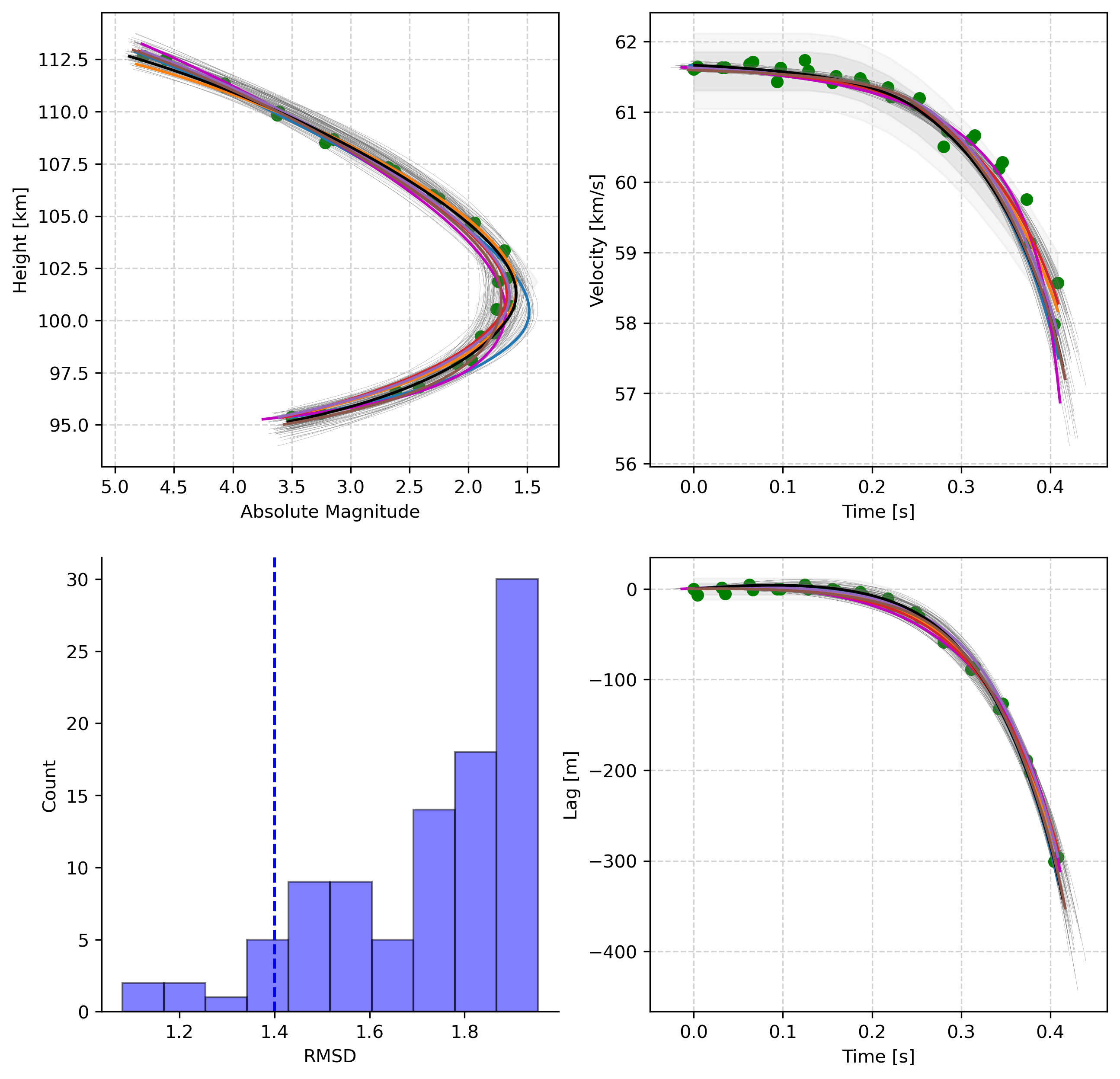}
\caption{The selected simulations with both normalized RMSD below 1.5 for the 20210813\_061452 EMCCD CAMO informed picks, showing the goodness of the fit with the lighcureve and deceleration data-points.}
\label{img:2021_obs}
\end{figure}

We applied the RMSD-based selection method (Section~\ref{subsec:RMSD_sel}) to derive a range of plausible physical parameters. By refining the EMCCD leading-edge picks using CAMO’s narrow-field measurements, we aligned the deceleration profiles as a function of height as illustrated in Figure~\ref{img:CAMO_EMCCD_nick}. This alignment required careful sub-pixel adjustments of the EMCCD picks to match the more precise deceleration curve observed by CAMO. The noise level, estimated via polynomial fitting, was about 0.1\,mag in brightness and 6.1\,m in lag.

\begin{figure}[h!]
\centering
\includegraphics[width=\linewidth]{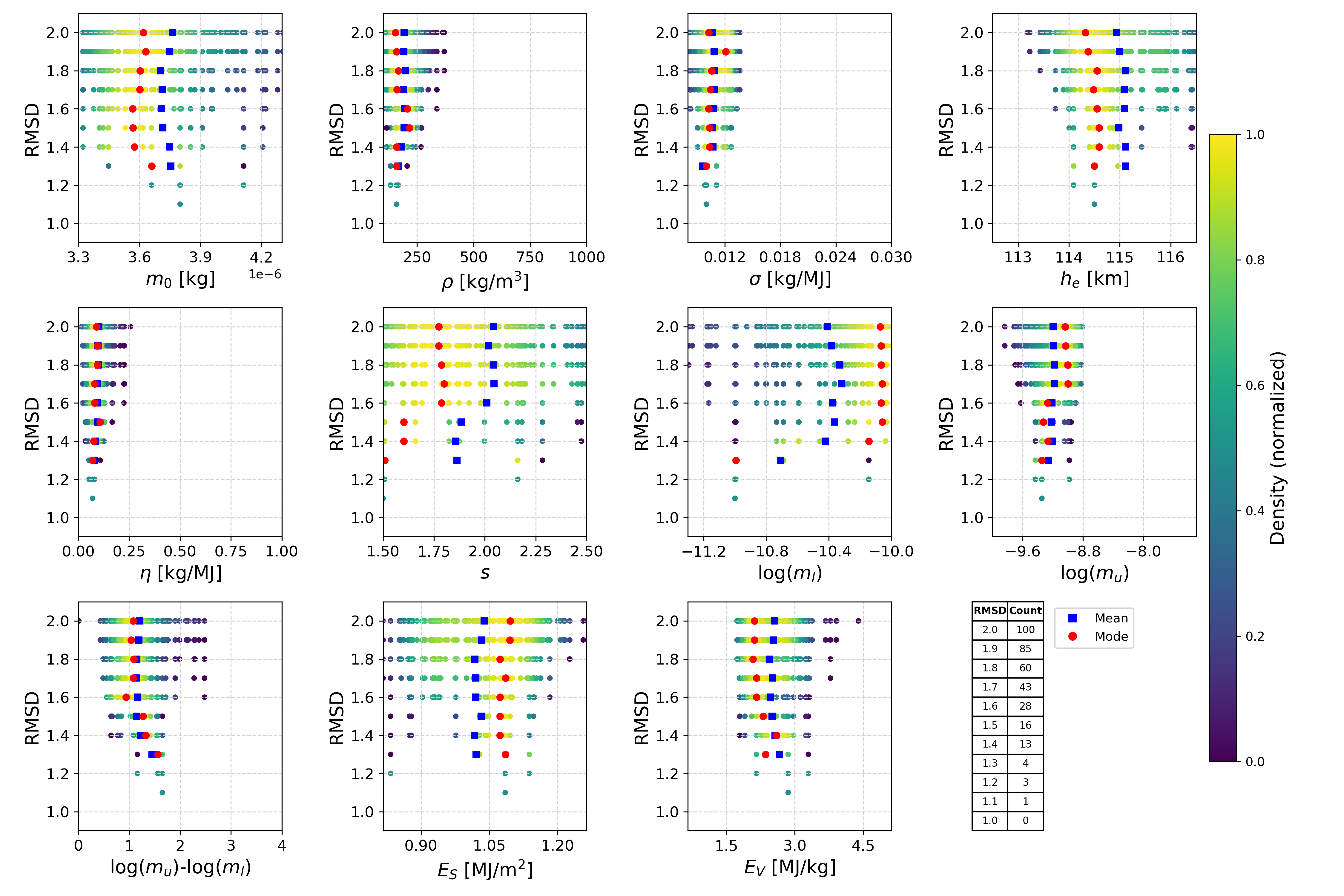}
\caption{Waterfall plots for the 20210813\_061452 meteor from the 100 RMSD normalized best results. In this case it is clear how the bulk density range is well constrained and is below 500 kg/m$^3$ as expected for a cometary meteoroid such as PER.}
\label{img:2021_water}
\end{figure}

Using an RMSD threshold of twice these noise estimates, we obtained a set of simulations that closely reproduced the measured light curve and lag data points, as shown in Figure~\ref{img:2021_obs}. The resulting 95\% confidence intervals for key physical parameters, summarized in Figure~\ref{img:2021_water} and Table~\ref{tab:comparison_2021}, indicate that the meteoroid’s bulk density is well constrained as being below 500\,kg\,m$^{-3}$ with a most probable value near 230\,kg\,m$^{-3}$ independent of the RMSD threshold used. This is consistent with its cometary origin.

\begin{figure}[h!]
\centering
\includegraphics[width=\linewidth]{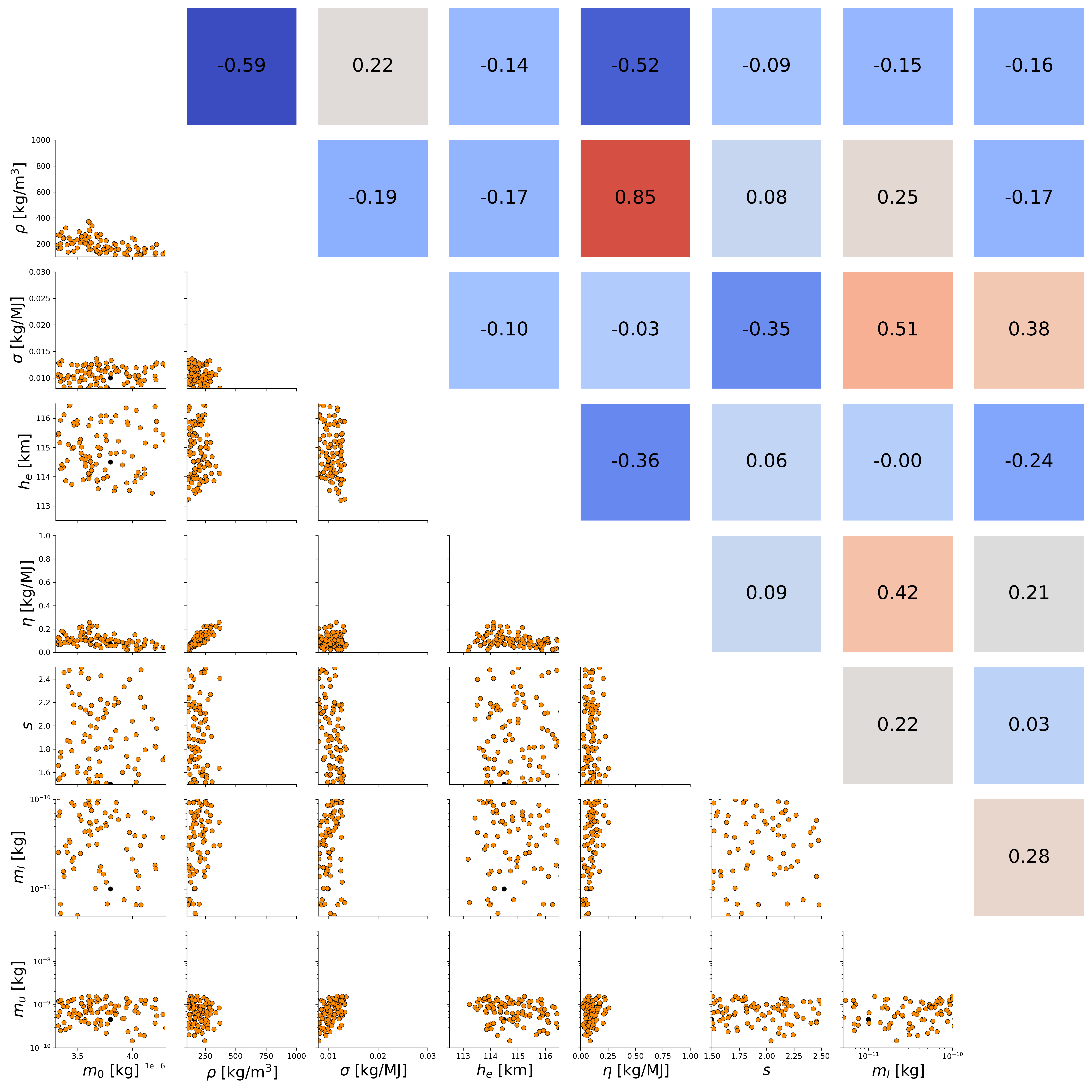}
\caption{Corner plot showing the relationships between key fitted physical parameters for the 20210813\_061452 meteor. The lower triangle displays the distribution of simulated solutions selected based on the RMSD threshold criterion in orange and the manual fit is shown black, while the upper triangle shows the correlation coefficients between each parameter pair.}
\label{img:corr_2021}
\end{figure}

Although most parameters are relatively well constrained, certain erosion-related quantities show broader allowed distributions. For instance, the mass distribution index $s$ it is varying drastically because the minimum erosion mass $\log(m_l)$ is just one order of magnitude from the maximum erosion mass $\log(m_u)$ as shown by $log(m_u)-log(m_l)$, making $s$ hard to precisely estimate. 
The parameter $E_s$ (energy per unit cross-section required to initiate erosion) is linked to the erosion height $h_e$ and so both remain dispersed but still converge at lower RMSD to plausible physical ranges. Overall, these results highlight the value of low noise cameras, like CAMO, achieving more precise density estimates for faint meteors.

\begin{table}[h!]
\centering
\caption{The 95\% confidence interval and associated mode and means in the inverted physical parameters derived from model fits to the 20210813\_061452 meteor. Here we use the 100 simulations which were found to be below twice the recovered RMSD of the measurements.}
\begin{tabular}{|c|c|c|c|c|}
\hline
Variables & 95\%CIlow & Mean & Mode & 95\%CIup \\
\hline
$m_0$ [kg] & 3.34 $\times$ $10^{-6}$ & 3.73 $\times$ $10^{-6}$ & 3.50 $\times$ $10^{-6}$ & 4.20 $\times$ $10^{-6}$ \\
\hline
$\rho$ [kg/m$^3$] & 136.7 & 227.3 & 192.9 & 405.8 \\
\hline
$\sigma$ [kg/MJ] & 0.008 & 0.010 & 0.009 & 0.0127 \\
\hline
$h_{e}$ [km] & 113.8 & 115.1 & 115.3 & 116.3 \\
\hline
$\eta$ [kg/MJ] & 0.048 & 0.112 & 0.0867 & 0.229 \\
\hline
$s$ & 1.528 & 1.942 & 1.649 & 2.449 \\
\hline
$m_{l}$ [kg] & 1.05 $\times$ $10^{-11}$ & 4.78 $\times$ $10^{-11}$ & 6.18 $\times$ $10^{-11}$ & 9.38 $\times$ $10^{-11}$ \\
\hline
$m_{u}$ [kg] & 2.65 $\times$ $10^{-10}$ & 7.09 $\times$ $10^{-10}$ & 5.82 $\times$ $10^{-10}$ & 1.36 $\times$ $10^{-9}$ \\
\hline
log($m_{u}$)-log($m_{l}$) & 0.548 & 1.18 & 1.40 & 1.72 \\
\hline
$E_{S}$ [MJ/m$^2$] & 0.88 & 1.02 & 0.98 & 1.16 \\
\hline
$E_{V}$ [MJ/kg] & 1.44 & 2.28 & 2.28 & 3.08 \\
\hline
\end{tabular}
\label{tab:comparison_2021}
\end{table}

The relationships between key fitted physical parameters for the 20210813\_061452 meteor are visualized in Figure~\ref{img:corr_2021}. 

A particularly strong and positive correlation is observed between the erosion coefficient and bulk density (correlation coefficient = 0.85), confirming the trend reported by \citet{borovivcka2009material} that denser meteoroids require higher erosion rates to reproduce observed behavior. This correlation reflects a known degeneracy in meteoroid modeling: in the absence of data before erosion begins, the erosion coefficient and bulk density cannot be independently constrained. Additional notable correlations include a moderate negative correlation between initial mass and both bulk density (-0.59) and erosion coefficient (-0.52), indicating that smaller, denser meteoroids are favored in the solution space. A moderate positive correlation is also seen between ablation coefficient and minimum erosion mass (0.51), as well as with maximum erosion mass (0.38), suggesting a tendency for brighter meteors to fragment into larger grain sizes when ablation is more efficient.

These relationships help constrain the physical interpretation of the meteoroid and illustrate the interdependence of key parameters in the erosion-ablation modeling framework.

\subsubsection{Meteor 20230811\_082649}\label{subsubsec:2023}

\begin{figure}[h!]
\centering
\includegraphics[width=\linewidth]{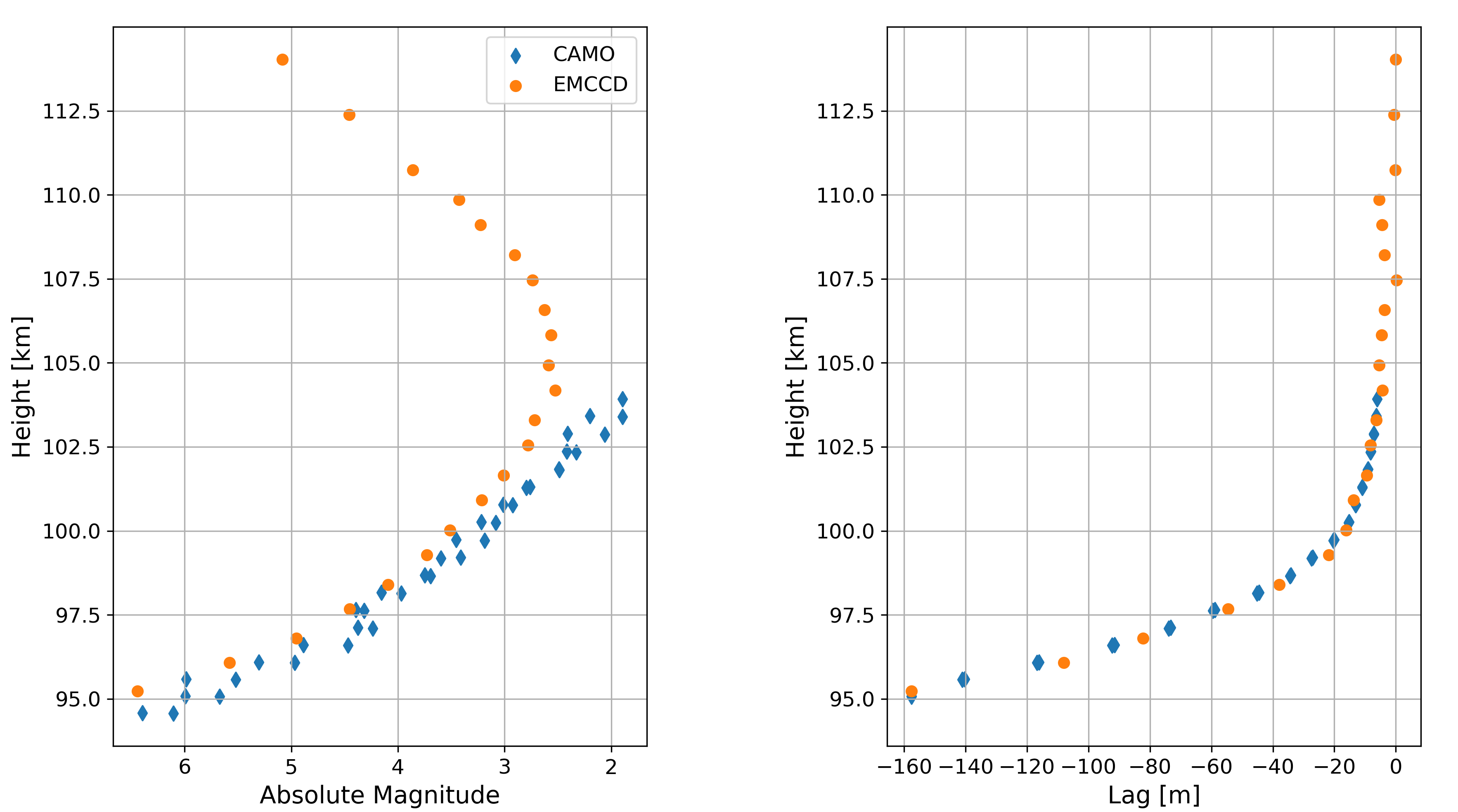}
\caption{A direct comparison of simultaneous measurements from the CAMO narrow-field system and EMCCD data for 20230811\_082649. Here the lag as a function of height from the two station EMCCD solution (orange circles) and the two station CAMO narrow field solution (blue diamonds) are directly compared in the left plot. The left plot shows the absolute magnitude measured by the narrow field CAMO and the EMCCD adjusted to the G-band. The apparent offset in brightness between the two datasets at the peak is primarily due to differences in spectral response between the two camera systems.}
\label{img:CAMO_EMCCD_2}
\end{figure}

Following the same procedure used for the previous meteor, we refined the EMCCD leading-edge picks for \texttt{20230811\_082649} by aligning them with CAMO narrow-field velocity data. As shown in Figure~\ref{img:CAMO_EMCCD_2}, this process required precise sub-pixel adjustments of the EMCCD data to match the more reliable deceleration curve recorded by CAMO. A polynomial fit to the residuals yielded noise estimates of approximately 0.1\,mag in brightness and 5\,m in lag. For the simulations, we fixed the zenith angle $z_c$ to 28.649°, based on the trajectory solution. All other fixed parameters are listed in Table~\ref{tab:fixed_params}.

\begin{figure}[h!]
\centering
\includegraphics[width=\linewidth]{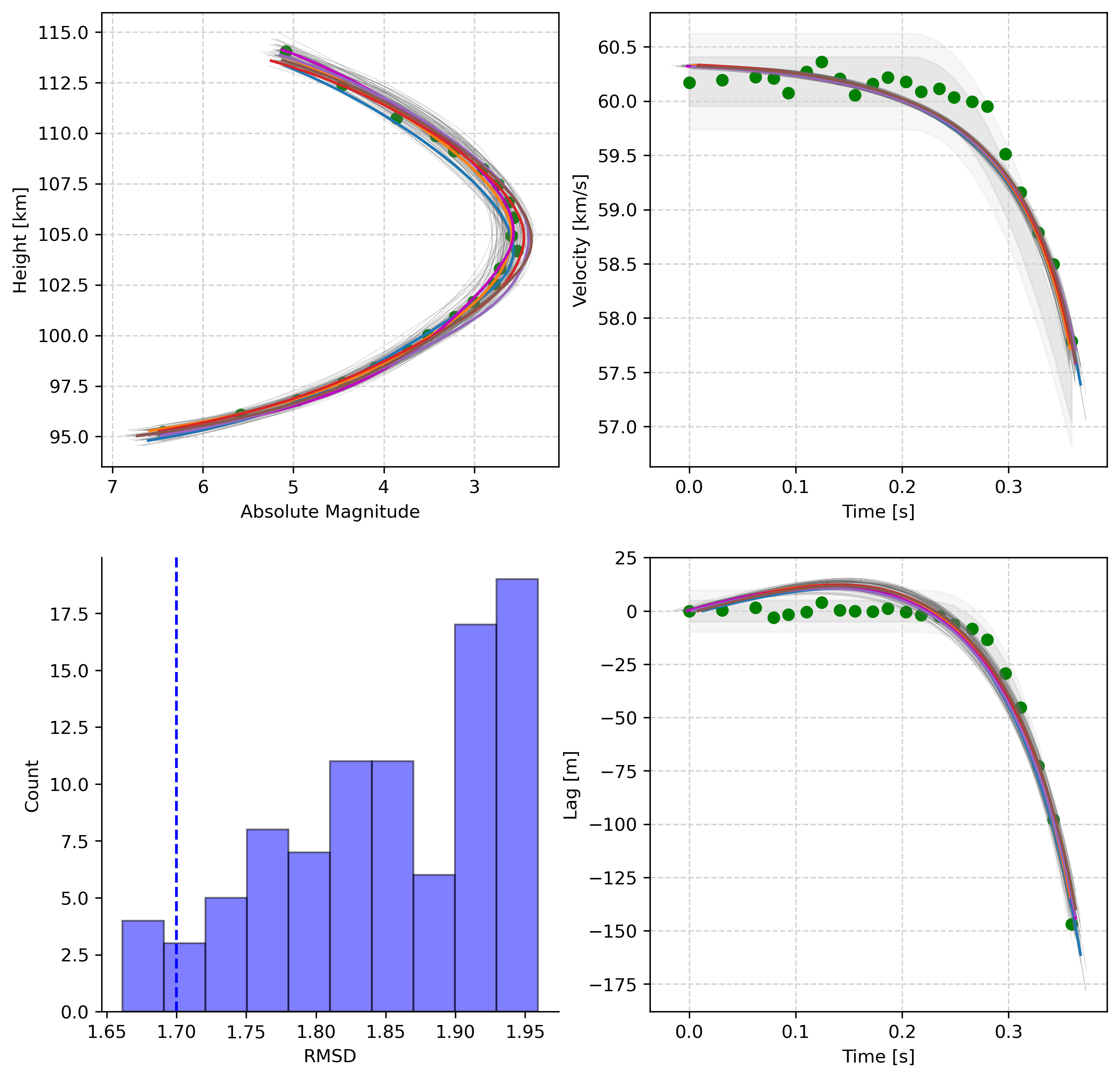}
\caption{The selected simulations with both normalized RMSD below 1.7 for the 20230811\_082648 EMCCD CAMO informed astrometry picks, showing a hump in the lag plot due to the choice of a higher initial velocity then the one automatically assumed for the detected meteors.}
\label{img:2023_obs}
\end{figure}

Using the RMSD-based method, we selected 100 candidate simulations whose light curves and deceleration profiles fell within twice the estimated RMSD of the observed data. While these simulations fit the magnitude measurements well, an artifact in the lag data caused a ``hump'' in the lag plots (Figure~\ref{img:2023_obs}). 

Despite exploring a wide range of physically plausible initial conditions, we were unable to find a single-fragmentation solution that could reproduce the flat velocity trend in the early trajectory without degrading the fit in magnitude. This discrepancy appears to stem from over-fitting low-resolution EMCCD data: in our initial analysis of 32 PER meteors reduced solely with EMCCD cameras, some lag curves were driven to artificially straight profiles and most had to be discarded as physically implausible (see subsection \ref{sec:Pitfalls}). We therefore caution that deceleration profiles derived exclusively from EMCCD fits of fast meteors may misrepresent true dynamics and recommend cross-validation—e.g., with higher-frequency CAMO data. For 20230811\_082649, the deceleration trend has been independently validated by CAMO observations, and tests in which we increased the assumed mass or bulk density could not reproduce the observed magnitude curve—higher values only worsened the mismatch. We thus interpret the perfectly straight early lag as a reduction artifact rather than a physical signature, and are confident that the remaining physical parameters accurately reflect the meteor’s true properties.

\begin{figure}[h!]
\centering
\includegraphics[width=\linewidth]{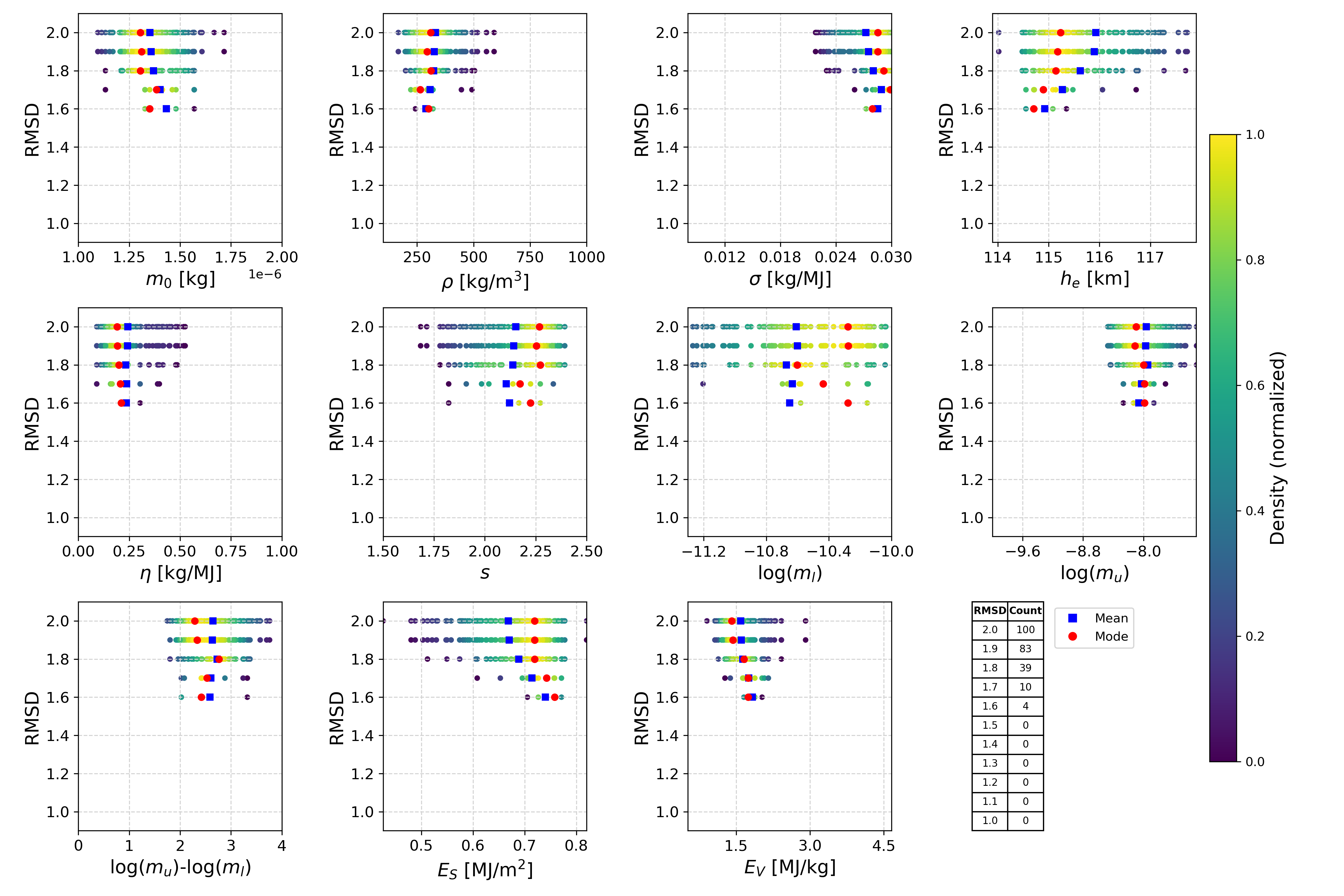} 
\caption{Waterfall plots for the 20230811\_082648 meteor for the 100 results based on the maximum normalized RMSD. The plot show the systematic error in lag that translates to a distribution that at best has a normalized density that is 1.6 times bigger than the polynomial RMSD fit.}
\label{img:2023_water}
\end{figure}

The 95\% confidence interval of the 100 simulations below twice RMSD results identified from comparison with the EMCCD observables encompassed the physical parameter values derived from both manual CAMO reductions, with and without wake data, as shown in Table \ref{tab:comparison_ci_updated_2023}. In Figure \ref{img:2023_water} it can be seen that like for 20210813\_061452 this meteor has a large allowed range in the lower erosion mass $log(m_l)$. This demonstrates the insensitivity to the lightcurve and lag by wide choices in the lower erosion mass compared to the upper erosion mass $m_u$ that is always well constrained in both cases. In contrast, because of the big difference between the lower and upper mass $log(m_u)-log(m_l)$ the mass distribution index $s$ is better determined.

Nonetheless, as shown in see Table~\ref{tab:comparison_ci_updated_2023}, the 95\% confidence interval for the physical parameters based on the 100 RMSD-selected simulations managed to constrain the possible density values to be still below 500 kg/m$^3$. As depicted in Figure~\ref{img:2023_water}, the lower erosion mass $\log(m_l)$ like for the upper erosion mass $\log(m_u)$ remain constrained in around a single order of magnitude, as they are both crucial for the fit. In this particular case, the large separation between $\log(m_u)$ and $\log(m_l)$ leads to better precision in the inversion of the mass distribution index $s$, illustrating the interplay between these erosion parameters in shaping the meteor’s deceleration curve.

\begin{figure}[h!]
\centering
\includegraphics[width=\linewidth]{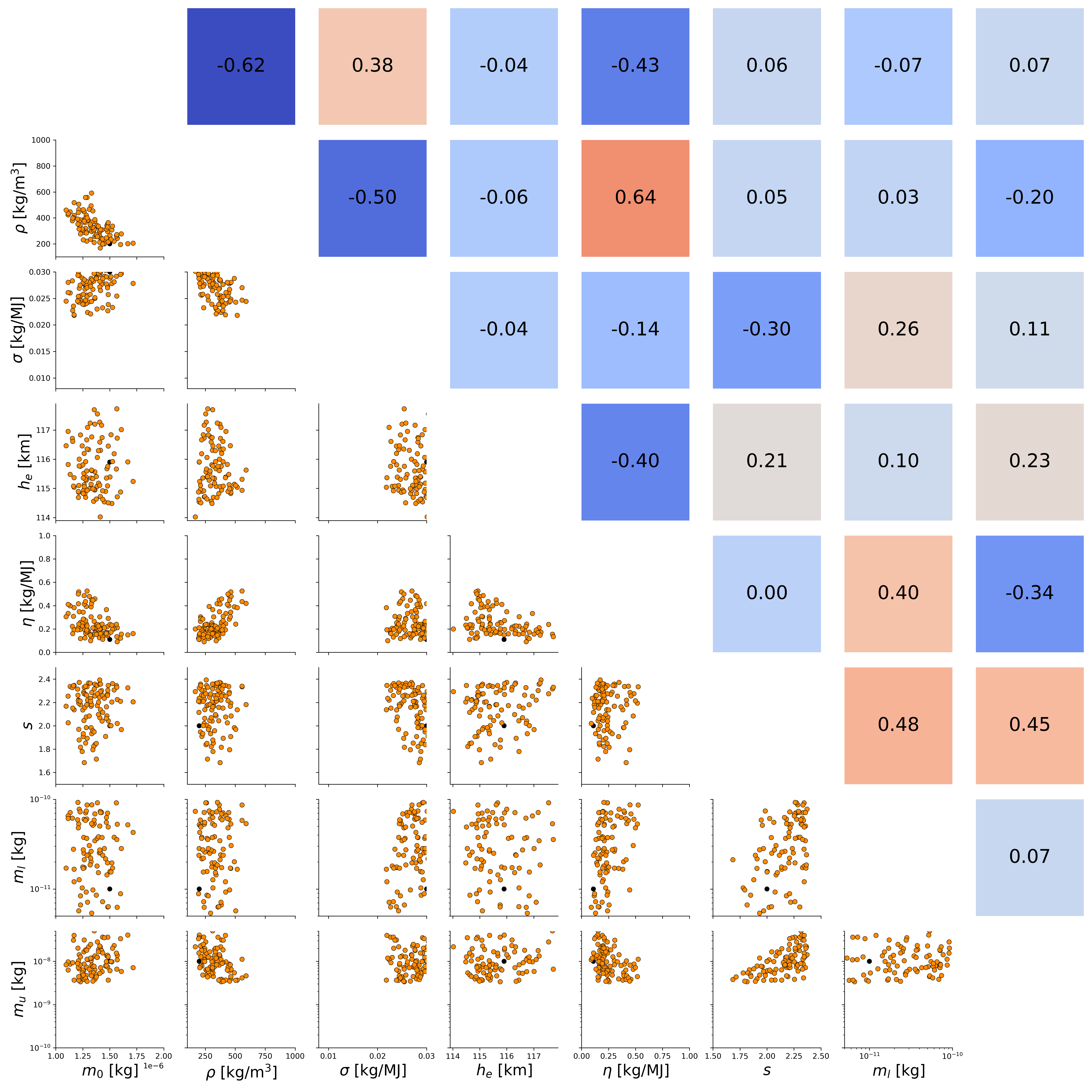}
\caption{Corner plot showing the relationships between key fitted physical parameters for the 20230811\_082648 meteor. The lower triangle displays the distribution of simulated solutions selected based on the RMSD threshold criterion in orange and the manual fit is shown black, while the upper triangle shows the correlation coefficients between each parameter pair. The lower correlations could be a symptom of this bad fit in lag and the likely second fragmentation.}
\label{img:corr_2023}
\end{figure}

Figure~\ref{img:corr_2023} shows the correlation structure between fitted physical parameters for the meteor 20230811\_082648.

Compared to the 20210813\_061452 event, the correlation structure for the 20230811\_082648 meteor is noticeably weaker and more diffuse. While key trends—such as the strong negative correlation between initial mass and bulk density, and the positive correlation between bulk density and erosion coefficient—persist, the overall strength of these relationships is reduced. Notably, the correlation between bulk density and erosion coefficient decreases from 0.85 in 2021 to 0.64 in 20230811\_082648. This reduction, along with generally lower inter-parameter correlations, may reflect the model’s inability to precisely capture the observed flat lag segment and sudden deceleration.

\begin{table}[h!]
\centering
\caption{The 95\% confidence interval and the related mode and mean of the 20230811\_082648 meteor for the 100 simulations below twice the recovered RMSD.}
\begin{tabular}{|c|c|c|c|c|}
\hline
Variables & 95\%CIlow & Mean & Mode & 95\%CIup \\
\hline
$m_0$ [kg] & 1.16 $\times$ $10^{-6}$ & 1.36 $\times$ $10^{-6}$ & 1.33 $\times$ $10^{-6}$ & 1.55 $\times$ $10^{-6}$ \\
\hline
$\rho$ [kg/m$^3$] & 200.4 & 347.3 & 333.1 & 550.5 \\
\hline
$\sigma$ [kg/MJ] & 0.022 & 0.026 & 0.025 & 0.029 \\
\hline
$h_{e}$ [km] & 114.7 & 115.7 & 115.2 & 117.1 \\
\hline
$\eta$ [kg/MJ] & 0.131 & 0.282 & 0.215 & 0.564 \\
\hline
$s$ & 1.999 & 2.25 & 2.35 & 2.446 \\
\hline
$m_{l}$ [kg] & 9.74 $\times$ $10^{-12}$ & 4.23 $\times$ $10^{-11}$ & 4.95 $\times$ $10^{-11}$ & 8.87 $\times$ $10^{-11}$ \\
\hline
$m_{u}$ [kg] & 4.54 $\times$ $10^{-9}$ & 1.27 $\times$ $10^{-8}$ & 6.18 $\times$ $10^{-9}$ & 3.35 $\times$ $10^{-8}$ \\
\hline
log($m_{u}$)-log($m_{l}$) & 1.86 & 2.48 & 2.50 & 3.20 \\
\hline
$E_{S}$ [MJ/m$^2$] & 0.58 & 0.68 & 0.71 & 0.75 \\
\hline
$E_{V}$ [MJ/kg] & 1.18 & 1.57 & 1.46 & 2.16 \\
\hline
\end{tabular}
\label{tab:comparison_ci_updated_2023}
\end{table}

\section{Discussion} \label{sec:discussion}



The primary motivation for this study was to measure and quantify the uncertainty in meteoroid bulk density, a critical parameter (alongside speed and mass) for assessing meteoroid impact risks on spacecraft. Because direct measurement of bulk density is difficult—particularly for sub-millimeter to millimeter-sized meteoroids—researchers often rely on indirect methods that frequently lack robust uncertainty quantification. Our goal was to automate the fitting process and improve the precision of density estimates by exhaustively sampling a wide range of physical parameters. A secondary goal was to estimate the range of other meteoroid physical parameters consistent with measurement precision including ablation and erosion coefficients, meteoroid mass, grain sizes and distribution and beginning erosion height. 

To address the challenge of estimating meteoroid bulk densities from meteor observations, we explored two modeling approaches. The first employs Principal Component Analysis (PCA) to match simulations with similar observable parameters; the second uses an RMSD-based method that directly tests how closely each simulation’s light curve and deceleration profile align with real measurements. Both approaches seek to derive reliable estimates of key meteoroid properties, especially density, while explicitly quantifying the associated uncertainties for cases with higher measurement precision (low noise meteor).

Our analysis shows that even sophisticated methods can fail to invert physical meteoroid properties when measurement noise is high. In particular, density is notoriously difficult to constrain for faint weakly decelerating meteors, as the low signal-to-noise ratio (light-curve range and the limited spatial/temporal resolution) can obscure the subtle light-curve and deceleration signatures that help distinguish different density values. Consequently, advanced inversion strategies and hardware options which maximize resolution/temporal sampling are essential for improving the reliability of density estimates.

\subsection{Limitations of Observable-Based Methods Compared to Fitting Methods}

A major finding of our study is that methods relying exclusively on a fixed set of observables—such as PCA—often struggle under significant measurement noise. Because the ablation process is highly nonlinear, the best match in observables (e.g., peak brightness or timing) may not translate into a faithful fit of the entire light curve or the deceleration profile. In such cases, the PCA-based approach can overlook the actual solutions that reproduce the observed data most accurately, thus failing to invert the key parameters with precision.

By contrast, our RMSD-based brute-force approach provides a more transparent and robust framework for handling noisy data. By continuously comparing the full simulated and observed light curves and deceleration profiles, it is possible to directly adjust the RMSD cutoff to refine (or relax) the set of acceptable solutions, while easily probing a broad parameter space. Although the PCA-based method occasionally yields good results, interpreting and mapping the PC distance to physical uncertainties can be ambiguous. Indeed, we observed cases where PCA converged on solutions with large systematic errors, even when no physically consistent solution was found by RMSD-based criteria.

In light of these findings, we conclude that direct fitting methods like our RMSD based approach are generally more resistant to noise and more reliable for defining uncertainty intervals. Observable-based methods offer potential advantages in computational speed or dimensionality reduction, but without careful calibration and cross-checking, they can misrepresent the true parameter space and produce misleading estimates of meteoroid density.

\subsection{Effect of Measurement Noise}
Our study highlights the central role of measurement noise in determining the precision and accuracy of the inverted parameters. For most camera systems, a 0.1\,mag brightness noise is simply unavoidable given the limitations of current camera technology. Lag noise, however, can vary between camera systems; high-precision instruments such as CAMO can reach a noise level in lag on average of 5 m while EMCCD results can have on average a noise of 40 m while wider field systems, such as CAMS or GMN, may have lag noise of order 100 m or more.

Smaller measurement error leads directly to better convergence on bulk density, whereas large noise levels can render the inversion problem nearly intractable. 

Another critical factor is the completeness of the light curve and the dynamic range in the magnitude and lag measurements. In this context, we define a light curve as “complete” when both the beginning and end of the luminous trajectory are fully captured within the field of view (FOV). This typically means that the meteor shows similar brightness values at both endpoints, ensuring that neither the rise nor the decay is clipped. Incomplete light curves—where the meteor enters or exits the FOV mid-flight—may show only a peak followed by a sharp rise or drop, as often seen in CAMO narrow-field data because of the tracking time. These incomplete observations limit our ability to constrain the meteoroid’s behavior before or after maximum brightness, thereby reducing the reliability of model fitting.

As a result, weakly decelerating and faint meteors—such as low-mass Perseids in the $10^{-7}$ to $10^{-6}$~kg range—produce the least robust inversion solutions. This is due to a combination of low signal-to-noise ratios, incomplete brightness profiles, and limited deceleration signatures, all of which degrade the model’s ability to resolve key physical properties.

\begin{figure}[h!]
\centering
\includegraphics[width=0.7\linewidth]{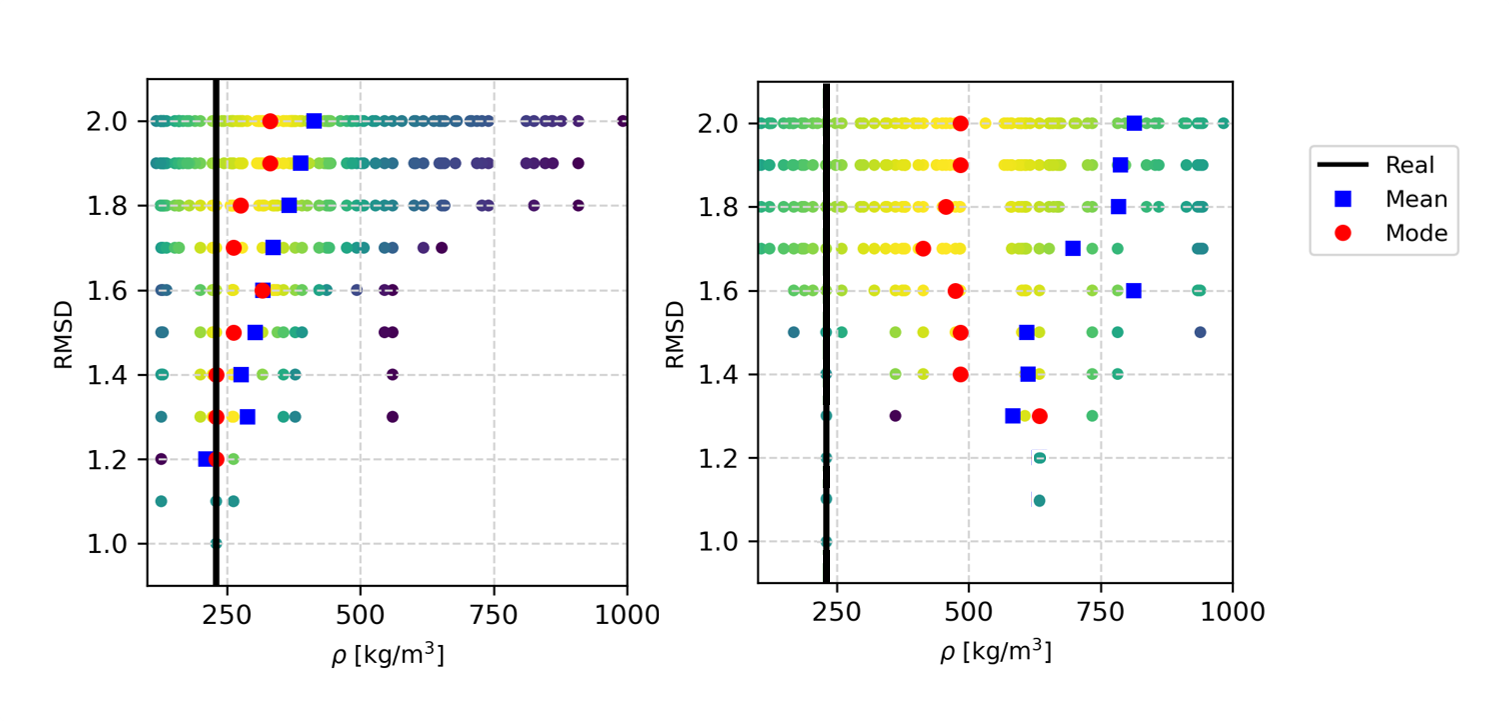}
\caption{Heavy test case with 0.1 magnitude and 40 m lag noise on the left.The same test case but with 0.25 magnitude and 40 m lag noise is shown on the right where the inversion no longer converges to the known value of density, illustrating the importance of noise in the inversion process.}
\label{img:mag_heavy_err}
\end{figure}

This consideration is of course valid for all meteoroids. By increasing the magnitude noise of the heavy test case to 0.25, for example, we find that the density will no longer converge and will be no longer constrained, as shown in the Figure \ref{img:mag_heavy_err}.

To improve the uncertainty precision when inverting the physical properties of meteoroids both a low noisiness level in lag using cameras like CAMO combined with cameras that can see the entire light curve like the EMCCD, produce optimal results.

\subsection{Mean and Mode results}

In our validation tests, we did not find a single best practice regarding whether the mean or the mode (using Kernel Density Estimation) of the parameter distribution offers a closer estimate of the true physical characteristics. Both statistics serve as useful descriptors of the underlying distribution of accepted simulations, but they convey different information:
\begin{itemize}
    \item \textbf{Mean:} Represents the average value over all selected solutions. It can be influenced by outliers and skewness in the distribution but generally reflects the ``center of mass'' for the parameter space.
    \item \textbf{Mode:} Represents the most frequently occurring value among all accepted solutions. It is less affected by outliers and can more directly capture the peak of the probability density. However, in highly multi-modal distributions, the concept of a single mode may be less meaningful. It also is more limited in utility when only a small number of model simulations are selected.
\end{itemize}

In our six test cases, we observed that when the parameter distributions are relatively narrow and symmetric, the mean and mode tend to coincide—and this coincident value often lies near the true parameter. Conversely, if the distribution is broad or skewed, these two statistics can diverge substantially, making it unclear which provides a better approximation of reality. In cases converging close to the real solution, both the mean and mode remained relatively stable and did not shift drastically for different choices of normalized RMSD cuts.

Practically, identifying when the mean and mode overlap can be an additional indicator of solution robustness: a narrow confidence interval and close agreement between these two measures usually signifies higher confidence in the parameter estimate. Reporting both the mean and the mode, alongside the minimum and maximum of the confidence interval, offers a more complete picture of the central trend in the solutions and the potential variability arising from measurement noise or model uncertainties.

\subsection{Pitfalls of Over-Fitting Fast Meteors with Low-Resolution Cameras}
\label{sec:Pitfalls}

Manual selection of the leading edge of a meteor can introduce significant observer-dependent biases, especially when analyzing high-speed events. As illustrated in Table~\ref{tab:per_meas_rmsd}, different users may yield divergent results, particularly in the derived deceleration trends. These discrepancies become especially pronounced when using only EMCCD data, which often lacks the resolution and the cadence to capture abrupt velocity changes in fast meteors.

Our initial survey of 32 Perseid (PER) meteors reduced solely with the EMCCD camera supports these concerns. In several cases, we observed inconsistencies where either the deceleration was unrealistically strong yielding light curves incompatible with any physical deceleration or, conversely, the initial velocity appeared excessively flat and prolonged (such as that discussed in Section~\ref{subsubsec:2023}), so the simulation found fail to match the data across the entire trajectory. These artifacts are particularly problematic when the EMCCD system is used in isolation, as its temporal resolution and resolution is insufficient to resolve sudden deceleration features critical for characterizing rapid meteor events.

This often results from the reduction process being implicitly guided by a desire to minimize apparent lag rather than accurately tracing the true kinematics. Consequently, physically implausible fragmentation behaviors or flat velocities may be inferred. We thus caution against relying solely on low-framerate, low-resolution systems for fast meteors and strongly recommend corroborating results with high-resolution data (e.g., CAMO systems) to ensure that measured deceleration trends reflect true meteoroid dynamics and are not artifacts of over-fitting.

\subsection{Future Directions}

Our work emphasizes the fact that the measurement noise is the fundamental limiting factor in physical meteoroid property inversion. Our approach underscores the value of coupling multiple data sources to minimize overall noise and the utility of using brute-force searches in the parameter space to define meteoroid physical uncertainties.

Future work should focus on scaling up computational resources to fully automate the reduction process. Although our code needs an initial guess for the mass and velocity, the broad ranges used to run the simulations shows that even a rough estimate of velocity at 180 km and a crude value of mass generally work to localize good model solutions.

Continued refinement and computational approaches, for instance expanding parameter studies for multiple fragmentation episodes and giving an initial estimate of range for each physical parameter using machine learning techniques, should improve the robustness of future studies using our approach.

More broadly, designing camera systems or refining automatic data-reduction strategies to minimize lag and magnitude noise remains the most direct path toward higher-confidence density determinations for faint meteors. In the interim, combining manually picked EMCCD observations with high-precision CAMO data and rigorously exploring extensive simulation ensembles represents the most viable approach to achieving reliable modeling outcomes.

\section{Conclusions}\label{sec:conclusions}

The goal of this work was to take the observed lag and brightness for two station optically recorded meteors and invert these using an entry model for the physical parameters of a meteoroid together with associated parameter uncertainty. While this has been done previously through forward modeling \citep[e.g.][]{borovivcka2007atmospheric, vojacek2019, vida2024first, buccongello2024physical} our aim was to automate the process of finding all physical solutions matching the observables to better allow for uncertainty estimation.  

The study focused in particular on the challenge of accurately estimating the bulk density of meteoroids, a critical factor for assessing the risk of impacts on spacecraft. Absolute measurement of bulk density is challenging, especially for sub-mm to mm sized meteoroids, necessitating reliable estimation methods. To meet this need, we developed 2 methods. One that relied on Principal Component Analysis (PCA) to quickly estimate the initial parameter space to generate simulations most likely to match observables within measurement uncertainty, and a brute force method that relied on RMSD as a cost function to find the best fitting simulated meteors that could match the light curve and the deceleration. 

Validation of this technique using six Perseid model test cases showed that the approach can recover the original physical properties of a meteoroid with good precision for bright long meteors. In contrast, faint meteors showing little deceleration require high precision cameras like CAMO to define sufficiently tight measurements to produce comparable parameter uncertainties. 

The method was successfully applied to 2 real PER meteors where the observables were found to be explained properly by the erosion model with suitable choices for the physical parameters. The resulting confidence interval of density of both meteors agrees with that reported by \cite{buccongello2024physical} (365 ± 134 kg/m\(^3\)). This provided further validation of our method, as both studies used the CAMO system and the same luminosity efficiency models.

The main conclusions from our study are:
\begin{itemize}
    \item \textbf{Limitations of Observable-Only Methods:} Relying solely on a linear combination of observable parameters (as is done with the PCA approach) can lead to inaccurate density estimates due to its susceptibility to noise. In contrast, fitting both light curve and lag profiles provides more reliable and accurate constraints on meteoroid density.
    \item \textbf{Influence of Noise on Density Constraints:} Our validation revealed that noise in magnitude and lag measurements strongly impacts density estimates—particularly for faint low decelerating meteors with PER characteristics. Higher-precision data (e.g., CAMO) and wider observation ranges in magnitude and lag significantly tighten density uncertainties.
    \item \textbf{Significance of Mean and Mode:} When the solution set was unimodal and relatively symmetric, the mean and mode of the parameter distribution tended to coincide, providing a stable and accurate estimate of the true parameter values. Broader or skewed distributions yielded larger differences between these statistics.
\end{itemize}

We present, to our knowledge, the first approach to objective uncertainty estimates for the physical properties of small, individual meteors. Our method explicitly allows for a comprehensive search in parameter phase space to identify the range of allowable solutions which match observations within uncertainty. This removes the issue of the degeneracy problem faced in forward modelling approaches.  

Overall, our results underscore that relying exclusively on observables can be misleading, particularly under high-noise conditions or for faint meteors with shallow deceleration. By contrast, an RMSD-based fitting of the entire light curve and deceleration profile delivers more robust density constraints. These findings mark a significant step toward reliably quantifying the bulk density of small meteoroids. Our methodology can be extended to other showers and observational setups, offering a clearer path to establishing practical, data-driven uncertainty measures for meteoroid risk assessment.

\subsection*{Note on Code Availability}

Implementation of all methods used in this work is published as open source on the following GitHub web page:
\begin{itemize}
    \item WesternMeteorPyLib: \\\url{https://github.com/wmpg/WesternMeteorPyLib}
\end{itemize}
Readers are encouraged to contact the lead author in the event they are not able to obtain the code on-line.

\section*{Acknowledgements}

Funding for this work was provided by the NASA Meteoroid Environment Office under cooperative agreement 80NSSC24M0060 and the European Space Agency (ESA) through Contract Number 4000145350, as part of the ESA Initial Support for Innovation (EISI) program.
The authors would like to thank Dr. Bill Cooke and Mark Millinger for providing insight and expertise that assisted the research. We thank Z. Krzeminski for help in optical data reduction.

\section*{Declaration of generative AI and AI-assisted technologies in the writing process}

During the preparation of this work the authors used ChatGPT in order to improve language and readability. After using this tool/service, the authors reviewed and edited the content as needed and take full responsibility for the content of the publication.

\bibliographystyle{Alphab-elsarticle-num-names} 
\bibliography{Bibliog.bib}




\newpage

\section*{Appendix A}\label{sec:Apx A}

Appendix A provides a comprehensive list of all inversion values obtained for the five additional model test cases that were not described in detail in the main text. These cases serve as supplemental examples, demonstrating the inversion differences of the two approaches. By including these results, we aim to offer a complete view of the performance and outcomes of our method across a broader range of test cases. Each test case is named according to the terminology used in Table \ref{tab:init_conditions}.

\textbf{Slow}

\begin{figure}[h!]
\centering
\begin{minipage}{0.45\linewidth}
    \centering
    \textbf{PCA} \\ 
    \includegraphics[width=\linewidth]{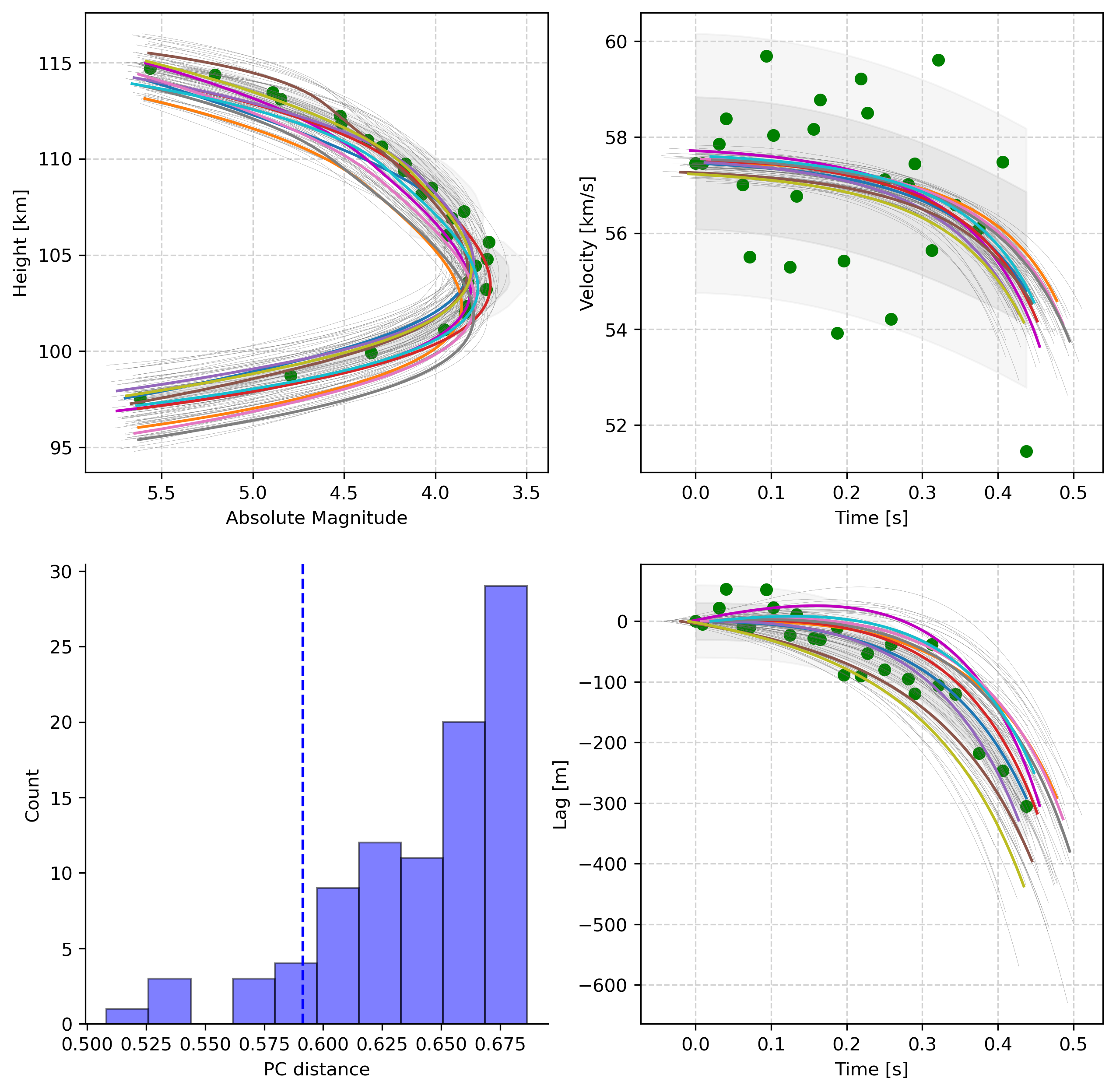}
\end{minipage}
\hfill
\vrule width 0.5pt
\hfill
\begin{minipage}{0.45\linewidth}
    \centering
    \textbf{RMSD} \\ 
    \includegraphics[width=\linewidth]{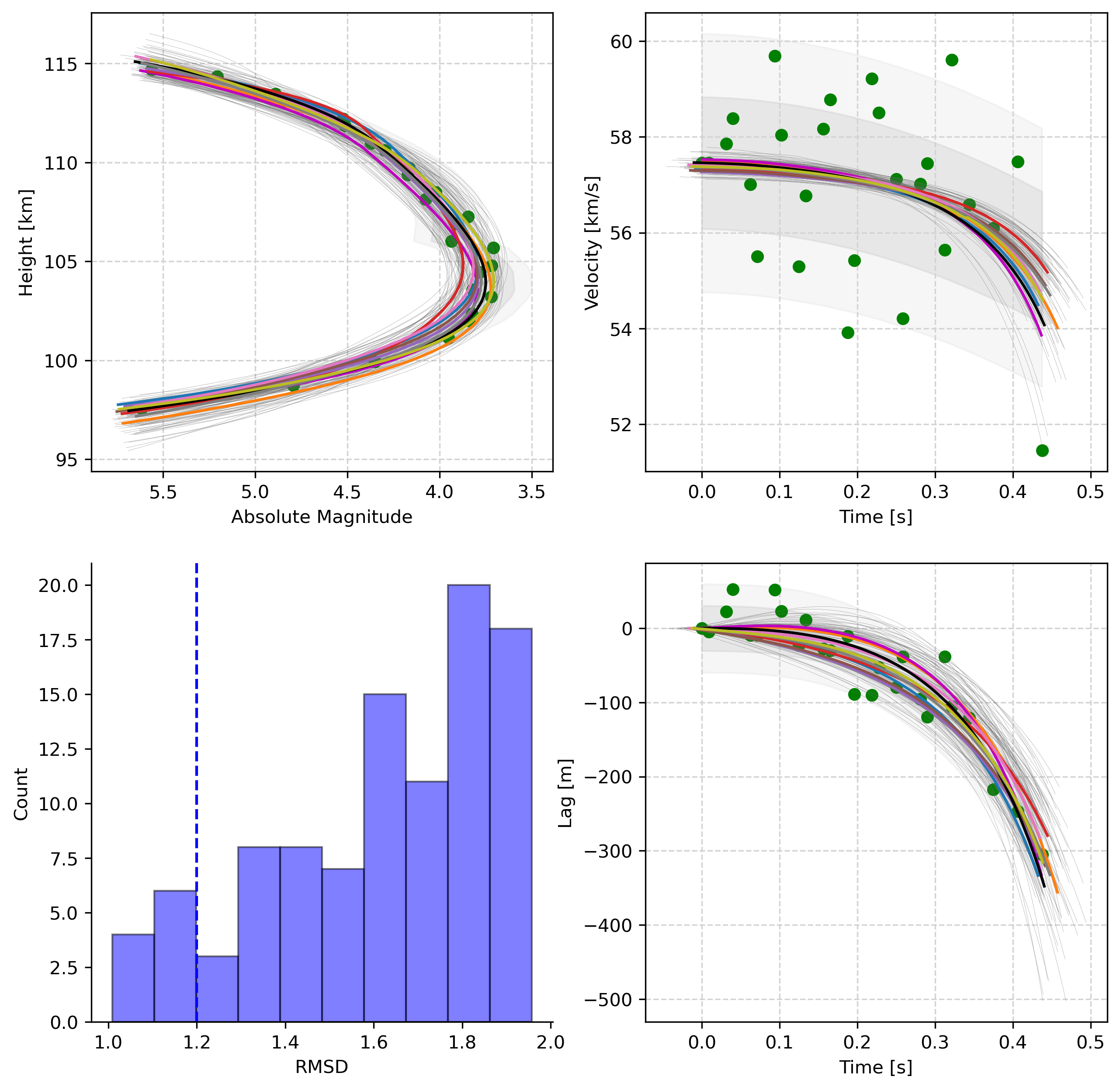}
\end{minipage}

\caption{Light curve, velocity, and lag profiles of the shallow test case for the top 0.1\% of PCA distances (left) and for brute-force simulations below an RMSD threshold of 1.2 (right). The 10 best RMSD solutions in the PCA set have an average PC distance of 0.79 that is above the 1 percentile further proving the bad fit.}
\label{img:obs_slow_1}
\end{figure}

\begin{figure}[h!]
\centering
\begin{minipage}{0.82\linewidth}
    \centering
    \textbf{PCA} \\ 
    \includegraphics[width=\linewidth]{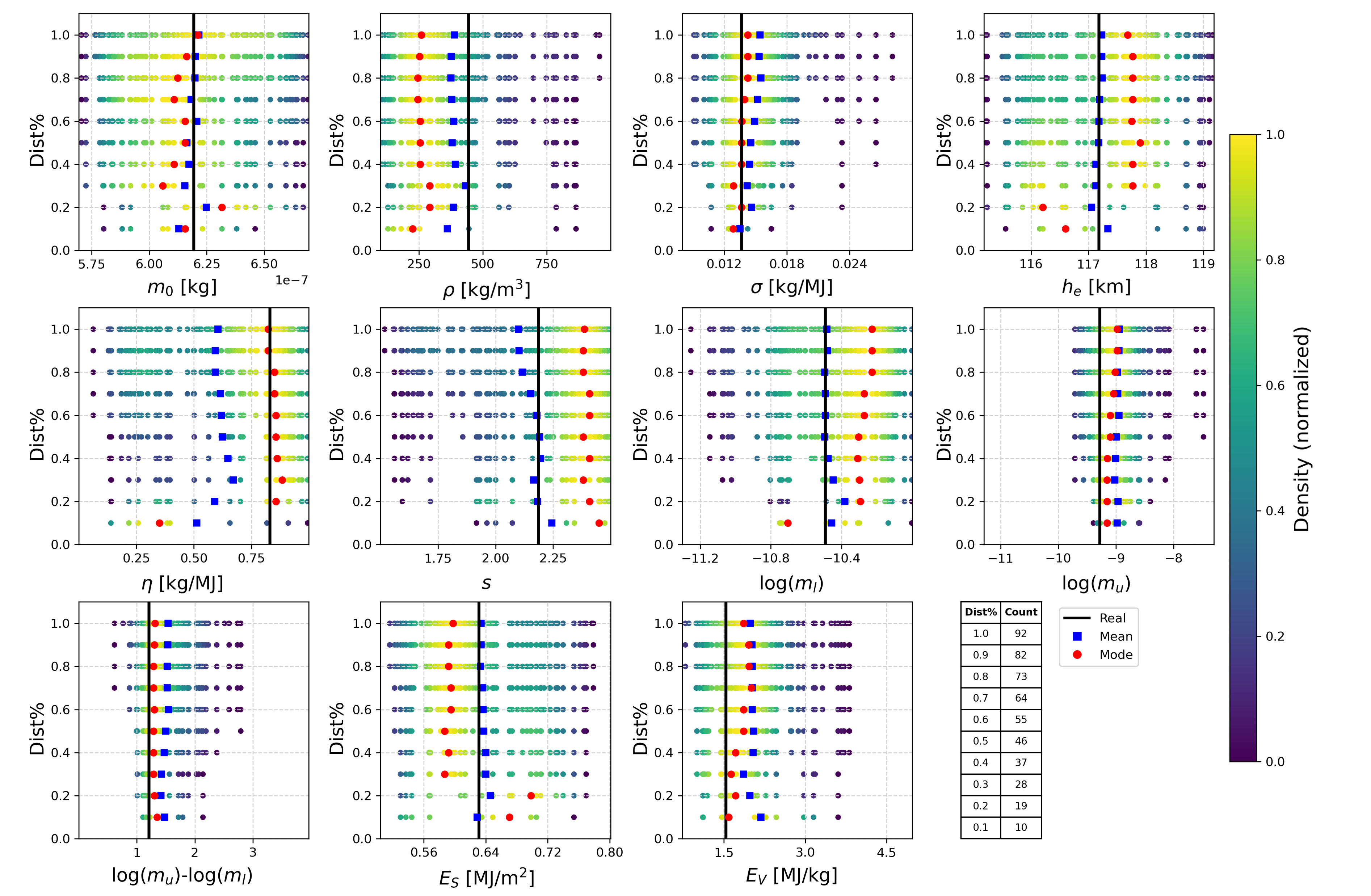}
\end{minipage}
\vspace{1em}
\begin{minipage}{0.82\linewidth}
    \centering
    \textbf{RMSD} \\ 
    \includegraphics[width=\linewidth]{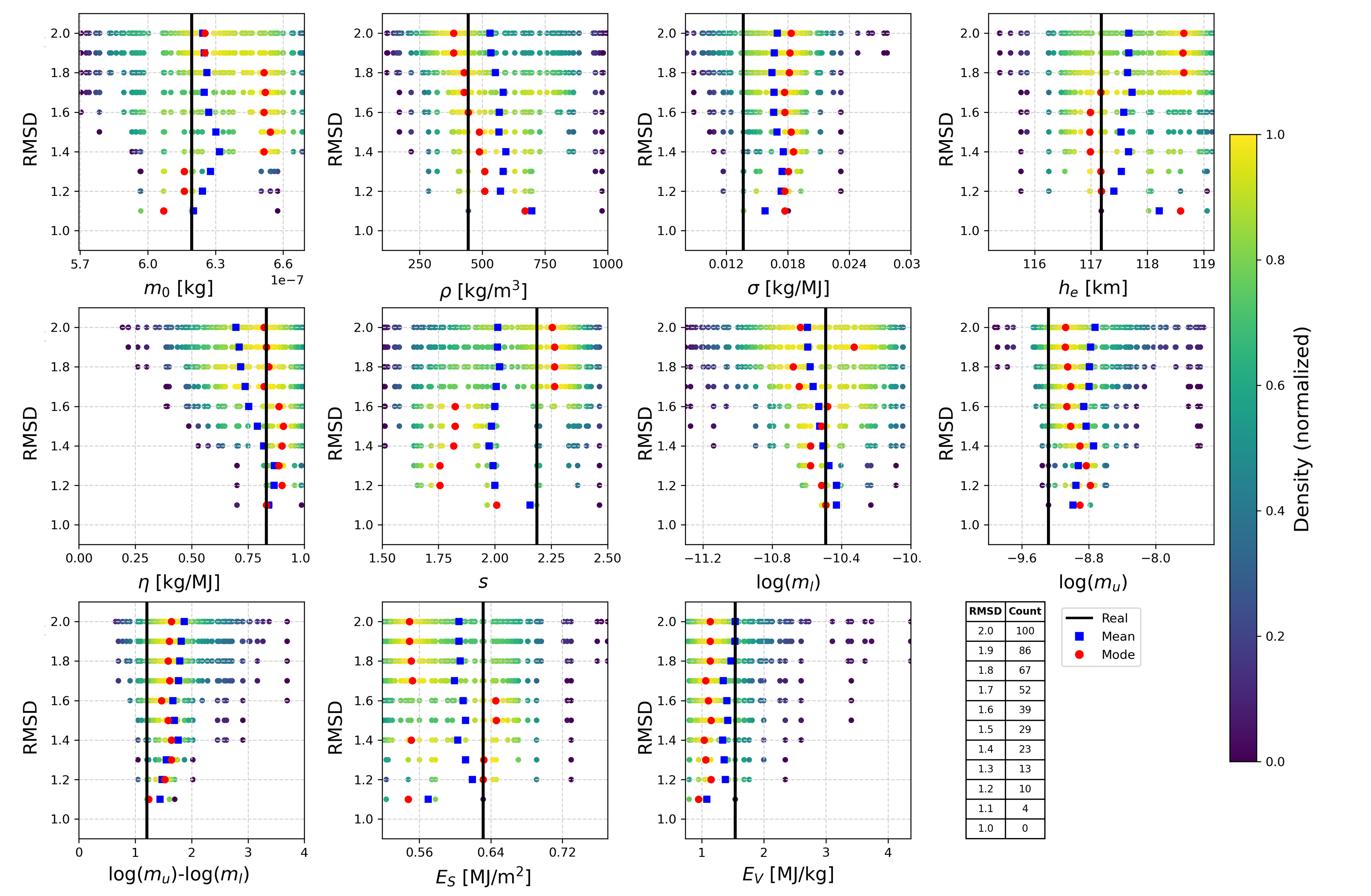}
\end{minipage}

\caption{Waterfall plot of the shallow test case for the PCA method (top) and the RMSD results (bottom). In both cases, all the physical parameters are not well constrained by either of the methods.}
\label{img:wat_slow_2}
\end{figure}

\newpage

\textbf{Shallow}

\begin{figure}[h!]
\centering
\begin{minipage}{0.45\linewidth}
    \centering
    \textbf{PCA} \\ 
    \includegraphics[width=\linewidth]{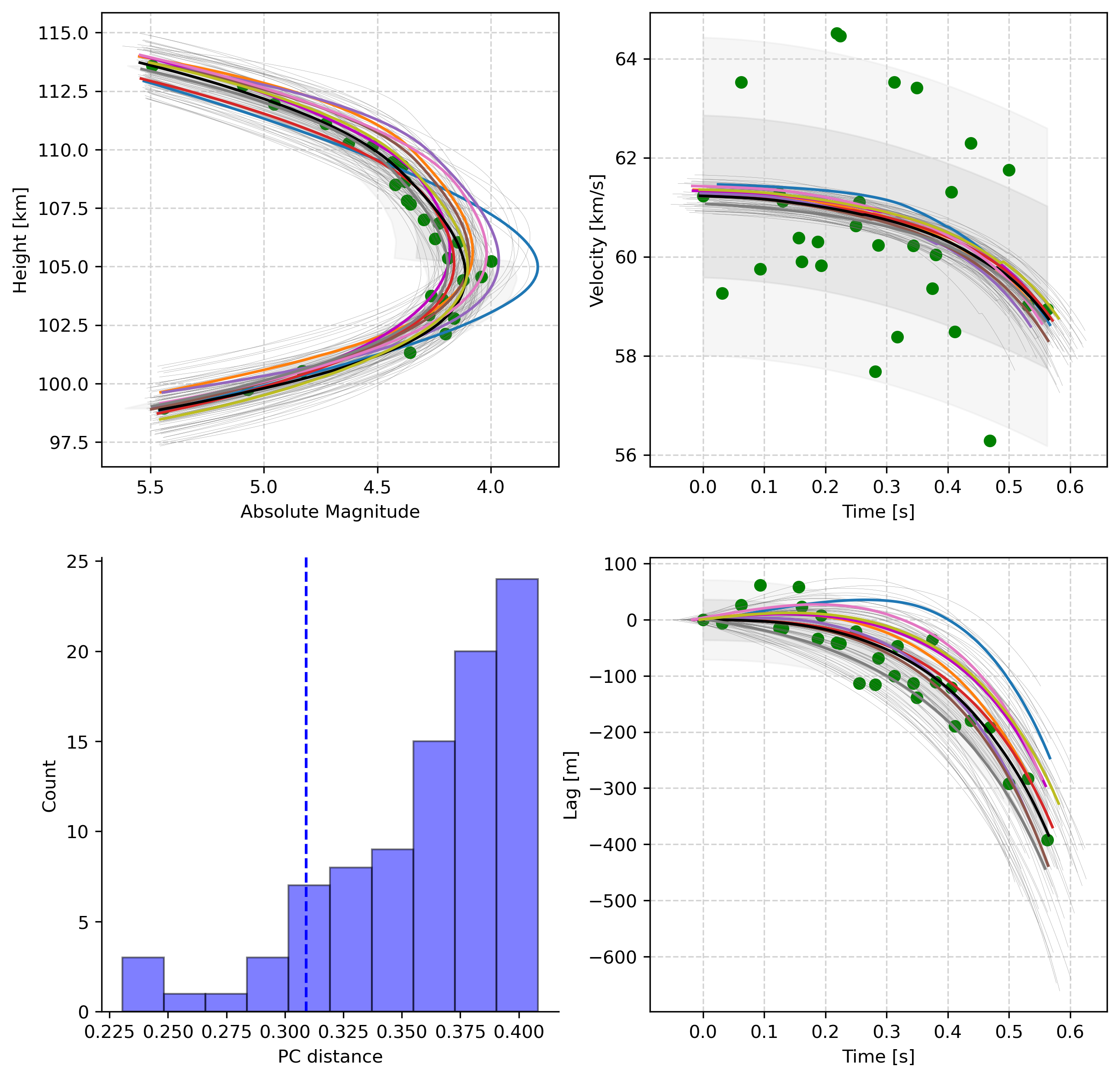}
\end{minipage}
\hfill
\vrule width 0.5pt
\hfill
\begin{minipage}{0.45\linewidth}
    \centering
    \textbf{RMSD} \\ 
    \includegraphics[width=\linewidth]{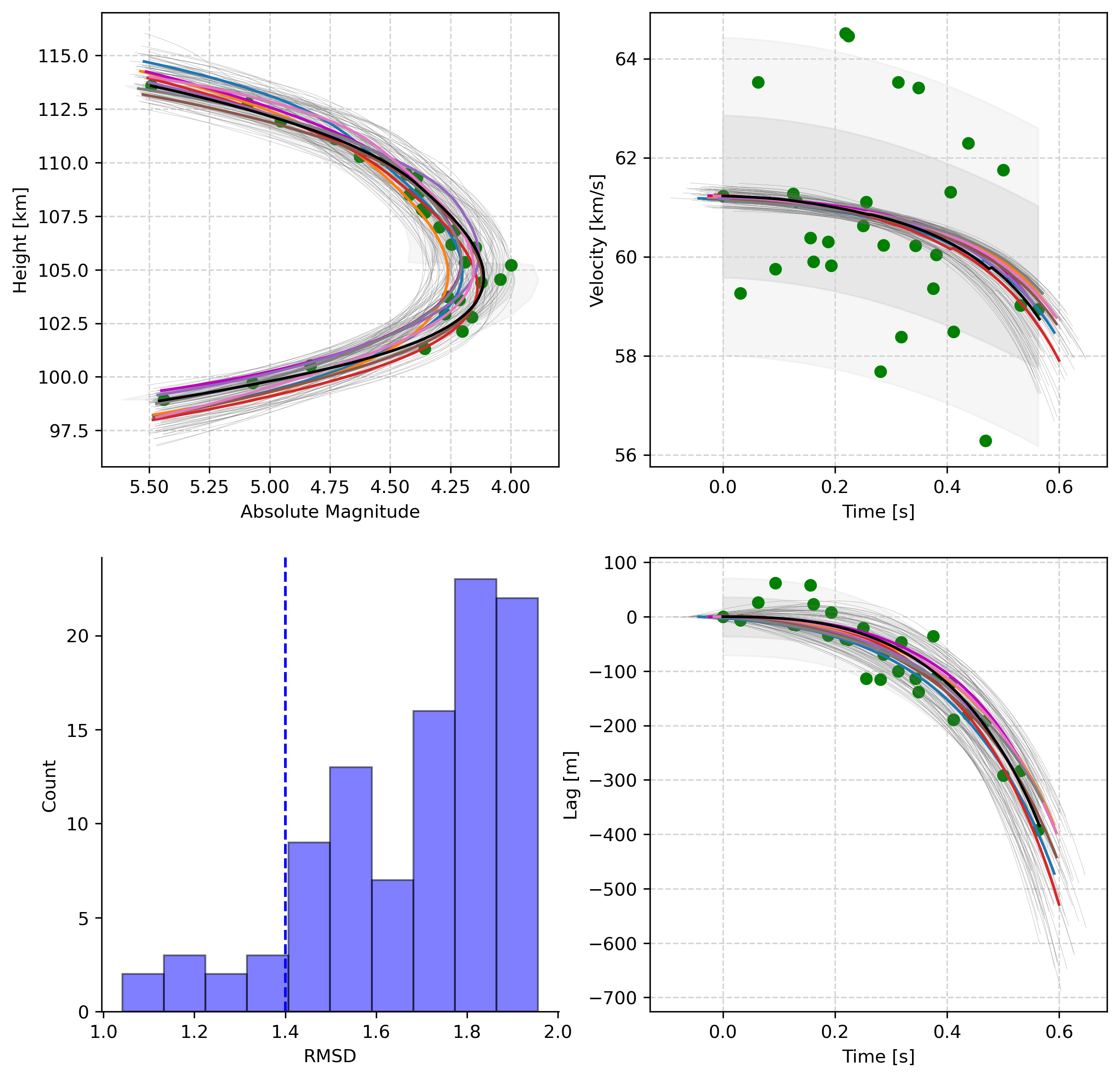}
\end{minipage}

\caption{Light curve, velocity, and lag profiles of the shallow test case for the top 0.1\% of PCA distances (left) and for brute-force simulations below an RMSD threshold of 1.4 (right). The 10 best RMSD solutions in the PCA set have an average PC distance of 0.44 that is above the 1 percentile further proving the bad fit.}
\label{img:obs_shallow_1}
\end{figure}

\begin{figure}[h!]
\centering
\begin{minipage}{0.82\linewidth}
    \centering
    \textbf{PCA} \\ 
    \includegraphics[width=\linewidth]{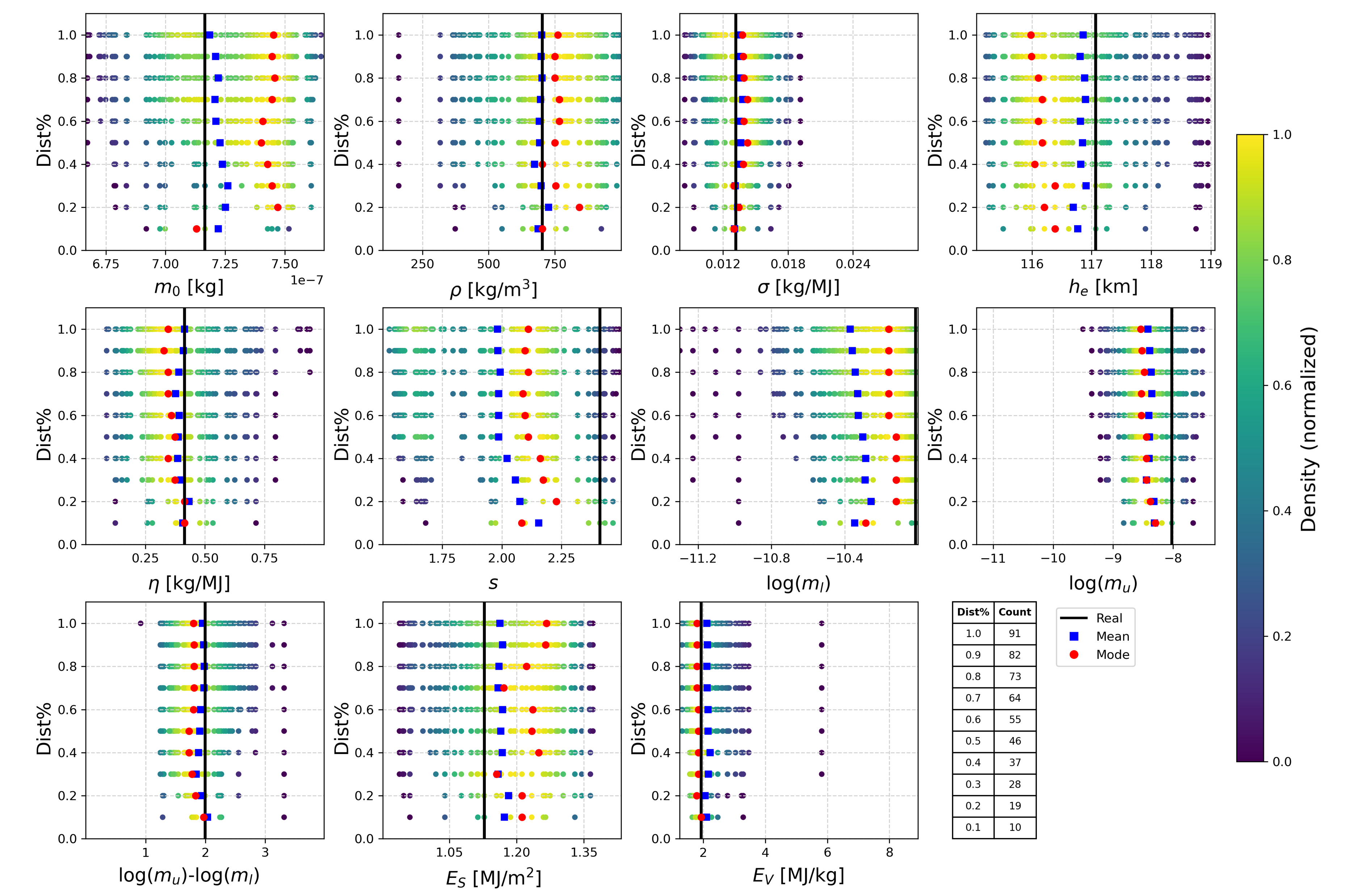}
\end{minipage}
\vspace{1em}
\begin{minipage}{0.82\linewidth}
    \centering
    \textbf{RMSD} \\ 
    \includegraphics[width=\linewidth]{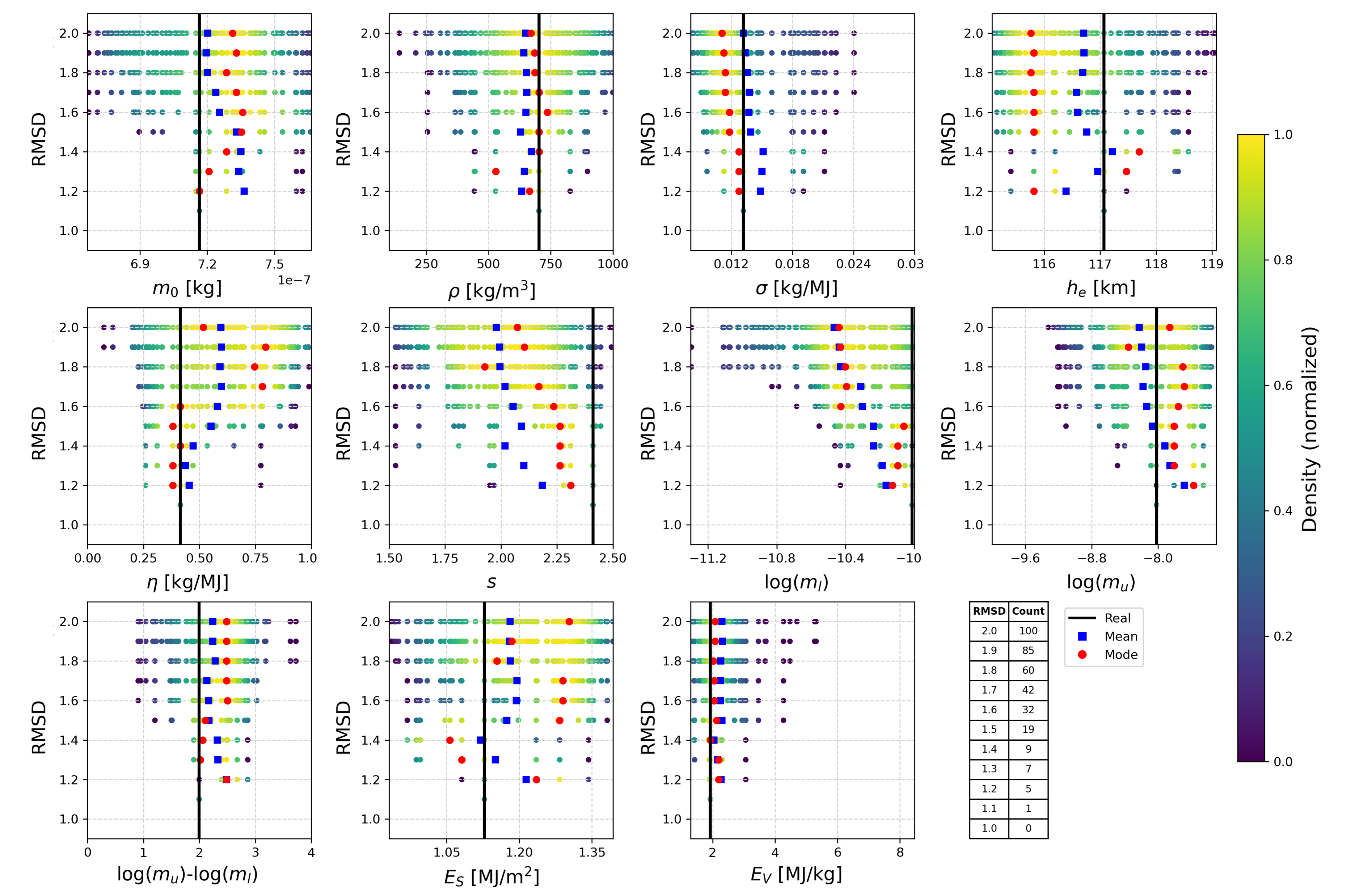}
\end{minipage}

\caption{Waterfall plot of the shallow test case for the PCA method (top) and the RMSD results (bottom). In both cases, even though the broad density range both mean and mode converge to the real value of density of the test case.}
\label{img:wat_shallow_2}
\end{figure}

\newpage

\textbf{Steep}

\begin{figure}[h!]
\centering
\begin{minipage}{0.45\linewidth}
    \centering
    \textbf{PCA} \\ 
    \includegraphics[width=\linewidth]{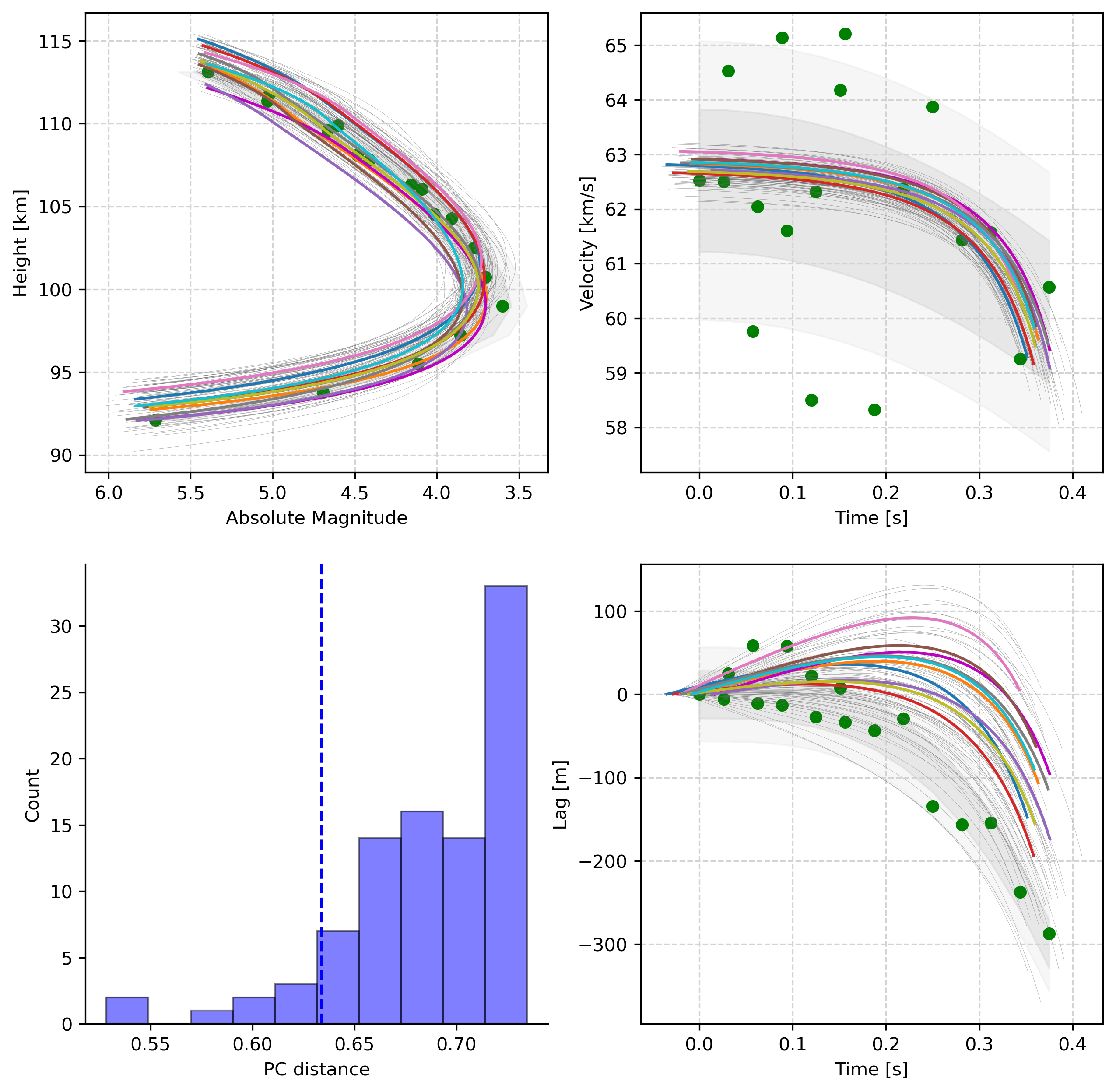}
\end{minipage}
\hfill
\vrule width 0.5pt
\hfill
\begin{minipage}{0.45\linewidth}
    \centering
    \textbf{RMSD} \\ 
    \includegraphics[width=\linewidth]{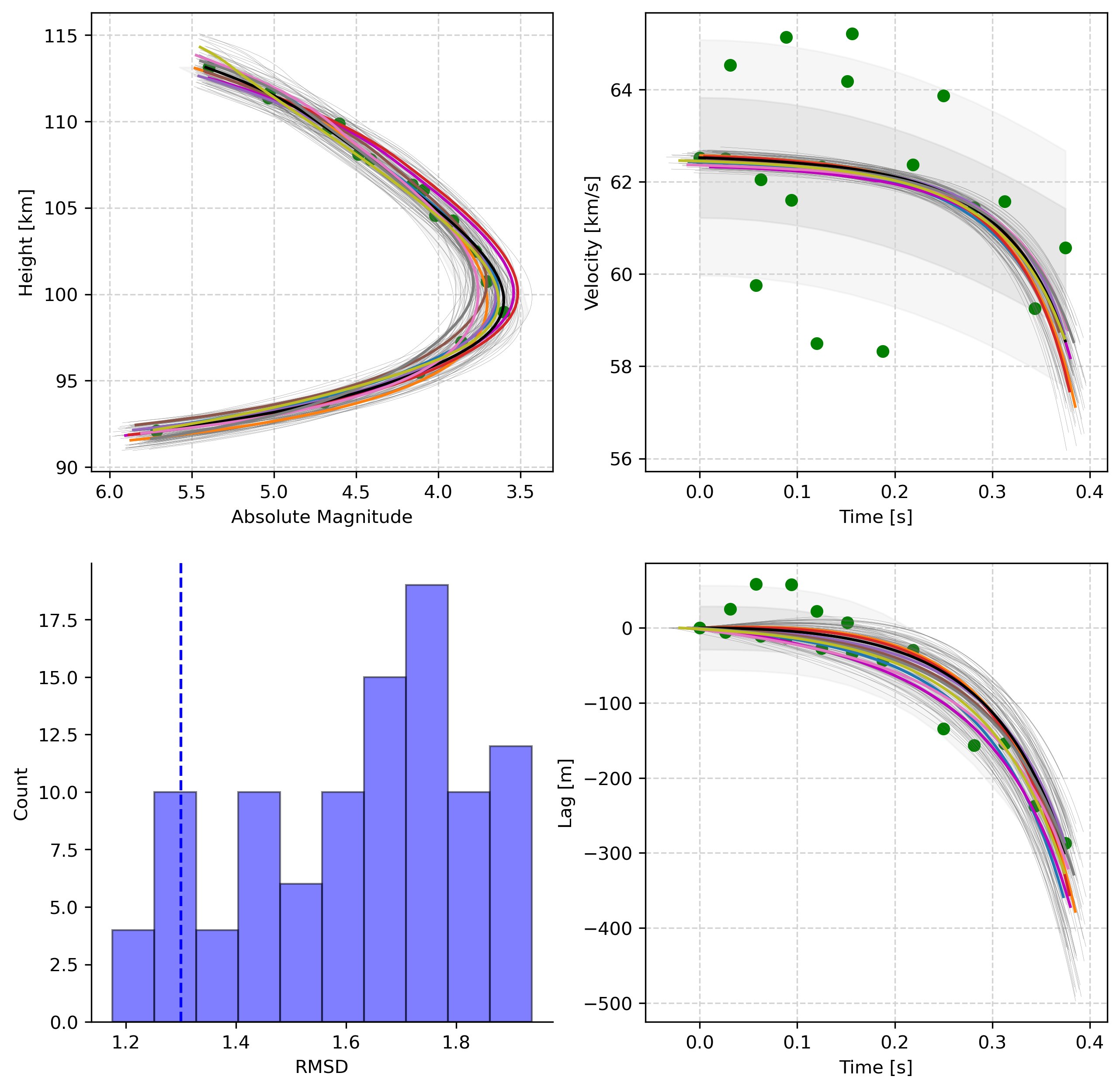}
\end{minipage}

\caption{Light curve, velocity, and lag profiles of the steep test case for the top 0.1\% of PCA distances (left) and for brute-force simulations below an RMSD threshold of 1.3 (right). The 10 best RMSD solutions in the PCA set have an average PC distance of 0.88 that is above the 1 percentile further proving the bad fit.}
\label{img:obs_steep_1}
\end{figure}

\begin{figure}[h!]
\centering
\begin{minipage}{0.82\linewidth}
    \centering
    \textbf{PCA} \\ 
    \includegraphics[width=\linewidth]{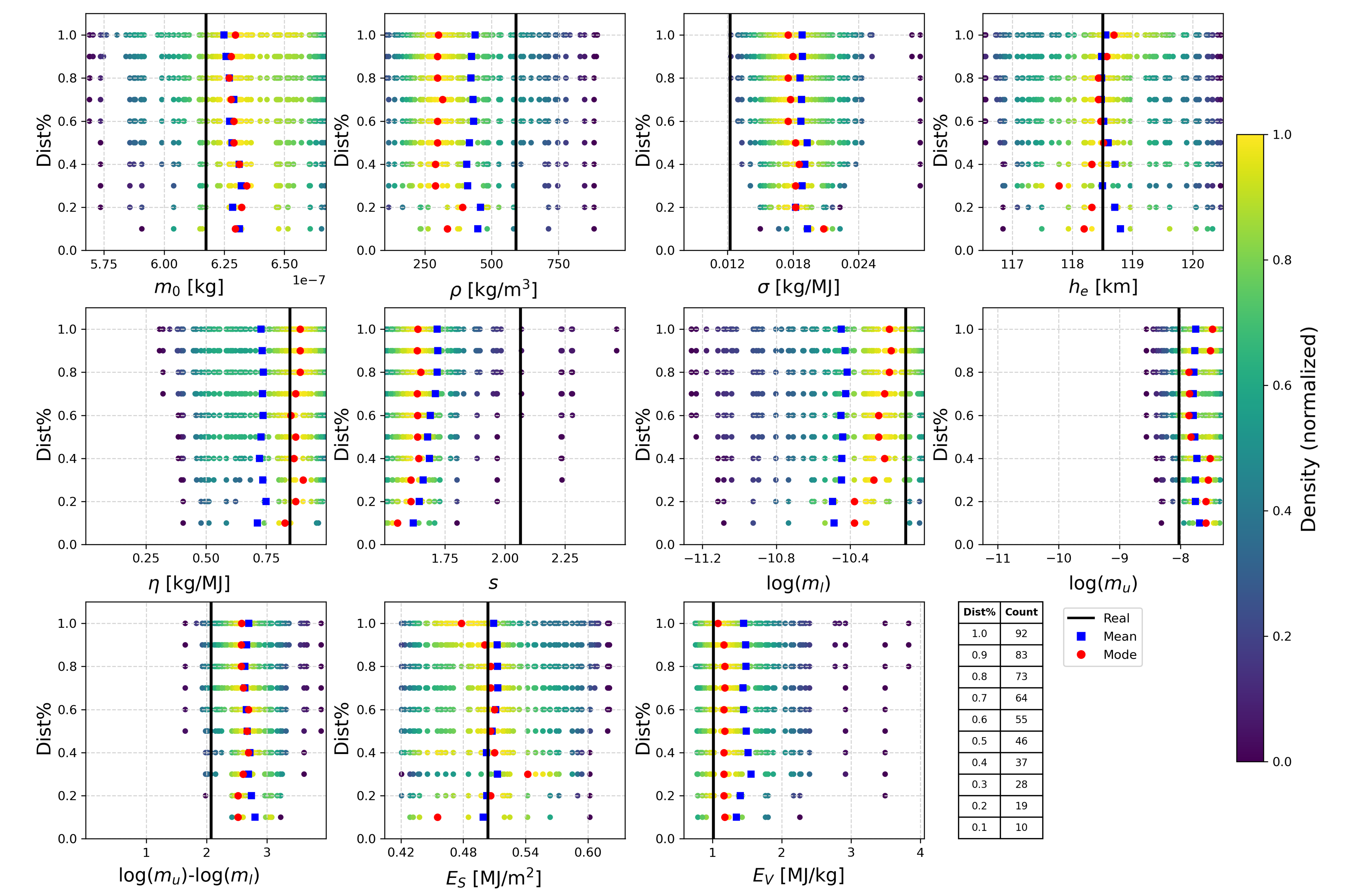}
\end{minipage}
\vspace{1em}
\begin{minipage}{0.82\linewidth}
    \centering
    \textbf{RMSD} \\ 
    \includegraphics[width=\linewidth]{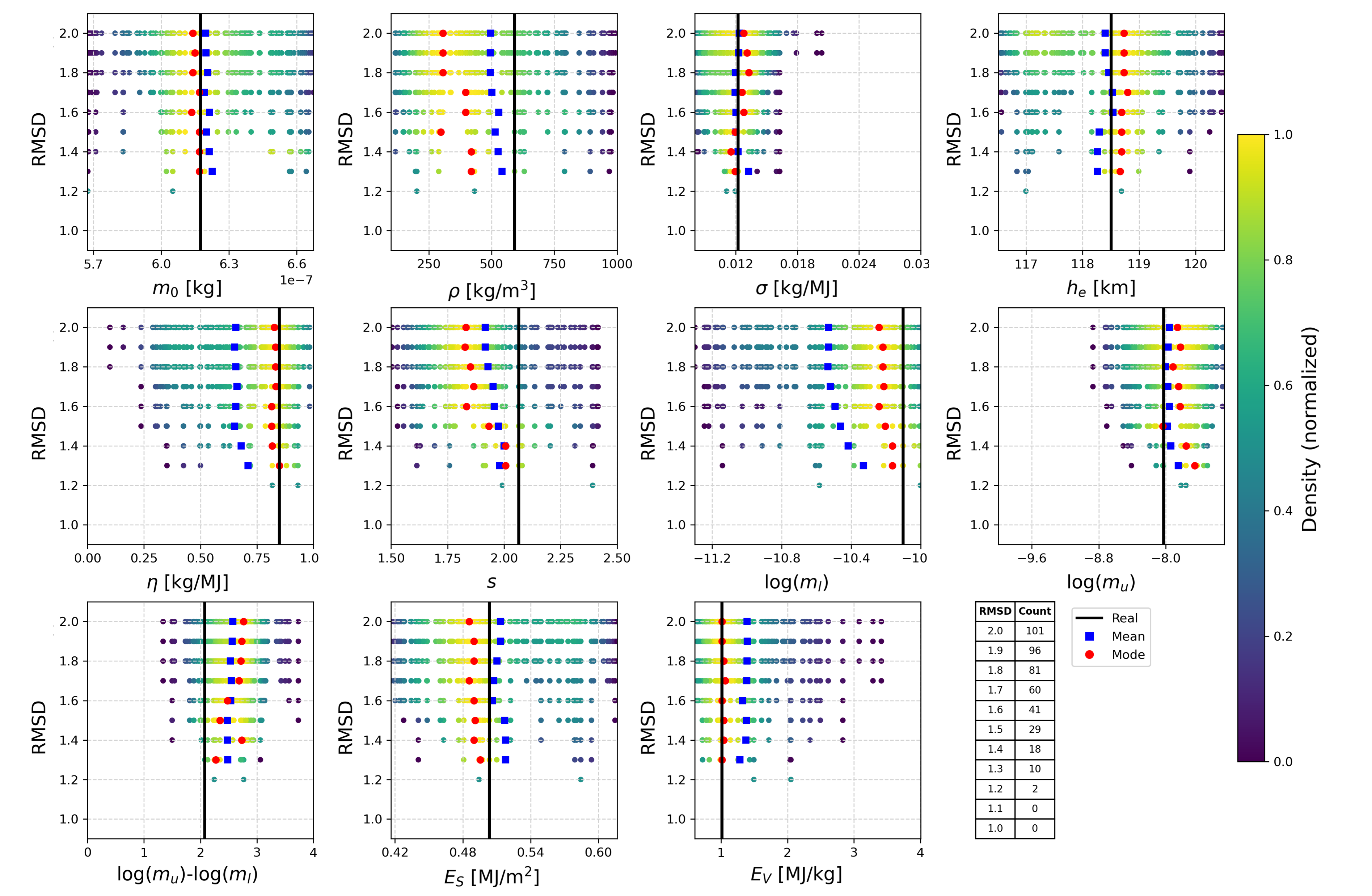}
\end{minipage}

\caption{Waterfall plot of the steep test case for the PCA method (top) and the RMSD results (bottom). In the PCA case the real ablation coefficient is outside from the limit of the confidence interval showing in this instance the PCA method converges to the wrong solution.}
\label{img:wat_steep_2}
\end{figure}

\newpage

\section*{Appendix B}\label{sec:Apx B}

Appendix B contains RMSD waterfall plots illustrating the properties of the two meteors \texttt{20210813\_061452} and \texttt{20230811\_082649} as processed independently by three different analysts. Presenting this information reveals how both EMCCD noise and analyst-based leading-edge picks can introduce biases that affect the calculated RMSD. 

The resulting wide confidence intervals highlight the critical need for high-precision cameras and unbiased, automated picking algorithms to reduce uncertainties in the derived physical characteristics of meteoroids.


\begin{figure}[h!]
\centering
\begin{minipage}{0.8\linewidth}
    \centering
    \textbf{20210813\_061452 Analyst A} \\ 
    \includegraphics[width=\linewidth]{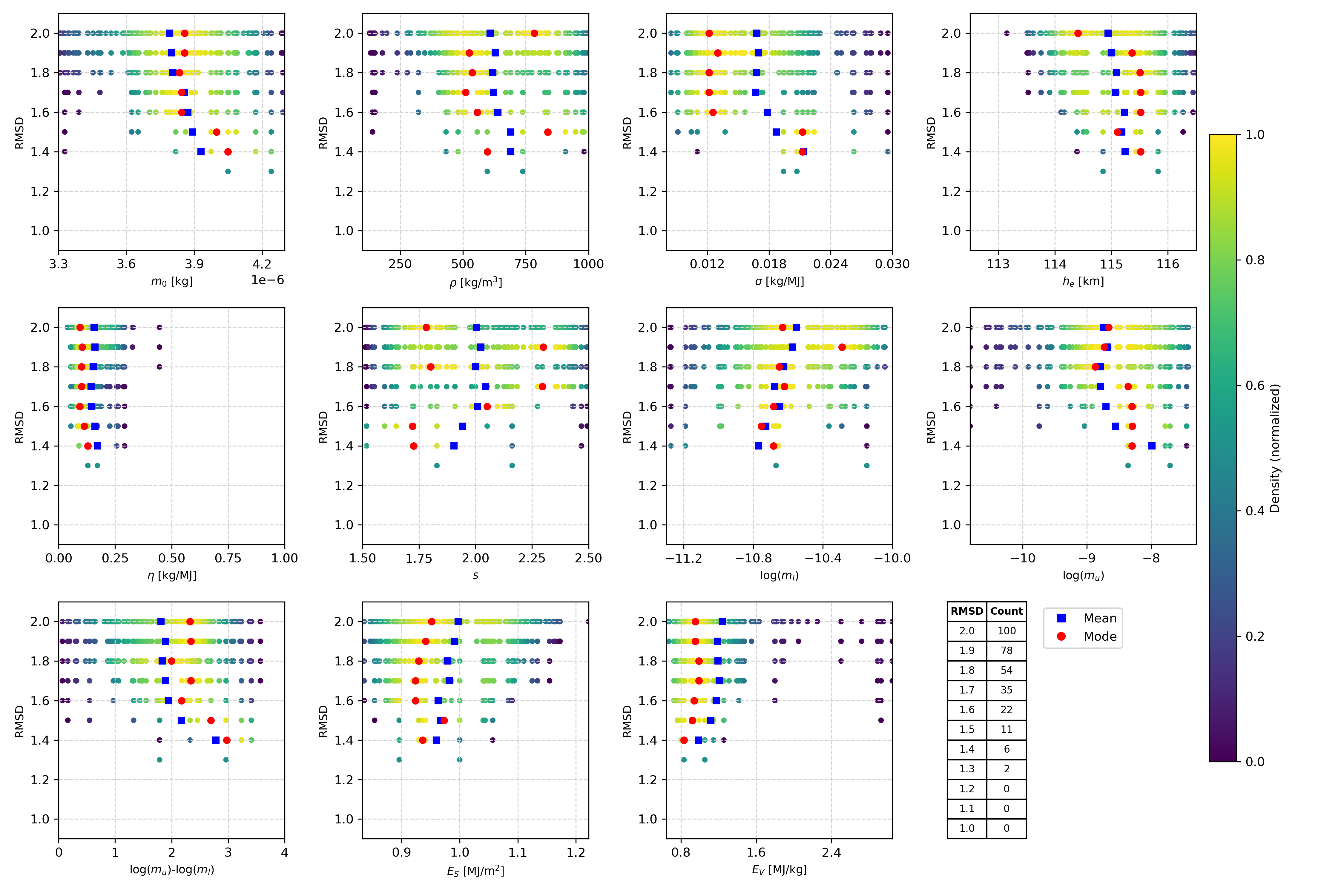}
\end{minipage}
\vspace{1em}
\begin{minipage}{0.8\linewidth}
    \centering
    \textbf{20230811\_082649 Analyst A} \\ 
    \includegraphics[width=\linewidth]{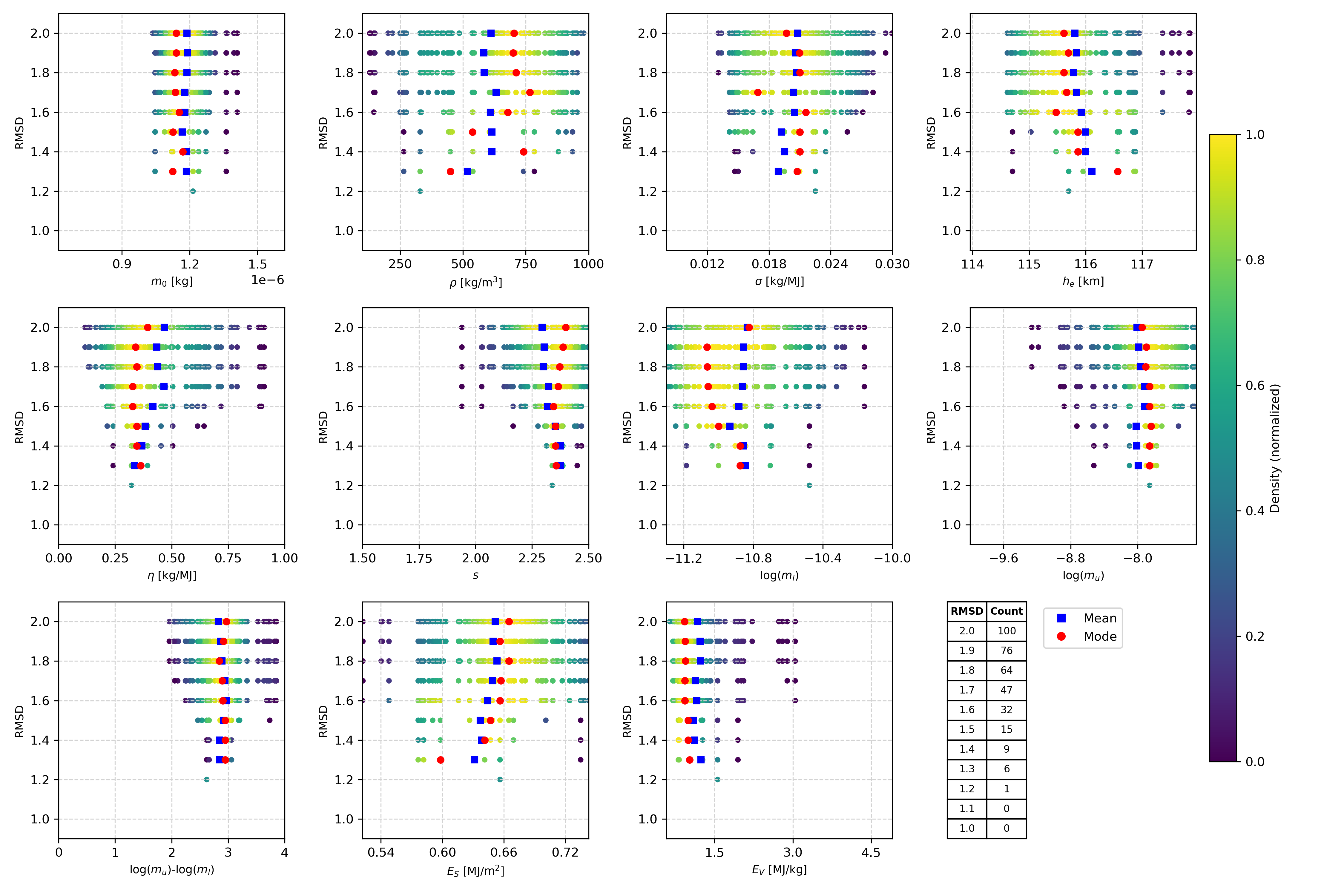}
\end{minipage}
\caption{RMSD-based waterfall plots for 100 selected simulations of meteors \texttt{20210813\_061452} (top) and \texttt{20230811\_082649} (bottom), both processed by Analyst~A using manual EMCCD picks.}
\label{img:wake2021}
\end{figure}

\begin{figure}[h!]
\centering
\begin{minipage}{0.8\linewidth}
    \centering
    \textbf{20210813\_061452 Analyst B} \\ 
    \includegraphics[width=\linewidth]{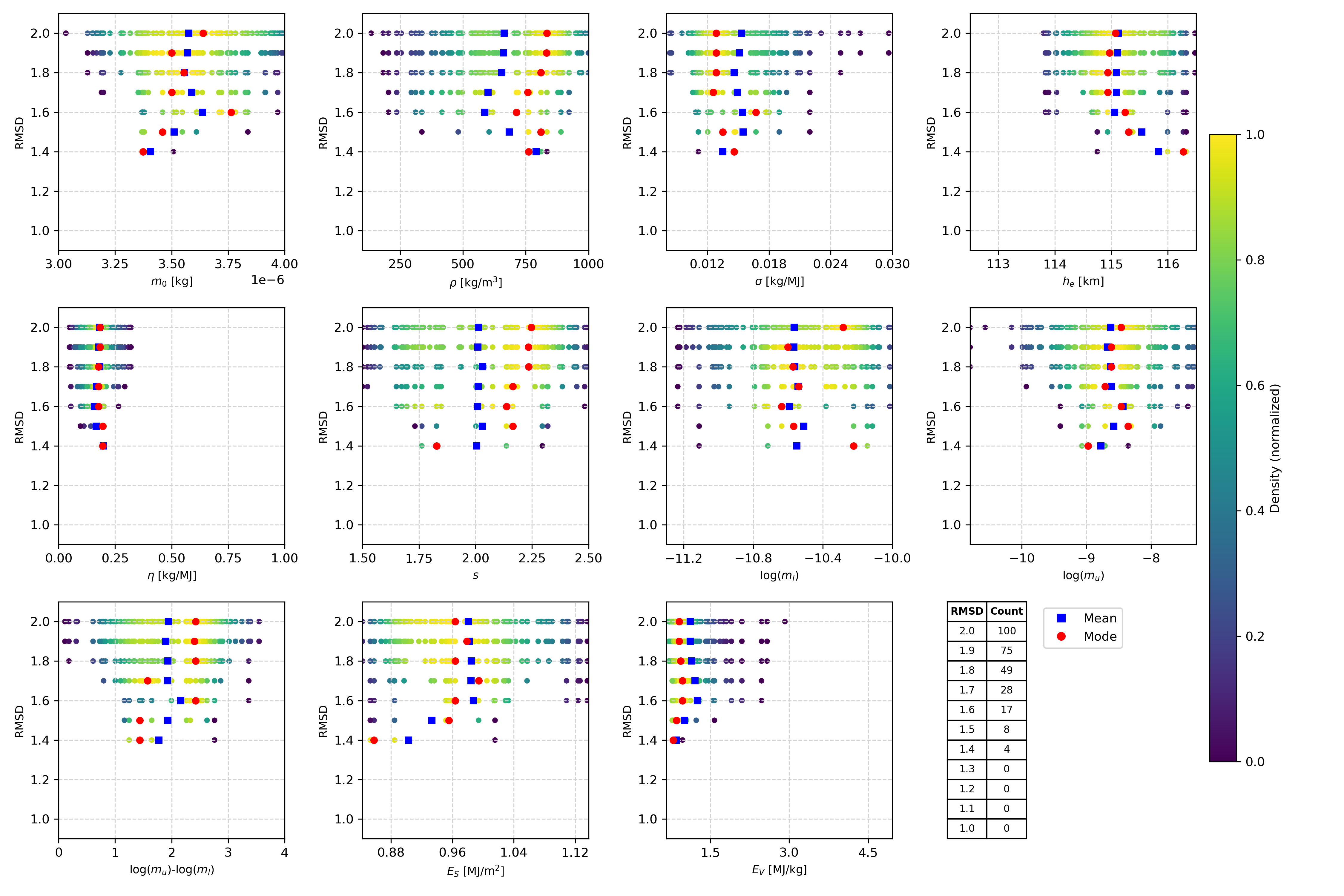}
\end{minipage}
\vspace{1em}
\begin{minipage}{0.8\linewidth}
    \centering
    \textbf{20230811\_082649 Analyst B} \\ 
    \includegraphics[width=\linewidth]{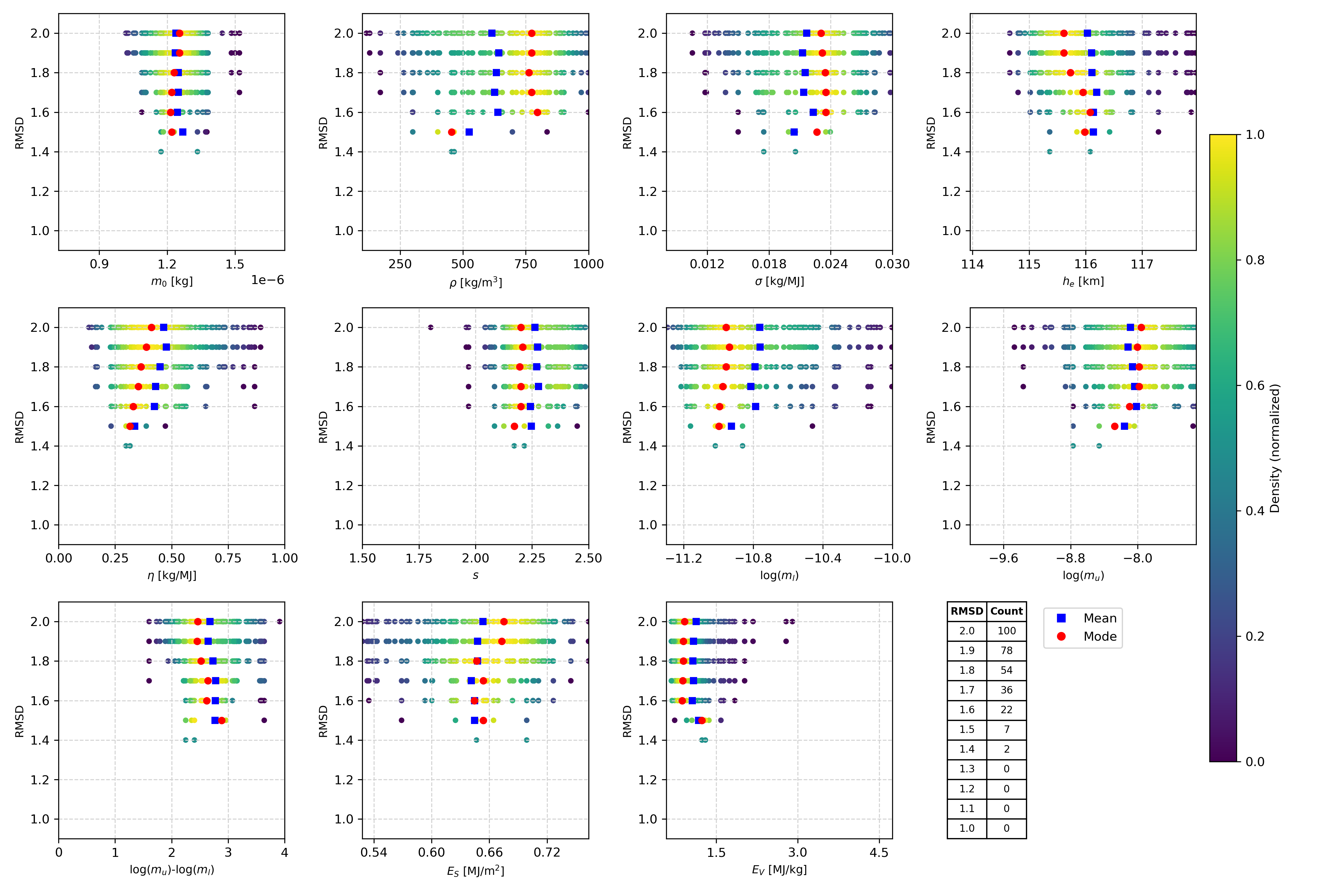}
\end{minipage}
\caption{RMSD-based waterfall plots for 100 selected simulations of meteors \texttt{20210813\_061452} (top) and \texttt{20230811\_082649} (bottom), both processed by Analyst~B using manual EMCCD picks.}
\label{img:wake2021}
\end{figure}

\begin{figure}[h!]
\centering
\begin{minipage}{0.8\linewidth}
    \centering
    \textbf{20210813\_061452 Analyst C} \\ 
    \includegraphics[width=\linewidth]{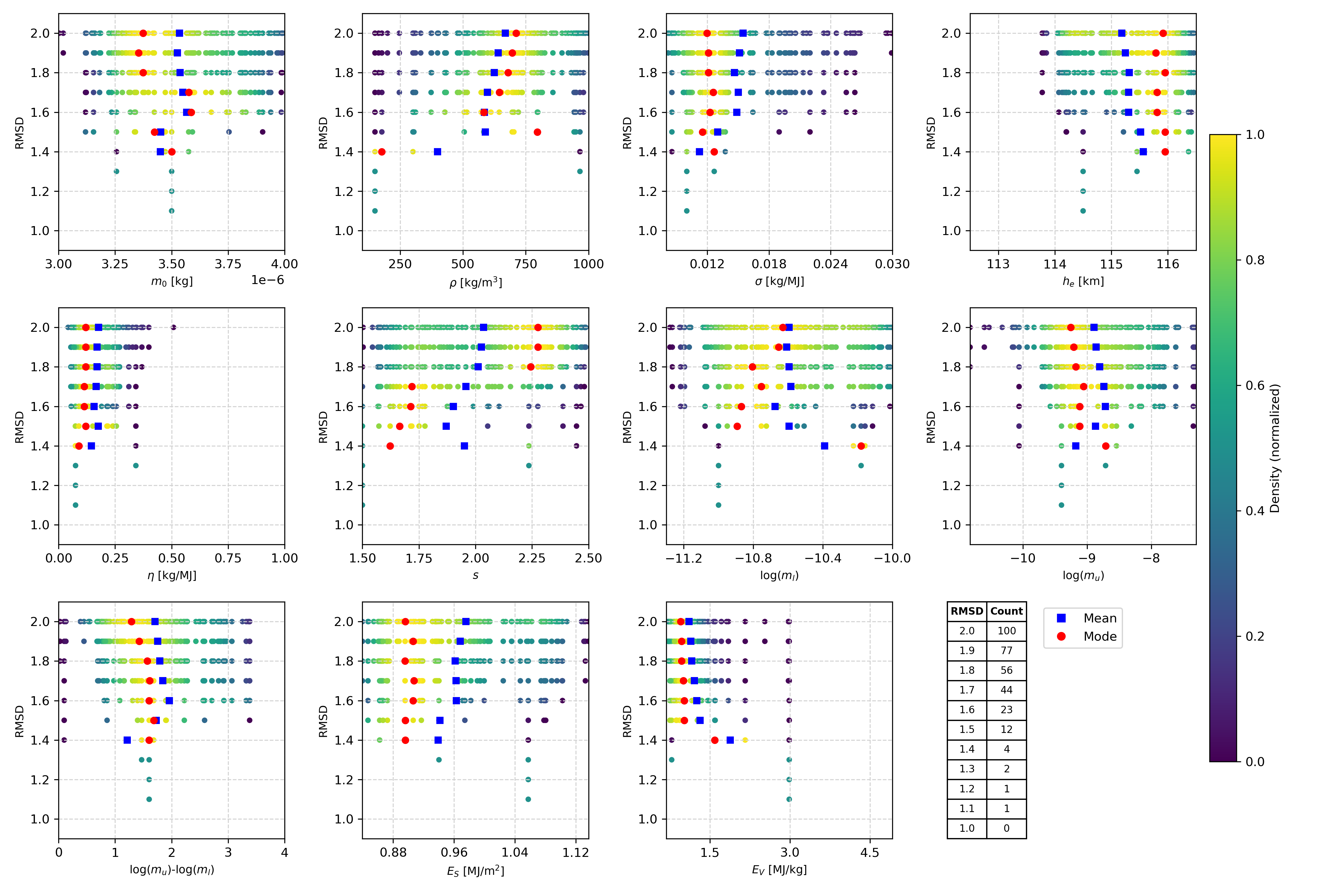}
\end{minipage}
\vspace{1em}
\begin{minipage}{0.8\linewidth}
    \centering
    \textbf{20230811\_082649 Analyst C} \\ 
    \includegraphics[width=\linewidth]{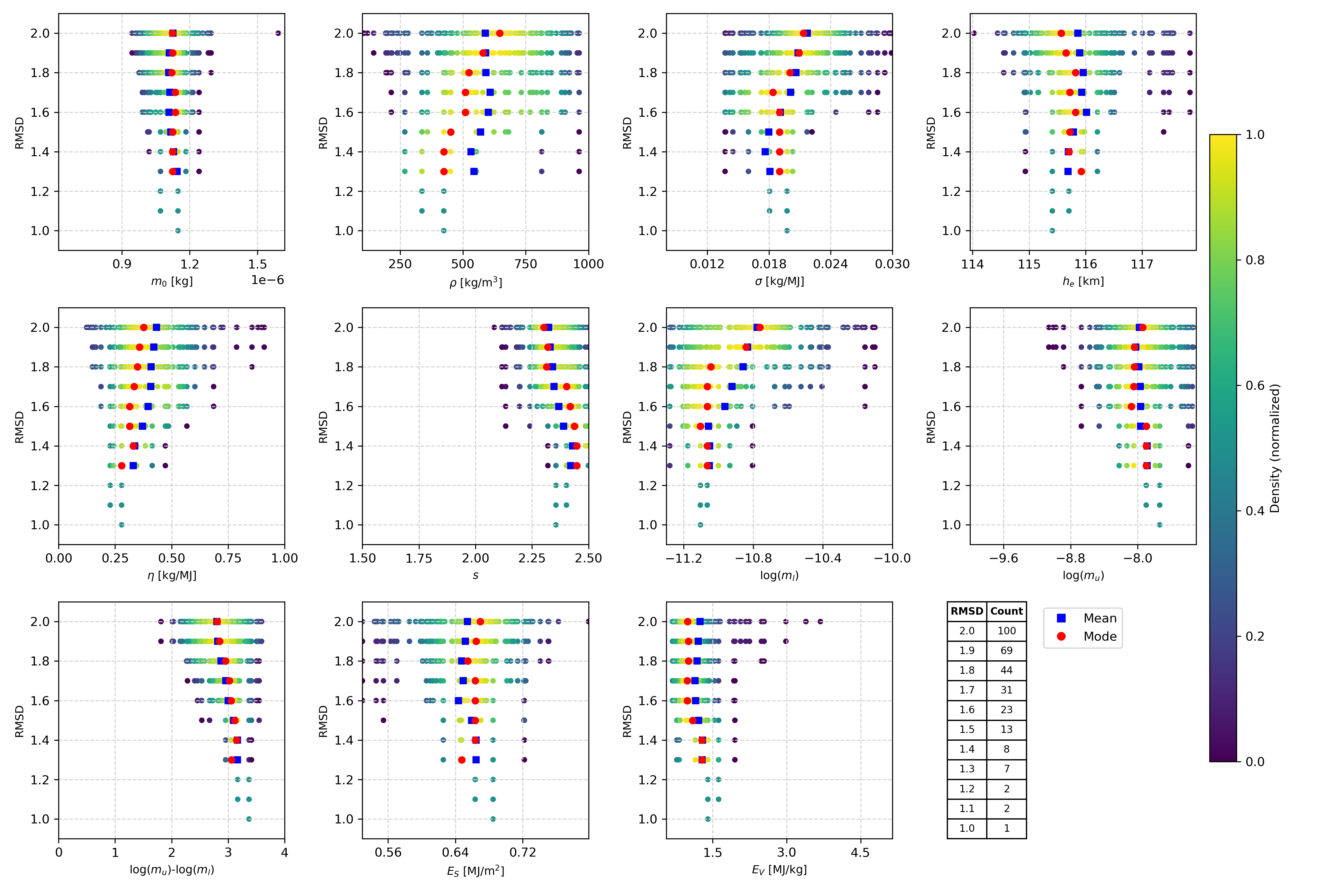}
\end{minipage}
\caption{RMSD-based waterfall plots for 100 selected simulations of meteors \texttt{20210813\_061452} (top) and \texttt{20230811\_082649} (bottom), both processed by Analyst~C using manual EMCCD picks.}
\label{img:wake2021}
\end{figure}

\newpage

\section*{Appendix C}\label{sec:Apx C}

Appendix C shows the wakes of the two best-fitting simulations, in terms of RMSD, for the CAMO meteors \texttt{20210813\_061452} and \texttt{20230811\_082649}. The two best simulations were selected from the 100 candidates by first normalizing each simulation's lag and magnitude RMSD by their respective measurement uncertainties, then summing those two normalized values. Each simulation was then ranked according to this total; the smaller the total, the better the overall fit for both lag and magnitude. 

In both cases, the simulated wakes closely match the observed CAMO data. This highlights the ability of our model to accurately capture meteoroid dynamical behavior even for observables not formally fit in the modeling process. 

\begin{figure}[h!]
\centering
\begin{minipage}{0.8\linewidth}
    \centering
    \textbf{20210813\_061452} \\ 
    \includegraphics[width=\linewidth]{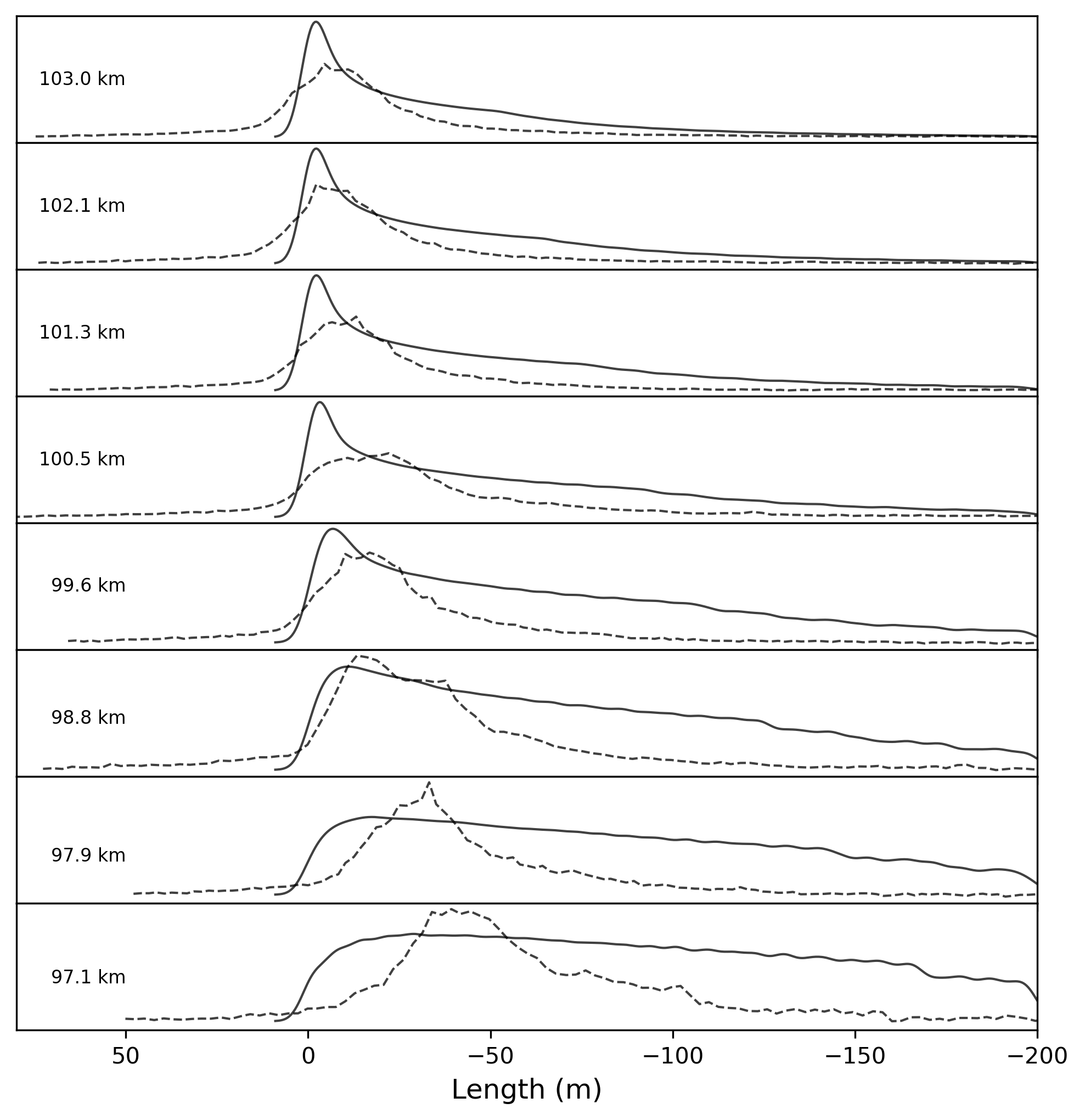}
\end{minipage}
\caption{Comparison between the CAMO-detected wake (dashed line) at various altitudes and the best-fitting simulated wake (solid line) from the RMSD-based selection for meteor \texttt{20210813\_061452}.}
\label{img:wake2021}
\end{figure}

\begin{figure}[h!]
\centering
\begin{minipage}{0.8\linewidth}
    \centering
    \textbf{20230811\_082649} \\ 
    \includegraphics[width=\linewidth]{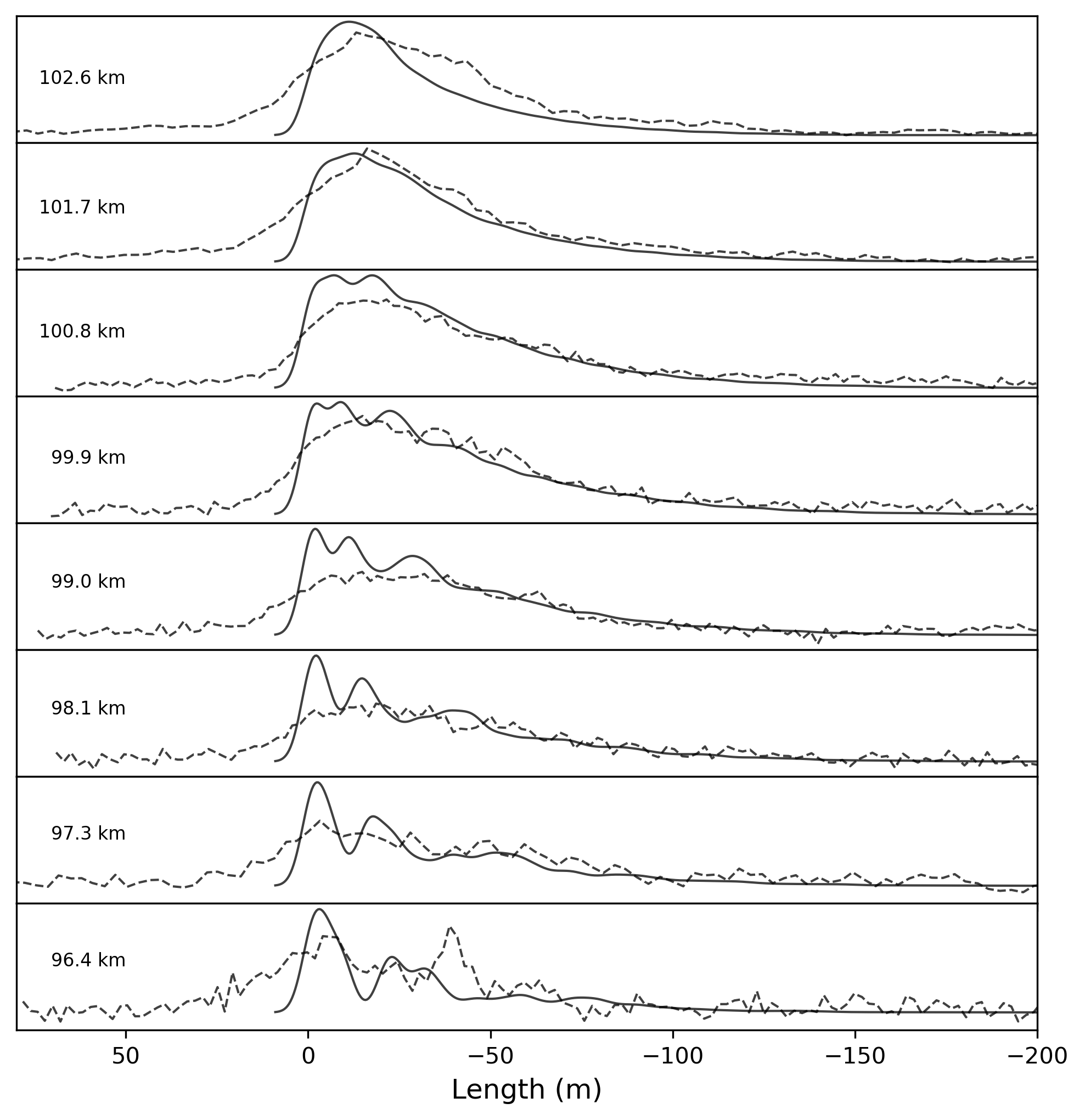}
\end{minipage}
\caption{Comparison between the CAMO-detected wake (dashed line) at various altitudes and the best-fitting simulated wake (solid line) from the RMSD-based selection for meteor \texttt{20230811\_082649}.}
\label{img:wake2023}
\end{figure}

\end{document}